\documentclass[authoryear, 11pt, review]{elsarticle}
\makeatletter
\def\ps@pprintTitle{%
   \let\@oddhead\@empty
   \let\@evenhead\@empty
   \let\@oddfoot\@empty
   \let\@evenfoot\@oddfoot
}
\makeatother
\usepackage{rotating}
\usepackage[letterpaper,hmargin=0.6in,vmargin=1in]{geometry}
\usepackage[flushleft]{threeparttable}
\usepackage{float,lscape}
\usepackage{mwe}
\usepackage{kantlipsum}

\usepackage[english]{babel}
\usepackage{xcolor}
\usepackage{subcaption}
\usepackage{lipsum}
\usepackage{pdfpages}
\usepackage{graphicx}
\usepackage[export]{adjustbox}
\usepackage{caption}
\usepackage{subcaption}
\usepackage{natbib}
\usepackage{enumerate}
\usepackage{amsmath,url}
\usepackage{amsmath,amsthm,amssymb}
\usepackage[colorlinks=true,citecolor=blue]{hyperref}
\usepackage{titletoc}
\usepackage{fancybox}
\usepackage{bm}
\usepackage{rotating}
\usepackage{tikz}
\makeatletter

\newcommand{\Rmnum}[1]{\expandafter\@slowromancap\romannumeral #1@}
\makeatother
\usepackage{tikz} 
\usetikzlibrary{positioning,shapes,shadows,arrows, shapes.geometric}
\newcommand{\beginsupplement}{%
        \setcounter{table}{0}
        \renewcommand{\thetable}{S\arabic{table}}%
        \setcounter{figure}{0}
        \renewcommand{\thefigure}{S\arabic{figure}}%
     }

\newtheorem{remark}{Remark}

\usepackage{etoolbox}
\makeatletter
\patchcmd{\@verbatim}
  {\verbatim@font}
  {\verbatim@font\small}
\makeatother

\input epsf.sty
\usepackage{lineno}
\linenumbers

\begin{document}

\title{A New Method to Determine the Presence of Continuous Variation in Parameters of Biological Growth Curve Models}

\author[1]{Md Aktar Ul Karim}
\ead{mdaktarulkarim829@gmail.com}
\author[1]{Supriya Ramdas Bhagat }
\ead{supriyabhagat71@gmail.com}
\author[1]{Amiya Ranjan Bhowmick\corref{cor1}}
\ead{amiyaiitb@gmail.com}
\cortext[cor1]{Corresponding author}
\fntext[1]{Department of Mathematics, Institute of Chemical Technology, Mumbai}

\begin{abstract}
Quantitative assessment of the growth of biological organisms has produced many mathematical equations and over time it has become an independent research area. Many efforts have been given on statistical identification of the correct growth model from a given data set. This generated several model selection criteria as well. Every growth equation is unique in terms of mathematical structures; however, one model may serve as a close approximation of the other by appropriate choice of the parameter(s). It is still a challenging problem to select the best estimating model from a set of models whose shapes are similar in nature. Our efforts in this manuscript are to reduce the efforts in model selection by utilizing an existing model selection criterion in an innovative way that reduces the number of model fitting exercises substantially. In this manuscript, we have shown that one model can be obtained from the other by choosing a suitable continuous transformation of the parameters. This idea builds an interconnection between many equations that are available in the literature. As the by product of this exercise, we also get several new growth equations, out of them large number of equations can be obtained from a few key models. Given a set of training data points and the key models, we utilized the idea of interval specific rate parameter (ISRP) proposed by \citet{Bhowmick2014} to obtain a suitable mathematical model for the data. The ISRP profile of the parameters of simpler models indicates the nature of variation in parameters with time, thus, enable the experimenter to extrapolate the inference to more complex models. Our proposed methodology significantly reduced the efforts involved in model fitting exercises. Connections have been built amongst many growth equations, which were studied independently to date by researchers. We believe that this work would be helpful for practitioners in the field of growth study. The proposed idea is verified by using simulated and real data sets. In addition, theoretical justifications have been provided by investigating the statistical properties of the estimators.   
\end{abstract}


\maketitle
\newpage
\tableofcontents
\section{Introduction}
Growth curve models serve as the mathematical framework for the qualitative studies of growth in many areas of applied science and due to its extensive use in the recent studies several new models were developed over a long period of time  \citep{TSOULARIS2002, Bhowmick2014}. There are many important applications of such models, viz. modelling of physiological age of animal or group of animals with respect to time \citep{Bridges2000}; study of growth comparison for different genotypes with respect to an expected growth behaviour \citep{perotto1992}; predicting the extinction pattern in natural populations \citep{BHOWMICK2015}. In the growth curve analysis the parameters in the model are assumed to be fixed but unknown quantity. So if $\frac{dX(t)}{dt} = f(X(t); \beta); t \geq 0$, represents a growth process, then the parameter $\beta$ (scalar or vector valued) is assumed to be a constant; $X(t)$ is size at time $t$. The parameters are estimated by using non-linear least squares method that provides the confidence interval based on $t$-distribution. However, it might happen that the parameter is not fixed and vary with respect to time \citep{Banks1994}. Now the problem is, if the experimenter has observed data over a time period then there is an uncertainty about the parameter being fixed or varying with time. Even if the information is known (from the biological theory) that a particular model parameter varies with time, but it's empirical estimation is a difficult task. The goal of this manuscript: \textbf{(1) to revisit the theory of parameter variation in the context of growth curve models and demonstrate its importance in growth curve analysis; (2) to develop statistical methods to detect whether any parameters of the growth model has undergone continuous variation with time by using the data; (3) To develop an appropriate way for the best model selection to obtain better insights about the growth process. }

To understand the research problem, studied in this paper, we consider the exponential model as the test bed. The exponential growth model is given by $\frac{dX(t)}{dt } = rX(t)$, where $r$ is the rate parameter. Now, suppose if we artificially simulate observations $X_1,X_2,\ldots,X_n$ using the gompertz growth model $\left(\frac{dX(t)}{dt } = r_0 e^{-\alpha t}X(t);~r_0,~ \alpha>0 \right)$ and plot the corresponding RGR values (logged difference in $X_i$'s) against time. The scatterplot would appear to be a monotonically decreasing function of time. Given the mathematical relationship between gompertz and exponential model, the scatterplot can be described ``as an exponential model with continuously decaying rate parameter $r$". Fortunately, the inter-relation between exponential and gompertz is known \citep{Banks1994, Kot2001}. Hence, it can be easily guessed that the parameter $r$ is decreasing with time. Thus, looking into the behaviour of RGR, we can guess whether the parameter $r$ is varying or not. The nature or pattern of variation in $r$ with time allow us to choose the correct continuous function $r(t)$. The problem is that such inter-relations are very difficult to be identified between growth models and may become mathematically cumbersome as well. Hence, it is difficult to identify the type of variation in the parameter empirically, and most importantly, difficulty in choosing the final transformed model after the type of variation being considered in the parameter. The problem becomes more difficult if multiple parameters are present in the model and one or few of them vary with time. In this paper, we attempt to solve this problem with the help of Interval Specific Rate Parameter (ISRP) proposed by \citet{Bhowmick2014}, which is briefly discussed later in this manuscript.


The organization of the rest of the paper is as follows: In Section~\ref{Literature_survey},  we briefly describe the development of extended families of growth models by varying parameters. Then in Section~\ref{Model_extension}, we introduce the concept of continuously varying parameters in four different models, namely exponential, logistic, theta-logistic and confined exponential. Also we explore possible connections between growth models by varying parameters continuously. The statistical properties of ISRP of the model parameters with a simulation study for parameters of logistic growth model and their continuous variation is described in Section~\ref{ISRP_and_statistical_identification}. In Section~\ref{Real-data_analysis}, the utility of proposed method has been demonstrated by using three real life data sets, namely (a) cattle growth data \citep{Kenward1987}, (b) cumulative sales of LCD-TV data \citep{TRAPPEY2008} and (c) cumulative number of COVID-19 cases for Germany (\href{https://ourworldindata.org/coronavirus}{https://ourworldindata.org/coronavirus}). Lastly we conclude the discussion with some remarks and possible future directions. The mathematical form of distribution of ISRP of the parameters for some growth models have been provided in the appendix.

\section{Literature survey} \label{Literature_survey}
In the literature, the idea of considering a continuously varying parameters with time probably dated back to the study of experimental data by \citet{Utida1957} in a host-parasitoid system. \citet{Turner1969} were the first one to consider an explicit time dependent function for the carrying capacity in the logistic growth model \citep{Verhulst1838}. The authors assumed the carrying capacity $k(t)$, at time $t$, to be an increasing function with a slower rate parameter $B$ and $k(t) \to K$ as $t \to \infty$, where $K$ is the maximum population size. Basically, the functional form of $k(t)$ is logistic with a slower rate of increase as compared to the rate parameter of underlying logistically growing population and the modified growth model was applied to the US population growth for illustration. \citet{TURNER1976} further proposed a ``generic growth model" from which many commonly known growth equations can derived as a special or limiting case. This is probably the first unification of existing growth equations, in the sense of \citet{Chakraborty2019}, that provided a compact representation of a class of growth functions (Fig. 1, \citet{TURNER1976}). Consideration of varying parameters with time was not only limited to the dynamics of single populations, but, to the predator prey models \citep{Cushing1977} and higher dimensional models as well \citep{IKEDA1980}. Using the bifurcation theory, \citet{Cushing1977} investigated the dynamics of predator-prey models by modifying the birth rate of prey and death rate of predator as periodic functions of time. 

It is interesting to note that in the literature, the logistic growth has received huge attention from the researchers and several studies have been carried out by parametrizing the two fundamental parameters, intrinsic growth rate ($r$) and the carrying capacity of the environment ($K$). \citet{COLEMAN1979}  was the first to consider the aperiodic time-dependent functional form for both $r$ and $K$ that essentially makes autonomous logistic growth equation to a non-autonomous one. The author basically considered two types of variation in $K$: in one situation, $K$ varies slowly with time and in the other scenario, $K$ fluctuates at an arbitrary rate, but remains close to a constant. Further investigation along this direction was carried out by \citet{HALLAM1981} who reparametrized the equations proposed by \citet{COLEMAN1979} to model the population with small growth rates in a deteriorating environmental conditions. \citet{BECK1982} considered logistically growing $K$ to model the population genetics of cystic fibrosis disease and probably, he was the first one to consider time-dependent form of $K$ to model diseases population genetics. Another application of varying parameter was considered by \citet{Ebart1997} to model the parasite growth in host body in which the authors considered logistically growing carrying capacity with time. In the context of habitat selection of marine fishes and associated ideal free distribution, \citet{shepard2004} discussed the biological implications of three different possibilities of variation in $r$ and $K$, namely: (1) constant $r$ and variable $K$, (2) variable $r$ and constant $K$ and (3) varying $r$ and $K$. Use of continuously varying parameters in growth modelling is not only limited to the biological systems, but, it has been used in other domains as well. \citet{TRAPPEY2008} analyzed 22 time series data of electronic products and showed that time varying logistic growth models (with $k(t) = 1-de^{-ct}$) gives 70\% better performance than simple logistic and gompertz models for short product life cycle datasets.

There are some important innovations in growth curve modelling. For example, \citet{SHARIF1981} combined the confined exponential model \citep{Banks1994} and logistic growth equation to describe the time pattern of the diffusion process for technological innovations. The author observed a serious limitation due to assumption of constant population of potential adopters (similar to the carrying capacity) in the model. The limitation was overcome by considering the number of potential adaptors to vary with time assuming some known functional form. According to \citet{MEYER1999}, the technological adoption process can be modelled by logistic growth equation. The author used the data on technological evolution from two countries, England and Japan, and showed that a sigmoid type function for $K$ was a better choice to analyze the data than the assumption of constant $K$. Following the idea of \citet{MEYER1999}, \citet{safuan2011} assumed the confined exponential function for $K(t)$ to study the growth of microbial biomass under occlusion of the skin. The author later obtained the exact solution of the non-autonomous logistic equation considering time dependent form of $K$ \citep{SAFUAN2013}. A schematic diagram of the literature is depicted in Table~\ref{table1}.

\begin{sidewaystable}
 \caption{The table demonstrates the evolution of growth curve models with continuously varying parameters available in literature.}
\begin{tabular}{ |p{3.5cm}p{3.5cm}p{4.3cm}p{3.8cm}|p{6.7cm}|
}
 \hline
Reference & Growth model & Varying parameter  & Type of variation & Biological justification or explanation\\
\hline
\citet{Utida1957} & Host-parasite model & Rate of reproduction ($R$) & Density dependent  & Impact of fluctuating environment on $R$\\
 \hline
\citet{Turner1969} & Generalized logistic model & Maximum population size ($k$) & Sigmoidally varying with time  & $k$ depends on technological advances such as housing and food sources\\
 \hline
\citet{nisbet1976} & Logistic delay model & Intrinsic growth rate ($r$), carrying capacity ($K$) & Periodically varying with time  & Impact of periodically varying environment\\
 \hline
\citet{Cushing1977} & Predator-prey model & Net birth rate of prey and predator ($b_1$ and $-b_2$) & Periodically varying with time  & Impact of oscillating behaviour of the environment\\
 \hline
\citet{COLEMAN1979} & Logistic model & Intrinsic growth rate ($r$), carrying capacity ($K$)  & Time dependent & Modelling the effect of environmental changes on populations\\
 \hline
\citet{IKEDA1980} & Theoretical model of fish population & Nutrient amount, carrying capacity, death rates & Increasing function of pollution load & impact of nutrient enrichment and pollution on fish population dynamics\\
 \hline
\citet{SHARIF1981} & Binomial innovation diffusion model & Population adaptors ($N$)  & Time dependent & Demands for innovation changes with time due to some factors\\
 \hline
\citet{HALLAM1981} & Logistic model & Intrinsic growth rate ($r$), carrying capacity ($K$) & Time dependent & Impact of deteriorating environment\\
 \hline
\citet{BECK1982} & Logistic model & Carrying capacity ($K$) & Logistically growing with time & Modelling the population genetics of cystic fibrosis\\
 \hline
\citet{arrigoni1985} & Logistic model & Carrying capacity ($K$) & Periodically varying with time & Impact of fluctuating environment\\
 \hline
\end{tabular}
\label{table1}
\end{sidewaystable}

\begin{sidewaystable}
\begin{tabular}{ |p{3.5cm}p{3.4cm}p{4.3cm}p{3.8cm}|p{6.9cm}|  }
 \hline
Reference & Growth model & Varying parameter  & Type of variation & Biological justification or explanation\\
\hline
\citet{Ebart1997} & Logistic model & Carrying capacity ($K$) & Sigmoidally varying with time & Modelling the parasite growth\\
\hline
\citet{MEYER1999} & Logistic model & Carrying capacity ($K$) & Sigmoidally varying with time & Carrying capacity of a system depends on invention and diffusion of technologies\\
\hline
\citet{LAKSHMI2003} & Malthusian growth model & Maximum sustainable population ($M$) & Periodically varying with time & Modelling the impact of oscillating population dynamics of a system\\
\hline
\citet{LEACH2004} & Verhulst model & Natural rate of replication ($r$), carrying capacity ($M$) & Periodically varying with time & To consider temporal variation of
carrying capacity and replication rate\\
\hline
\citet{shepard2004} & Population density models, Constant density models and  Basin models & Intrinsic growth rate ($r$), carrying capacity ($K$) & Density dependent & Impact of habitat selection, ideal free distribution and environmental effects\\
\hline
\citet{LAKSHMI2005} & Malthusian growth model & Maximum sustainable population ($M$) & Periodically varying with time & Oscillating population model\\
\hline
\citet{ROGOVCHENKO2009} & Pearl–Verhulst model & Intrinsic growth rate and carrying capacity & Periodically varying with time & Modelling the effect of periodic environmental fluctuations\\
\hline
\citet{safuan2011} & Logistic model & Carrying capacity ($K$) & Confined exponential function of time & Modelling the drastically changes in cutaneous  bacteria population\\
\hline
\citet{SAFUAN2013} & Logistic model & Carrying capacity ($K$) & Time dependent & Impact of environmental changes\\

\hline
\end{tabular}\\

\footnotesize{The stochastic variation of parameters is not being considered here.}
\end{sidewaystable}

The time dependent variations in the model parameters can be classified into two broad categories: periodic and non-periodic. Till now we have discussed the literature available for non periodic variations of time only. We now mention some literature in which the parameters were assumed as periodic function of time. \citet{nisbet1976} was the first to consider periodic functional form in the single population dynamics. Considering time varying periodic functional form for both $r$ and $K$ in logistic delay model with time delay $\tau$, the author investigated the dynamics numerically. \citet{arrigoni1985} further investigated the logistic growth model considering periodically varying functional form of $K$. \citet{LAKSHMI2003} considered the time varying parameter `maximum sustainable population' (which is similar to carrying capacity $K$), $M$ as: $M(t) = \sin{t}$, in the Malthusian growth model \citep{Freedman} to model the oscillating population dynamic system. Prompted by the work done by \citet{LAKSHMI2003}, \citet{LEACH2004} also considered periodic time-varying parameters to introduce the possibility of chaos into the Verhulst model \citep{Verhulst1838}. Also considering various forms of $M$ and $r$, the author solved the model and investigated the behaviour, stability and properties of solution with comparison of the solution obtained by \citet{LAKSHMI2003}. Following \citet{LEACH2004}, \citet{LAKSHMI2005} extended her proposed work in \citet{LAKSHMI2003} and showed that $W(t) = P(t) + \frac{M(t)}{2} $ is also periodic for any periodic form of $M(t)$, where $P(t)$ is the population at given time $t$. \citet{ROGOVCHENKO2009} modified the work done by \citet{LAKSHMI2003}, \citet{LEACH2004} and \citet{LAKSHMI2005} and briefly examined the periodic variation in intrinsic growth rate ($r$). They assumed the model framework proposed by \citet{HALLAM1981} and examined the existence of a unique positive asymptotically stable periodic solution of the system for $r(t) > 0$ for all $t$. 

The literature on growth curves is not only limited to the variation in parameters with respect to time; several studies are also available in which the size dependent variations are also explicitly considered. For example, \citet{LOPEZ2004} used density dependent variations in parameters in a predator-prey system. Also there are evidence of stochastic variation in  parameters as well \citet{Dubkov2008, MENDEZ2010, anderson2015, Yoshioka2019}. However we do not delve deep into the stochastic analogue in this manuscript.


\section{Model extension by continuous variation of parameters} \label{Model_extension}
\begin{sidewaysfigure}
\vskip5in
\begin{tikzpicture}
\centering
 \includegraphics[scale=0.65]{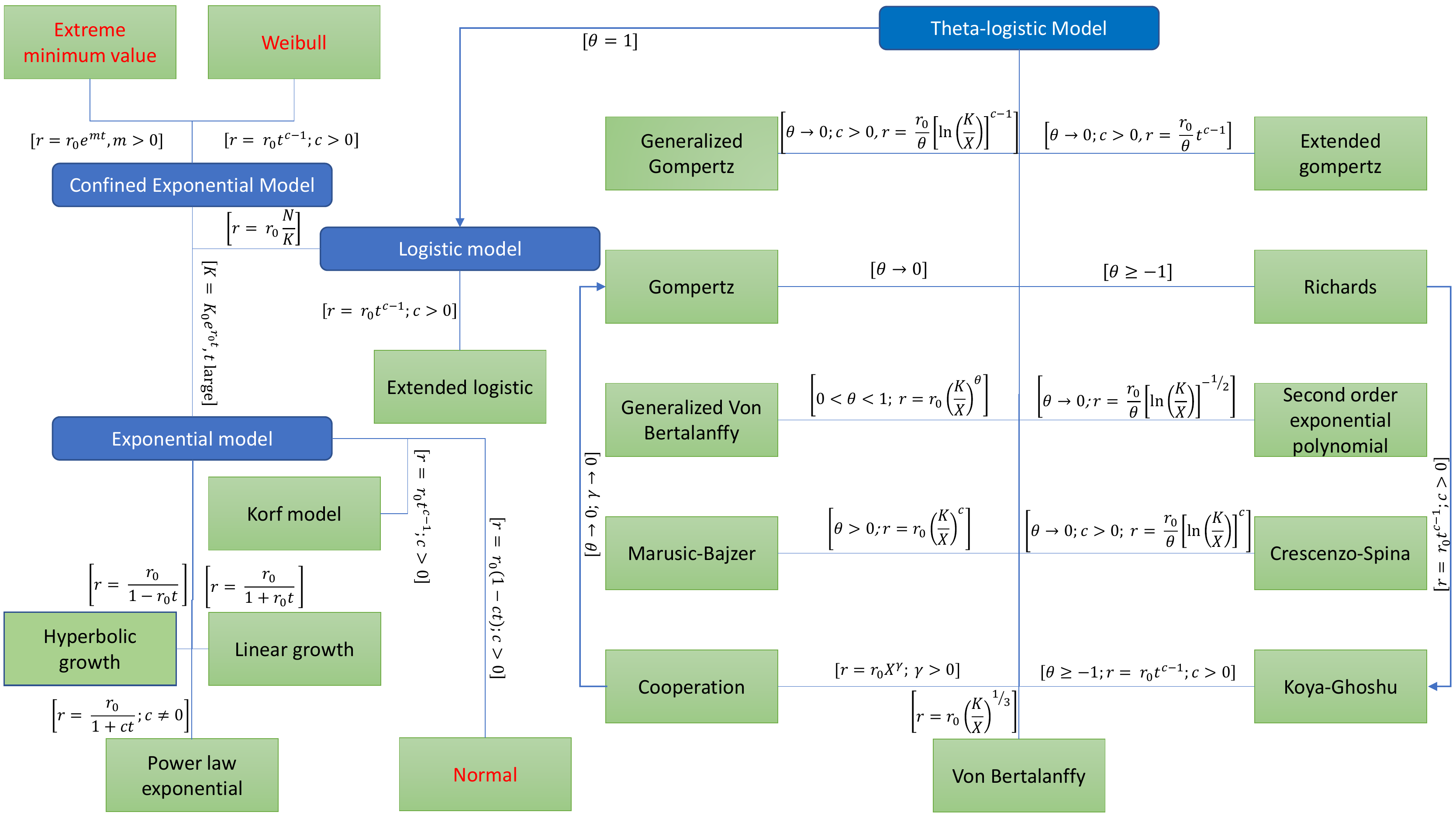}
 \end{tikzpicture}
\caption{The flowchart demonstrates a road-map how we can interchange between models only considering the continuous variation in the parameters. Here taking four growth models (a) exponential model, (b) logistic model, (c) confined exponential model and (d) theta-logistic model as the parent models and varying the parameters continuously in various ways we demonstrate the road-map. In this manner, we are also able to find some well known distribution functions, which are represented by red lettering in the above flowchart.}
 \label{Relation_flowchart_between models}
\end{sidewaysfigure}

\citet{Banks1994} discussed a number of models in which the parameters varied continuously. In this section, we find the possible relationship between two growth models by varying parameter with time (continuously) and density. We have considered four different models as basic models, namely: (a) exponential growth model, (b) logistic growth model, (c) theta-logistic growth model and (d) confined-exponential growth model. The details analysis is given below and the findings are also represented diagrammatically in Fig.~\ref{Relation_flowchart_between models}.

\begin{sidewaystable}
\caption{Different time dependent functions for the parameter $r$ of exponential growth model ($c$ and $\omega > 0$).}
\begin{tabular}{ |p{0.6cm}|p{3.6cm}p{5.6cm}p{4.3cm}p{6.5cm}|  }
 \hline
 \multicolumn{5}{|c|}{Parent Model: Exponential Model} \\
 \hline
 Srl. & $r$  & Analytical Solution & Asymptotic size   & Model identification\\
 \hline
 $1$. & $r_0(1-ct)$ & $X=X_0\exp\left({r_0t(1-\frac{ct}{2})}\right)$ & $0$ & Normal Distribution \citep{Banks1994}\\
 
$2$. & $\frac{r_0}{1+ct}$  &  $X=X_0(1+ct)^{\frac{r_0}{c}}$ & $\infty$ & Power Law Exponential \citep{Banks1994}\\
 
$3$. & $\frac{r_0}{1+r_0t}$  &   $X=X_0(1+r_0t)$ & $\infty$  & Linear Model \citep{Banks1994}\\
 
$4$. & $\frac{r_0}{1-r_0t}$  &   $X=X_0\left(\frac{1}{1-r_0t}\right)$ & $0$ & Hyperbolic Model \citep{Banks1994}\\
 
$5$. & $r_0e^{-ct}$  &   $X=X_0\exp\left({\frac{r_0}{c}(1-e^{-ct})}\right)$ & $X_0\exp{\left(\frac{r_0}{c}\right)}$ & Gompertz Model \citep{Banks1994}\\
 
 $6$. & $r_0t^{c-1}$   &   $X=X_0\exp\left({r_0\frac{t^c}{c}}\right)$ & $\infty$  & Korf Model \citep{korf1939}\\
 
 $7$. & $r_0(1+ct)$ &  $X=X_0\exp\left({r_0t(1+\frac{ct}{2})}\right)$ & $\infty$ & Linearly increasing $r$\\ 
 
 $8$. & $r=r_0+c\sin({\omega t})$  &  $X=X_0\exp\left({r_0t-\frac{c}{\omega \cos({\omega t})}}\right)$ & $\infty$ & Hyperbolically varying $r$ \\
 
 $9$. & $r=r_0+c\cos({\omega t})$  &  $X=X_0\exp\left({r_0t+\frac{c}{\omega \sin({\omega t})}}\right)$ & $\infty$ &Hyperbolically varying $r$ \\
 \hline
\end{tabular}
\label{table2}
\end{sidewaystable}

\begin{sidewaystable}
\caption{Different time dependent functions for the parameter $r$ and $K$ of logistic growth model ($c$ and $\omega > 0$).}
\begin{tabular}{ |p{0.6cm}|p{2.2cm}p{1.2cm}p{7.8cm}p{2.4cm}p{6.6cm}|  }
 \hline
 \multicolumn{6}{|c|}{Parent Model: Logistic Model} \\
 \hline
Srl.  & $r$ & $K$   & Analytical Solution & Asymptotic size & Model identification\\
 \hline
$1$. & $r$ & $K_0e^{r_0t}$ & $X=K_0\left[\left(rt+\frac{K_0}{X_0}\right)e^{-rt}\right]^{-1}$  & $\infty$ & Exponential Model \citep{Banks1994}\\
 
$2$. & $r$ & $\frac{K_0}{1+ct}$  &  $X=K_0\left[\left(1+ct-\frac{c}{r}\right)+\left(\frac{K_0}{X_0}-1+\frac{c}{r}\right)e^{-rt}\right]^{-1}$ & $0$ & Hyperbolically Varying $K$\citep{Banks1994}\\
 
$3$. & $r_0(1+ct)$ & $K$  &  $X=K\left[1+\left(\frac{K}{X_0}-1\right)e^{-r_0t\left(1+\frac{ct}{2}\right)}\right]^{-1}$ & $K$ & Linearly Growing $r$ \\

$4$. & $r_0(1-ct)$ & $K$  &  $X=K\left[1+\left(\frac{K}{X_0}-1\right)e^{-r_0t\left(1-\frac{ct}{2}\right)}\right]^{-1}$ & $0$ & Linearly Decaying $r$ \\
 
$5$. & $r_0t^{c-1}$ & $K$  &  $X=K\left[1+\left(\frac{K}{X_0}-1\right)e^{-\frac{r_0}{c}t^c}\right]^{-1}$ & $K$ & Extended Logistic Model \citep{CHAKRABORTY2017}\\
 
$6$. & $r_0e^{-ct}$ & $K$  &  $X=K\left[1+\left(\frac{K}{X_0}-1\right)e^{-\frac{r_0}{c}(1-e^{-ct})}\right]^{-1}$ & $\frac{K}{1+\left(\frac{K}{X_0}-1\right)e^{-\frac{r_0}{c}}}$ & Exponentially Decaying $r$ \\
 
$7$. & $r_0e^{ct}$ & $K$  &  $X=K\left[1+\left(\frac{K}{X_0}-1\right)e^{-\frac{r_0}{c}(e^{ct}-1)}\right]^{-1}$ & $K$ & Exponentially Growing $r$ \\
 
$8$. & $\frac{r_0}{1+ct}$ & $K$  &  $X=K\left[1+\left(\frac{K}{X_0}-1\right)(1+ct)^{-\frac{r_0}{c}}\right]^{-1}$ & $K$ & Hyperbolically Varying $r$ \\

$9$. & $r_0+c\sin({\omega t})$ & $K$  &  $X=K\left[1+\left(\frac{K}{X_0}-1\right)e^{-\left[r_0t+\frac{c}{\omega}(1-\cos{\omega t})\right]}\right]^{-1}$ & $K$ & Periodically Varying $r$ \\
 
$10$. & $r_0+c\cos({\omega t})$ & $K$  &  $X=K\left[1+\left(\frac{K}{X_0}-1\right)e^{-\left[r_0t+\frac{c}{\omega}(\sin{\omega t})\right]}\right]^{-1}$ & $K$ & Periodically Varying $r$ \\
 
 \hline
\end{tabular}
\label{table3}\\
\footnotesize{We did not consider the variation $r=\frac{r_0}{1-ct}$, $c>0$ (used in \citet{Banks1994}) as it does not give biologically realistic model.}
\end{sidewaystable}

\begin{sidewaystable}
\caption{Different density and time dependent functions for the parameter $r$ with various limiting choices of the parameter $\theta$ of theta-logistic growth model ($c>0$).}
\begin{tabular}{ |p{0.6cm}|p{1.8cm}p{0.4cm}p{1.7cm}p{7.4cm}p{3cm}p{6cm}|  }
 \hline
 \multicolumn{7}{|c|}{Parent Model: theta-logistic Model} \\
 \hline
 Srl. & $r$ & $K$ & $\theta$  & Analytical Solution & Asymptotic size   & Model identification\\
 \hline
$1$. &  $r$ & $K$ & $1$ & $X=K\left[1+\left(\frac{K}{X_0}-1\right)e^{-rt}\right]^{-1}$ & $K$ &Logistic Model \citep{Verhulst1838}\\

$2$. &  $\frac{r_0}{\theta}$ & $K$ & $\theta \to 0$ & $X=K\exp({e^{-r_0t}\ln{\frac{X_0}{K}}})$ & $K$ & Gompertz Model \citep{Gompertz1825}\\

$3$. &  $r$ & $K$ & $\theta \geq -1$ & $X=K\left[1+\left(\left(\frac{K}{X_0}\right)^{\theta}-1\right) e^{r_0\theta t}\right]^{-\frac{1}{\theta}}$ & $0$ $(-1<\theta<0)$ and $K$ $(\theta \geq 0)$ & Richards Model \citep{Richards1959}\\

$4$. &  $r_0t^{c-1}$ & $K$ & $\theta > 0$ & $X=K\left[1+\left(\left(\frac{        k}{X_0}\right)^{\theta}-1\right)e^{r_0\theta \frac{t^{c}}{c}}\right]^{-\frac{1}{\theta}}$ & $K$ & Koya-Goshu Model \citep{koya2013}\\

$5$. &  $r_0(1+ct)$ & $K$ & $\theta$ & $X=K\left[1+\left(\left(\frac{K}{X_0}\right)^{\theta}-1\right)e^{-r_0\theta (t+\frac{ct^2}{2})}\right]^{-\frac{1}{\theta}}$ & $K$ & Linearly Increasing $r$\\

$6$. &  $r_0(1-ct)$ & $K$ & $\theta$ & $X=K\left[1+\left(\left(\frac{K}{X_0}\right)^{\theta}-1\right)e^{-r_0\theta (t-\frac{ct^2}{2})}\right]^{-\frac{1}{\theta}}$ & $0$ & Linearly Decreasing $r$\\

$7$. &  $\frac{r_0}{\theta}t^{c-1}$ & $K$ & $\theta \to 0$ & $X=K\exp({e^{-r_0\frac{t^{c}}{c}}\ln{\frac{X_0}{K}}})$ & $K$ & Extended Gompertz Model \citep{Bhowmick2014}\\

$8$. &  $r_0\left(\frac{K}{X}\right)^{\theta}$ & $K$ & $\theta = \frac{1}{3}$ & $X=K\left[1+\left(\left(\frac{X_0}{K}\right)^{\frac{1}{3}}-1\right)e^{\frac{-r_0t}{3}}\right]^{3}$ & $K$ & Von Bertalanffy Model \citep{Von1949}\\

$9$. &  $r_0\left(\frac{K}{X}\right)^{\theta}$ & $K$ & $0<\theta <1$ & $X=K\left[1+\left(\left(\frac{X_0}{K}\right)^{\theta}-1\right)e^{-r_0\theta t}\right]^{\frac{1}{\theta}}$ & $K$ & Generalized Von Bertalanffy Model \citep{von1960}\\

$10$. & $\frac{r_0}{\theta}\left(\ln{\frac{K}{X}}\right)^{c-1}$ & $K$ & $\theta \to 0$ & $X=K\exp{\left[\left (1+\frac{r_0t}{c-1}\left(\ln{\frac{K}{X_0}}\right)^{c-1}\right)^{\frac{-1}{c-1}} \ln{\frac{X_0}{K}}\right]}$ & $0$ $(0<\theta<1)$ ~and $K$ $(\theta\geq 1)$
& Generalized Gompertz Model \citep{CHAKRABORTY2017}\\

$11$. & $\frac{r_0}{\theta}\left(\ln{\frac{K}{X}}\right)^{c}$ & $K$ & $\theta \to 0$ & $X=K\exp{\left[\left (1+\frac{r_0t}{c}\left(\ln{\frac{K}{X_0}}\right)^{c}\right)^{-
\frac{1}{c}} \ln{\frac{X_0}{K}}\right]}$ & $K$
& Crescenzo-Spina  Model \citep{CRESCENZO2016}\\

$12$. & $\frac{r_0}{\theta}\left(\ln{\frac{K}{X}}\right)^{-\frac{1}{2}}$ & $K$ & $\theta \to 0$ & $X=K\exp{\left[\left (1-2r_0t\left(\ln{\frac{K}{X_0}}\right)^{-\frac{1}{2}}\right)^{2} \ln{\frac{X_0}{K}}\right]}$ & $0$
& Second-order Exponential Polynomial\citep{Chakraborty2019}\\

\hline
\end{tabular}
\label{table4}
\end{sidewaystable}

\begin{sidewaystable}
\caption{Different time dependent functions for the parameter $r$ and $K$ of confined exponential growth Model ($c,\omega> 0$).}

\begin{tabular}{ |p{0.6cm}|p{2.3cm}p{1.7cm}p{6.6cm}p{2.8cm}p{7.6cm}|  }
 \hline
 \multicolumn{6}{|c|}{Parent Model: Confined Exponential Model} \\
 \hline
 Srl. & $r$ & $K$  & Analytical Solution & Asymptotic size  & Model identification\\
 \hline
$1$. & $r_0e^{ct}$ & $K$ & $X=K-(K-X_0)\exp({-\frac{r_0}{c}(e^{ct}-1)})$ & $K$ & Extreme Minimal Value Distribution  \citep{Banks1994}\\
 
$2$.  & $r_0t^{c-1}$ & $K$  & $X=K-(K-X_0)\exp({-r_0\frac{t^c}{c}})$ & $K$  & Weibull Distribution \citep{weibull1951}\\
 
$3$.  & $\frac{r_0}{K}X$ & $K$  &  $X=K\left[1+\left(\frac{K}{X_0}-1\right)e^{-r_0t}\right]^{-1}$ & $K$  & Logistic Model \citep{Verhulst1838}\\

$4$.  & $r_0(1-ct)$ & $K$  &  $X=K-(K-X_0)\exp\left({-r_0t\left(1-\frac{ct}{2}\right)}\right)$  & $-\infty$  & Linearly Decaying $r$ \\
 
$5$.  &  $r_0(1+ct)$ & $K$  &  $X=K-(K-X_0)\exp\left({-r_0t\left(1+\frac{ct}{2}\right)}\right)$ & $K$  & Linearly Increasing $r$ \\
 
$6$.  &  $r_0e^{-ct}$ & $K$  &  $X=K-(K-X_0)\exp({\frac{r_0}{c}(e^{-ct}-1)})$ &  $K-\frac{K-X_0}{e^{\frac{r_0}{c}}}$ & Exponentially Decaying $r$ \\
 
$7$.  &  $\frac{r_0}{1+ct}$ & $K$  & $X=K-(K-X_0)(1+ct)^{\frac{-r_0}{c}}$ &   $K$ & Hyperbolically Varying $r$ \\
 
$8$.  &  $r_0+c\sin({\omega t})$ & $K$  &  $X=K-(K-X_0)e^{-\left[r_0t+\frac{c}{\omega}(1-\cos{\omega t})\right]}$  & $K$ & Periodically Varying $r$ \\

$9$.  &  $r_0+c\cos({\omega t})$ & $K$  &  $X=K-(K-X_0)e^{-\left[r_0t+\frac{c}{\omega}(\cos{\omega t})\right]}$  & $K$ & Periodically Varying $r$ \\
 
$10$.  & $r$ & $K_0(1+ct)$  &  $X=X_0e^{-rt}+K_0(1+ct-\frac{c}{r})(1-e^{-rt})$  & $\infty$ & Linearly Increasing $K$ \\

$11$.  & $r$ & $K_0e^{ct}$  &  $X=X_0e^{-rt}+\frac{rK_0}{r+c}(e^{ct}-e^{-rt})$  & $\infty$ & Exponentially Increasing $K$ \\

$12$.  & $r$ & $K_0e^{-ct}$  &  $X=X_0e^{-rt}+\frac{rK_0}{r-c}(e^{-ct}-e^{-rt})$  & $0$ & Exponentially Decaying $K$ \\
\hline
\end{tabular}
\label{table5}\\
\footnotesize{We did not consider the variation $r=\frac{r_0}{1-ct}$, $c>0$ (used in \citet{Banks1994}) as it does not give biologically realistic model.}
\end{sidewaystable}

\begin{sidewaystable}
\caption{The new models obtained by variation in the parameters for which analytical solution does not exist.}
\begin{tabular}{ |p{0.6cm}|p{2.45cm}p{1.1cm}p{1.9cm}p{1cm}p{5.1cm}p{2.7cm}p{5.7cm}|  }
 \hline
Srl.  & Parent Model & $r$ & $K$ & $\theta$  & New Model & Asymptotic size  & Model identification\\
 \hline
$1$. & Logistic & $r$ & $K_0(1+ct)$ & - &
$\frac{dX(t)}{dt}=rX\left(1-\frac{X}{K_0(1+ct)}\right)$  & $\infty$ & Linearly Varying $K$ \\

$2$. & Theta-Logistic & $r_0X^{\gamma}$ & $K$ & $\theta$ & $\frac{dX(t)}{dt}=r_0X^{\gamma +1}\left(1-\left(\frac{X}{K}\right)^{\theta}\right)$  & $K$ & Co-operation Model \citep{BHOWMICK2015}\\

$3$. & Theta-Logistic & $r_0\left(\frac{X}{K}\right)^{c}$ & $K$ & $\theta>0$ &
$\frac{dX(t)}{dt}=r_0X\left(\frac{X}{K}\right)^{c}\left(1-\left(\frac{X}{K}\right)^{\theta}\right)$  & $K$ & Marusic-Bajzer Model \citep{MARUSIC1993}\\

$4$. & Theta-logistic & $r_0e^{ct}$ & $K$ & $\theta$ &
$\frac{dX(t)}{dt}=r_0e^{ct}X\left(1-\left(\frac{X}{K}\right)^{\theta}\right)$  & $K$ & Exponentially  increasing $r$ ($c>0$) \\
\hline
\end{tabular}
\label{table6}
\end{sidewaystable}

\citet{Banks1994} showed that after varying intrinsic growth rate ($r$) in exponential model, one can build connections with normal distribution, power law exponential model, linear model, hyperbolic model and gompertz model. In this paper, the extended models which are obtained by continuous transformation of the growth coefficient ($r$) is depicted in Table~\ref{table2}. In Table~\ref{table3}, we considered continuously variation in intrinsic growth rate $r$ and carrying capacity $K$ in logistic growth model. Considering continuous variation in $K$, \citet{Banks1994} built some connections between logistic model with other existing growth models. We mainly focus on the variation in the parameter $r$ and built connection with extended logistic model studied by \citet{CHAKRABORTY2017}. The exercises have also been carried out on the theta-logistic and confined exponential model and the derived connections are depicted in Table~\ref{table4} and Table~\ref{table5} respectively. Connections have been found between theta-logistic model with logistic model, gompertz model, Richards model, Koya-Goshu model, extended gompertz model, Von Bertalanffy model, generalized Von Bertalanffy model, generalized gompertz model, Crescenzo-Spina model and second order polynomial model. 

Noteworthy to mention that by applying the above transformations many new growth equations are obtained. It is to be noted that for each of the previous cases, the final differential equation (after replacing $r$ by $r(t)$ or $K$ by $K(t)$) can be solved analytically or solutions are available using some special functions. The differential equations must be solved numerically to obtain the size profile in case the analytical solution is not available. In Table~\ref{table6}, we consider few cases in which the analytical expression for the size variable $X(t)$ is not available. The detail discussion for each extended growth equation is discussed in the supporting information.

\section{ISRP and statistical identification of varying parameters} \label{ISRP_and_statistical_identification}
\citet{Bhowmick2014} proposed the concept of Interval Specific Rate Parameter (henceforth, ISRP) which performed better in selecting the true model more accurately than the Fisher's RGR \citep{PAL2018}. For every model the authors have identified a key parameter $b$ (called the rate parameter), and obtained its interval specific estimates based on the longitudinal data set up. If the underlying model is true, then the rate parameter should remain same over different time intervals. The advantage of using ISRP is, it can detect crucial intervals where the growth process is erratic and unusual. It may help experimental scientists to study more closely the effect of the parameters responsible for the growth of the organism/population under study.\\

\subsection{Asymptotic distribution of the estimator}
In this sub section we shall derive the distribution of ISRP estimator of $r(t)$ and $K(t)$ under some assumptions using Multivariate Delta Method \citep{wasserman2004,casella2002}. This distribution will be the key mechanism to identify whether the parameter has undergone any continuous variation over time.  

\subsubsection{The data structure and model set up}
To derive this distribution, we consider the data matrix $\bm{\mathrm{X}}$, where
\[ \bm{\mathrm{X}}=\begin{pmatrix}
    X_{1t_{1}} & X_{1t_{2}}  & \dots  & X_{1t_{q}} \\
    X_{2t_{1}} & \ddots & \ddots  & X_{2t_{q}} \\
    \vdots & \ddots & \ddots  & \vdots \\
    X_{nt_{1}} &X_{nt_{2}} & \dots  & X_{nt_{q}}
\end{pmatrix},\] whose rows are assumed to be independent and identically distributed (iid) random variables following $q$-variate normal distribution with mean vector $\bm{\mu}=(\mu_{t_{1}},\mu_{t_{2}},\ldots,\mu_{t_{q}})'$ and variance covariance matrix $\bm{\Sigma}_{q\times q}=\left(\sigma^2 \rho^{|i-j|}\right)_{i,j=1}^{q}$. If we take $\bm{X_i}=(X_{it_{1}},X_{it_{2}},\ldots,X_{it_{q}})'$, then $\mbox{E}(\bm{X_i})=\bm{\mu}$ and $\mbox{Var}(\bm{X_i})=\bm{\Sigma}$, for all $i=1,2,\ldots,n$; $A'$ denotes the transpose of a matrix $A$. We have considered ${1,2,\ldots,q}$ time points. We consider logistic growth model as test bed model for simulation study and under this model \begin{equation} \label{mean_function}
    \mu_{t}=\left[\frac{K}{1+\left(\frac{K}{\mu_{0}}-1\right)e^{-rt}}\right].
\end{equation}
As existence of second order moments is assumed, so taking $\overline{\bm{X}}=\left(\overline{X}_1,\overline{X}_2,\ldots,\overline{X}_q\right)'$, where $\overline{X}_j = \frac{1}{n}\sum_{i=1}^n X_{it_j}$, we can apply Multivariate central limit theorem \citep{Timm2002} by which we get
\begin{equation} \label{delta_metod}
    \sqrt{n}\left(\bm{\overline{X}}-\bm{\mu}\right) \overset{\text{d}}{\to}\mathcal{N}_{q}(\bm{0},\bm{\Sigma}).
\end{equation}

Now the estimator of ISRP of $r(t)$ based on time points $\{t_j, t_{j+1}, t_{j+2}\}$, $j=1,2,\ldots,q-2$, denoted as $\widehat{{r}_{j}(\Delta t)}$, is given by \citep{PAL2018},
\begin{equation}\label{eqn_r_hat}
    \widehat{{r}_{j}(\Delta t)} =\frac{1}{h}\ln{\left[\frac{\frac{1}{\overline{X}_j}-\frac{1}{\overline{X}_{j+1}}}{\frac{1}{\overline{X}_{j+1}}-\frac{1}{\overline{X}_{j+2}}}\right]}= \phi (\overline{X}_{j},\overline{X}_{j+1},\overline{X}_{j+2}) ~~~(\mbox{say}),
\end{equation}
We aim to obtain the sampling distribution of $\phi\left(\overline{X}_{j},\overline{X}_{j+1},\overline{X}_{j+2}\right)$ using Multivariate delta method \citep{wasserman2004}. In terms of variables $x$, $y$ and $z$ the function $\phi$ can be written as
\begin{equation} \label{eqn:r_hat_delta_t}
\phi(x,y,z)=\frac{1}{h}\ln{\left[\frac{\frac{1}{x}-\frac{1}{y}}{\frac{1}{y}-\frac{1}{z}}\right]}    
\end{equation}
By using Multivariate central limit theorem  for $\left(\overline{X}_{j},\overline{X}_{j+1},\overline{X}_{j+2}\right)$, we obtain
\[ \sqrt{n} \left[\begin{pmatrix}
   \overline{X}_{j} \\
    \overline{X}_{j+1}  \\
    \overline{X}_{j+2}
\end{pmatrix}
- \begin{pmatrix}
   \mu_{j} \\
     \mu_{j+1}  \\
     \mu_{j+2}
\end{pmatrix}\right]
\xrightarrow{\text{d}}\mathcal{\bm{N}}\left(0,\bm{\Sigma}_{3 \times 3}\right)\]
Here the mean function $\mu_t$ under logistic growth law is given by eqn.~(\ref{mean_function}) and $\phi \colon \mathbb{R}^{3}\to \mathbb{R}$ is a differentiable function at $\bm{\mu}=(\mu_{j},\mu_{j+1},\mu_{j+2})'\in \mathbb{R}^{3}$. Now we have to find the $\nabla\phi$ at the point $\bm{\mu}$ to apply Multivariate delta method. Taking partial derivatives of $\phi(x,y,z)$ with respect to $x$, $y$ and $z$ and evaluate the above partial differential equations at the point $\bm{\mu}$, we obtain
\begin{eqnarray} \nonumber
&&\frac{\partial \phi}{\partial x}\bigg|
_{\bm{\mu}}= -\frac{1}{h}\frac{\mu_{j+1}}{\mu_{j}(\mu_{j+1}-\mu_{j})} \\ \nonumber
&&\frac{\partial \phi}{\partial y}\bigg|_{\bm{\mu}} =\frac{1}{h}\frac{\mu_{j+2}-\mu_{j}}{(\mu_{j+1}-\mu_{j})(\mu_{j+2}-\mu_{j+1})} \\ \nonumber
&&\frac{\partial \phi}{\partial z}\bigg|_{\bm{\mu}}= -\frac{1}{h}\frac{\mu_{j+1}}{\mu_{j+2}(\mu_{j+2}-\mu_{j+1})}~~
,\end{eqnarray} where the values of $\mu_i$'s, $i=j,j+1,j+2$ are provided in eqn.~(\ref{mean_function}). Using matrix notation, we obtain 
\[ \nabla \phi \big|_{\bm{\mu}}= \begin{pmatrix}
  -\frac{1}{h}\frac{\mu_{j+1}}{\mu_{j}(\mu_{j+1}-\mu_{j})}  \\
 \frac{1}{h}\frac{\mu_{j+2}-\mu_{j}}{(\mu_{j+1}-\mu_{j})(\mu_{j+2}-\mu_{j+1})}  \\
   -\frac{1}{h}\frac{\mu_{j+1}}{\mu_{j+2}(\mu_{j+2}-\mu_{j+1})}
\end{pmatrix}.\]
 Now by using Multivariate delta method, the distribution of $\widehat{{r_{j}}(\Delta t)}$ is given as
 \[ \sqrt{n} \left[\phi \begin{pmatrix}
   \overline{X}_{j} \\
    \overline{X}_{j+1}  \\
    \overline{X}_{j+2}
\end{pmatrix}
- \phi \begin{pmatrix}
   \mu_{j} \\
     \mu_{j+1}  \\
     \mu_{j+2}
\end{pmatrix}\right]
\xrightarrow{\text{d}}\mathcal{\bm{N}}\left(0,\bm{\nabla'}\phi|_{\bm{\mu}} \Sigma \bm{\nabla}\phi|_{\bm{\mu}}\right). \]

The estimator of ISRP of the parameter $K$ based on the triplet $\{t_j, t_{j+1}, t_{j+2}\}$, $j=1,2,\ldots,q-2$, denoted by $\widehat{{K}_{j}(\Delta t)}$, is given by \citep{PAL2018},
\begin{equation}
    \widehat{{K}_{j}(\Delta t)} =\left[\frac{1}{\overline{X}_0}-\frac{\frac{1}{\overline{X}_j}-\frac{1}{\overline{X}_{j+1}}}{\exp\left({-\widehat{{r}_{j}(\Delta t)}}t_j\right)\left[1-\exp\left({-\widehat{{r}_{j}(\Delta t)}}h\right)\right]}\right]^{-1}= \psi (\overline{X}_{j},\overline{X}_{j+1},\overline{X}_{j+2}) ~~~(\mbox{say}).
\end{equation} Since $\widehat{{K}_{j}(\Delta t)}$ is a function of the random variable $\widehat{{r}_{j}(\Delta t)}$, one may argue that the distribution can be derived by using the delta method with taking some function of the form $g\left(\overline{X}_{j},\overline{X}_{j+1}, \widehat{{r}_{j}(\Delta t)}\right)$. However, note that the co-variance structure between $\widehat{{r}_{j}(\Delta t)}$ and $(\overline{X}_{j},\overline{X}_{j+1})$ is not known. Hence, the function $\psi$ was chosen using the original random variables whose covariance structure is known. In terms of $(\overline{X}_{j},\overline{X}_{j+1},\overline{X}_{j+2})$, the function $\psi$ is given as
\begin{equation} \label{eqn:K_hat_delta_t}
\psi(\overline{X}_{j},\overline{X}_{j+1},\overline{X}_{j+2})= \left[\frac{1}{\overline{X}_{0}}-\frac{\left(\overline{X}_{j+1}-\overline{X}_{j}\right)^{\frac{t_j}{h}+2}~\left(\overline{X}_{j+2}\right)^{\frac{t_j}{h}+1}}{\left(\overline{X}_{j+2}-\overline{X}_{j+1}\right)^{\frac{t_j}{h}}~\overline{X}_{j+1}~\left(\overline{X}_{j}\right)^{\frac{t_j}{h}+1}~\left[\overline{X}_{j+2}\left(\overline{X}_{j+1}-\overline{X}_{j}\right)-\overline{X}_{j}\left(\overline{X}_{j+2}-\overline{X}_{j+1}\right)\right]}\right]^{-1}  
\end{equation} Taking partial derivatives of $\psi (x,y,z)$ with respect to the real variables $x$, $y$ and $z$ and evaluating it at the point $\bm{\mu}$, we obtain
\begin{eqnarray} \nonumber
&&\frac{\partial \psi}{\partial x}\bigg|_{\bm{\mu}} =\frac{\zeta}{\eta^2}\left[-\frac{\frac{t_j}{h}+2}{\mu_{j+1}-\mu_{j}} -\frac{\frac{t_j}{h}+1}{\mu_{j}}+\frac{2\mu_{j+2}-\mu_{j+1}}{ {\mu}_{j+2}({\mu}_{j+1}-{\mu}_{j})-{\mu}_{j}({\mu}_{j+2}-{\mu}_{j+1})}\right] \\ \nonumber
&&\frac{\partial \psi}{\partial y}\bigg|_{\bm{\mu}} =\frac{\zeta}{\eta^2}\left[\frac{\frac{t_j}{h}+2}{\mu_{j+1}-\mu_{j}} +\frac{\frac{t_j}{h}}{\mu_{j+2}-\mu_{j+1}}-\frac{1}{\mu_{j+1}}-\frac{\mu_{j+2}+\mu_{j}}{ {\mu}_{j+2}({\mu}_{j+1}-{\mu}_{j})-{\mu}_{j}({\mu}_{j+2}-{\mu}_{j+1})}\right] \\ \nonumber
&&\frac{\partial \psi}{\partial z}\bigg|_{\bm{\mu}}
=\frac{\zeta}{\eta^2}\left[\frac{\frac{t_j}{h}+1}{\mu_{j+2}} -\frac{\frac{t_j}{h}}{\mu_{j+2}-\mu_{j+1}}+\frac{2\mu_{j}-\mu_{j+1}}{ {\mu}_{j+2}({\mu}_{j+1}-{\mu}_{j})-{\mu}_{j}({\mu}_{j+2}-{\mu}_{j+1})}\right],
\end{eqnarray} where the values of $\mu_i$'s, $i=j,j+1,j+2$ are provided in eqn.~(\ref{mean_function}). Using matrix notation, we get
\[ \nabla \psi \big|_{\bm{\mu}}=\left(\frac{\zeta}{\eta^2}\right)\bigg|_{\bm{\mu}} \begin{pmatrix}
  -\frac{\frac{t_j}{h}+2}{\mu_{j+1}-\mu_{j}} -\frac{\frac{t_j}{h}+1}{\mu_{j}}+\frac{2\mu_{j+2}-\mu_{j+1}}{ {\mu}_{j+2}({\mu}_{j+1}-{\mu}_{j})-{\mu}_{j}({\mu}_{j+2}-{\mu}_{j+1})} \\
 \frac{\frac{t_j}{h}+2}{\mu_{j+1}-\mu_{j}} +\frac{\frac{t_j}{h}}{\mu_{j+2}-\mu_{j+1}}-\frac{1}{\mu_{j+1}}-\frac{\mu_{j+2}+\mu_{j}}{ {\mu}_{j+2}({\mu}_{j+1}-{\mu}_{j})-{\mu}_{j}({\mu}_{j+2}-{\mu}_{j+1})} \\
  \frac{\frac{t_j}{h}+1}{\mu_{j+2}} -\frac{\frac{t_j}{h}}{\mu_{j+2}-\mu{j+1}}+\frac{2\mu_{j}-\mu_{j+1}}{ {\mu}_{j+2}({\mu}_{j+1}-{\mu}_{j})-{\mu}_{j}({\mu}_{j+2}-{\mu}_{j+1})}
\end{pmatrix},\] 
where $\zeta|_{\bm{\mu}}=\frac{(\mu_{j+1}-\mu_{j})^{\frac{t_j}{h}+2}(\mu_{j+2})^{\frac{t_j}{h}+1}}{(\mu_{j+2}-\mu_{j+1})^{\frac{t_j}{h}} \mu_{j+1} (\mu_{j})^{\frac{t_j}{h}+1}[\mu_{j+2}(\mu_{j+1}-\mu_{j})-\mu_{j}(\mu_{j+2}-\mu_{j+1})]}$ and $\eta|_{\bm{\mu}}=\left(\frac{1}{\mu_0}-\zeta\right)$. By using Multivariate delta method, the distribution of $\widehat{{K_{j}}(\Delta t)}$ is given as 
\[ \sqrt{n} \left[\psi \begin{pmatrix}
   \overline{X}_{j} \\
    \overline{X}_{j+1}  \\
    \overline{X}_{j+2}
\end{pmatrix}
- \psi \begin{pmatrix}
     \mu_{j} \\
     \mu_{j+1} \\
     \mu_{j+2}
\end{pmatrix}\right]
\xrightarrow{\text{d}}\mathcal{\bm{N}}\left(0,\bm{\nabla'}\psi|_{\bm{\mu}} \Sigma \bm{\nabla}\psi|_{\bm{\mu}}\right). \]
The expression for partial derivatives which are required for the computations of interval specific estimators for all the models are given in Appendix~\ref{app:partial}.

\subsection{Simulation study}
\begin{figure}[H]
  \begin{subfigure}{9cm}
    \includegraphics[width=9cm]{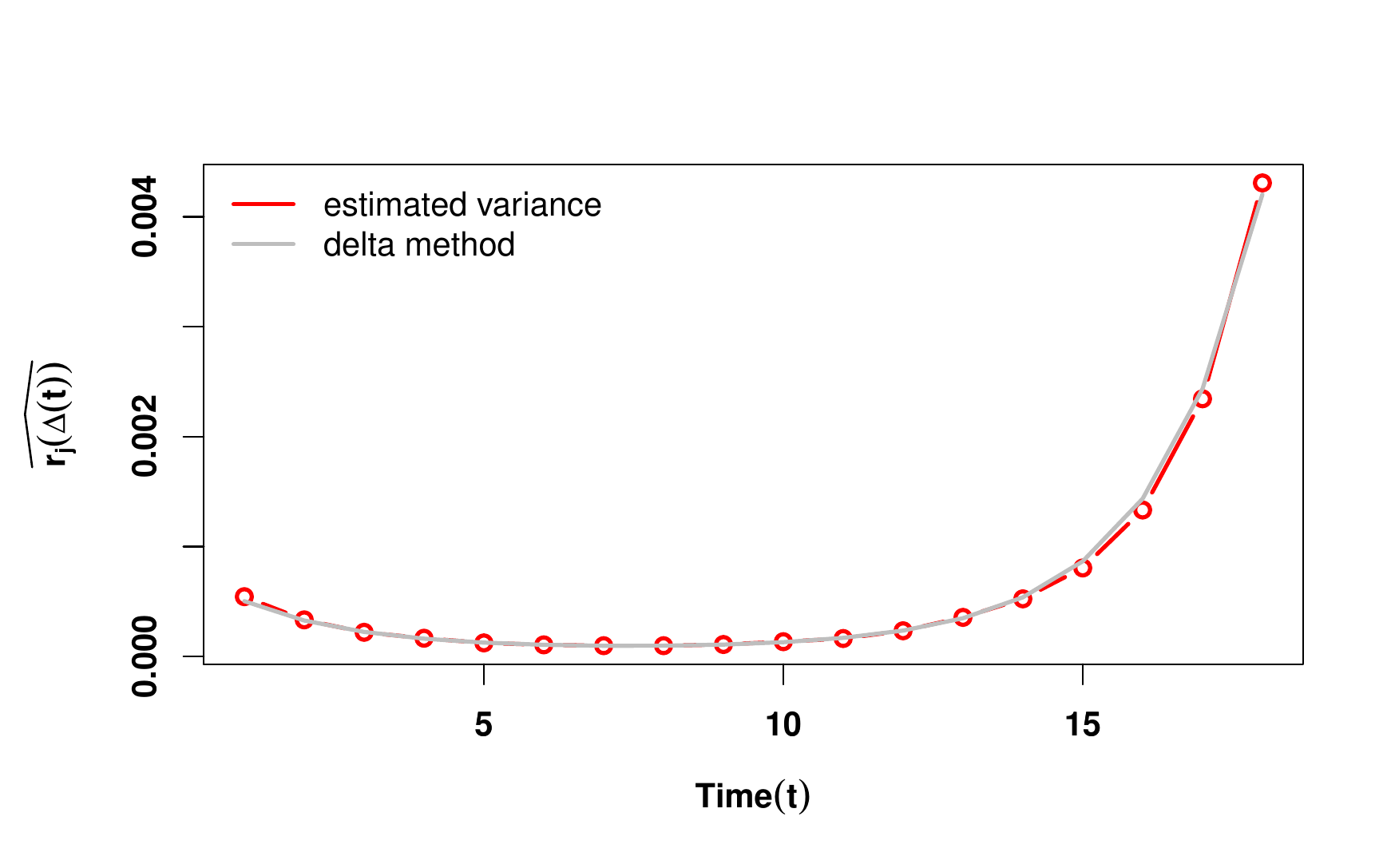}
    \caption{}
    \label{r_hat_delta_method_a}
  \end{subfigure}
  \begin{subfigure}{9cm}
\includegraphics[width=9cm]{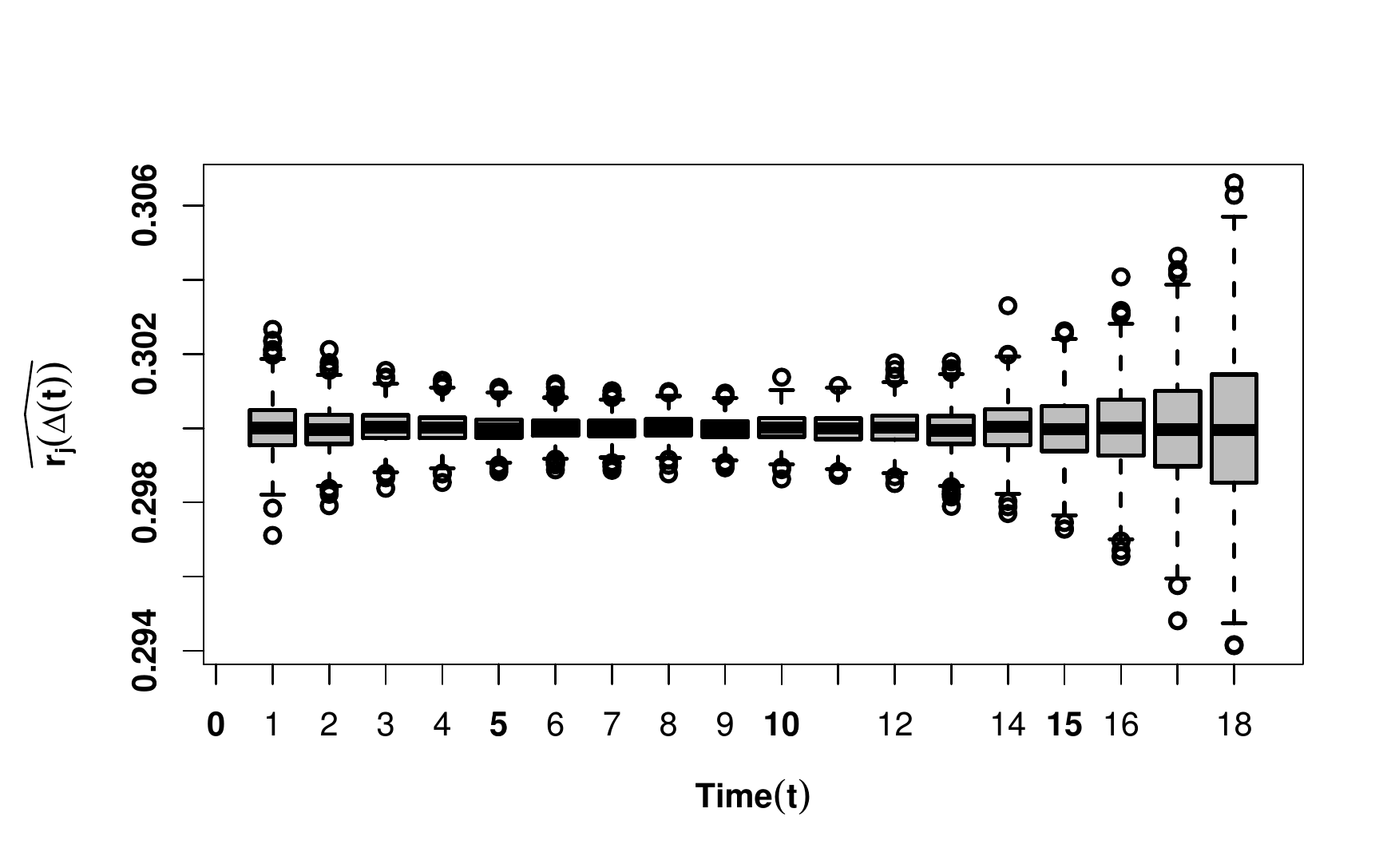}
    \caption{}
    \label{r_hat_delta_method_b}
  \end{subfigure}
  \begin{subfigure}{9cm}
    \includegraphics[width=9cm]{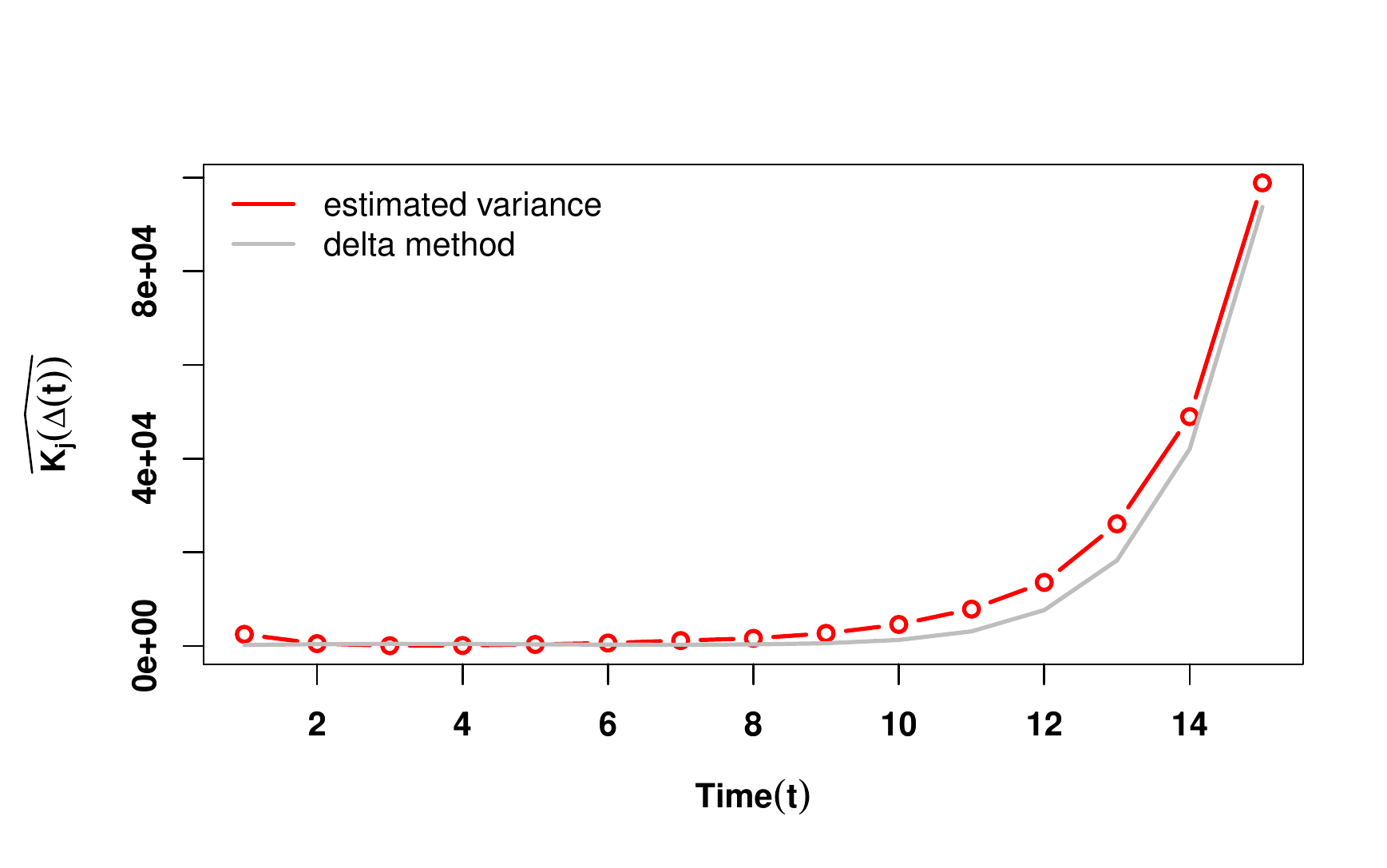}
    \caption{}
    \label{K_hat_delta_method_a}
  \end{subfigure}
  \begin{subfigure}{9cm}
\includegraphics[width=9cm]{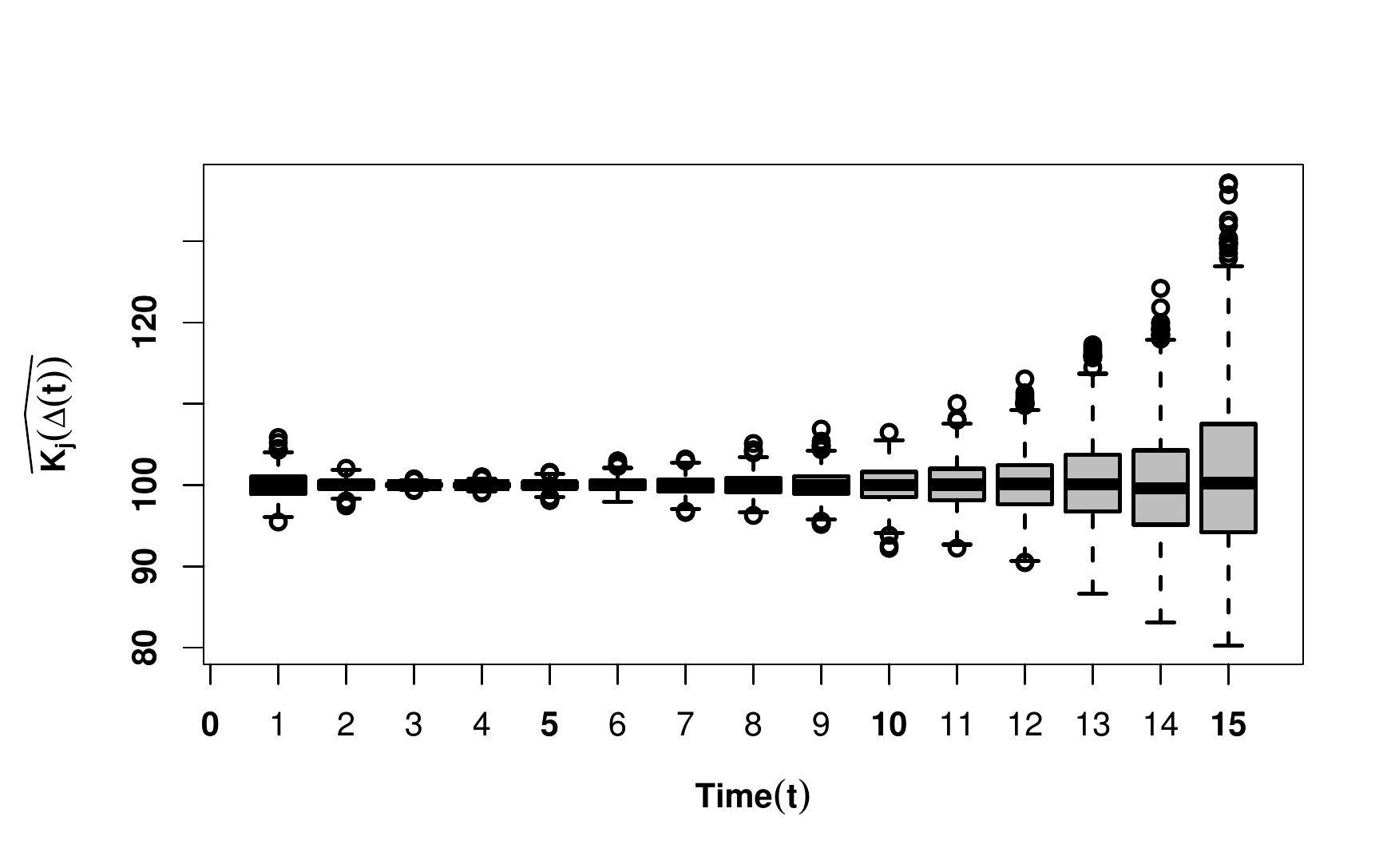}
    \caption{}
    \label{K_hat_delta_method_b}
  \end{subfigure}
\label{Boxplot_and_estimation_of_logistic}
\caption{(a): The red curve shows the estimated variance of $\widehat{r_j(\Delta t)}$ based on 1000 simulations for $j=1,2,\ldots, 18$. The grey curve indicates the variance obtained from the delta method. It is evident that delta method gave an accurate approximation of the variance of ISRP. (b): Simulated realizations of $\widehat{r_j(\Delta t)}$ are visualized by boxplot based on 1000 simulation. (c): The red curve shows the estimated variance of $\widehat{K_j(\Delta t)}$ based on 1000 simulations for $j=1,2,\ldots, 18$. The grey curve indicates the variance obtained from the delta method. It is evident that delta method gave an accurate approximation of the variance of ISRP. ; (d): Simulated realizations of $\widehat{K_j(\Delta t)}$ are visualized by boxplot based on 1000 simulation. For all the cases the simulation study was carried out using the logistic growth model as underlying model with the parameter set up as: $r = 0.3$, $K=100$, $\mu_0 = 10$, $\rho = 0.1$ and $\sigma^2=0.001$. }
\end{figure}

In this section the theoretical results are verified by using the simulation study. To check the accuracy of delta method is approximating the sampling distribution of $\widehat{r_j(\Delta t)}$ for $j=1,2,\ldots,q-2$, we used computer simulation. We simulated the growth trajectories for $n=1000$ individuals for $20$ times points where each trajectory was generated from the multivariate normal distribution with logistic mean function and variance-covariance matrix with Koopman structure \citep{Koopmans}. Based on this data, we obtained the estimate of $r_j(\Delta t)$ which acts as a single realization from the sampling distribution of $\widehat{r_j(\Delta t)}$. The process was replicated 1000 times to obtain 1000 realizations from the distribution of $\widehat{r_j(\Delta t)}$ to visualize the distribution by using the histograms. The histograms obtained from simulation study clearly suggested the agreement with normal distribution with estimated mean and and variance obtained from the delta method. The close agreement between the approximate distribution by using delta method and simulated sampling distribution is depicted in Fig.~\ref{r_hat_delta_method_a}, \ref{r_hat_delta_method_b} for $\widehat{r_j(\Delta t})$ and in Fig.~\ref{K_hat_delta_method_a}, \ref{K_hat_delta_method_b} for $\widehat{K_j(\Delta t)}$. The parameter choices are kept as $K=100,~\mu_0 = 10,~\sigma^2 = 0.001,~\rho = 0.1$, and the covariance matrix is the Koopman structure. 

\begin{figure}[H]
    \centering
  \begin{subfigure}{9cm}
    \includegraphics[width=9cm]{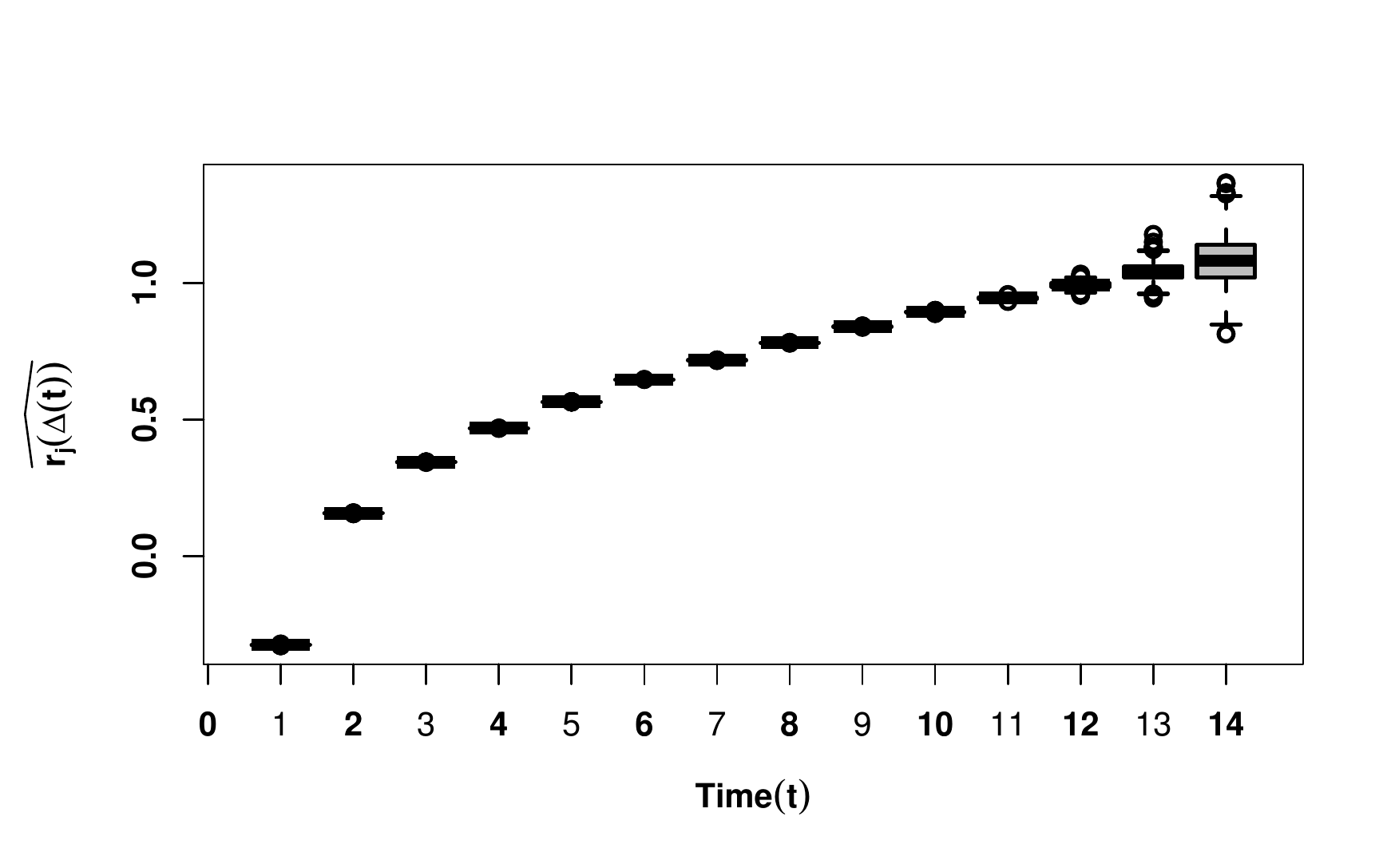}
    \caption{$r(t) = r_0 t^{c-1}$,~~$c=1.5$}
    \label{r_hat_extended_quadratic_boxplot}
  \end{subfigure}
  \begin{subfigure}{9cm}
\includegraphics[width=9cm]{sim_delta_r_lin_grow_logistic_boxplot.pdf}
    \caption{$r(t) = r_0(1+ct),~~c = 0.5$}
    \label{r_hat_extended_linear_boxplot}
  \end{subfigure}
 \caption{The figure shows the boxplot of $\widehat{r_j(\Delta t)}$ obtained from 1000 simulated realizations by using logistic growth model with continuously varying parameter $r$. It is to be noted that the assumed variation in $r$ is reflected in the ISRP profile. For both the cases the simulation study was carried out using the following parameter set up as: $r_0 = 0.3$, $K=100$, $\mu_0 = 10$, $\rho = 0.1$ and $\sigma^2=0.001$. The reason for depicting only first few time points are mentioned in the main text (Discussion).}
\label{Boxplot}
\end{figure}

In addition, even if the data were simulated from the other growth model, the normal approximation of the sampling distribution of the estimators remain unchanged. For example, we carried out the same simulation study using the extended version of the logistic growth model with $r(t)=r_0t^{c-1}$ and $r(t) = r_0(1+ct)$ and evaluated the $\widehat{r_j(\Delta t)}$ under the assumption of the logistic model. The associated patterns in ISRP profiles are depicted in Fig.~\ref{r_hat_extended_quadratic_boxplot} (power function) and Fig.~\ref{r_hat_extended_linear_boxplot} (linear function). The simulation study was carried out in software \texttt{R}. All the necessary codes are being provided in the online supporting material.

\begin{remark}
For simulation, we have considered the co-variance matrix $\Sigma$ having Koopmans \citep{Koopmans} correlation structure. According to this structure the correlation between $X_{t_i}$ and $X_{t_j}$ will be small if the time points $t_i$ and $t_j$ are far apart. This is quite natural in growth processes. Similar structure is considered by many other researchers \citep{PAL2018, Chakraborty2019}.
\end{remark}

\section{Real data analysis} \label{Real-data_analysis}

 In this section we establish the utility of the discussed method using some real data sets from different domains. The examples have been selected here from three different domains for illustration purpose only. Analysis of domain specific data sets and providing meticulous analysis is beyond the scope of our current study. We have considered representative examples from animal growth, retail marketing and epidemiology.

\subsection{Case study - I}

\begin{figure}[H]
    \centering
  \begin{subfigure}{9cm}
    \includegraphics[width=9cm]{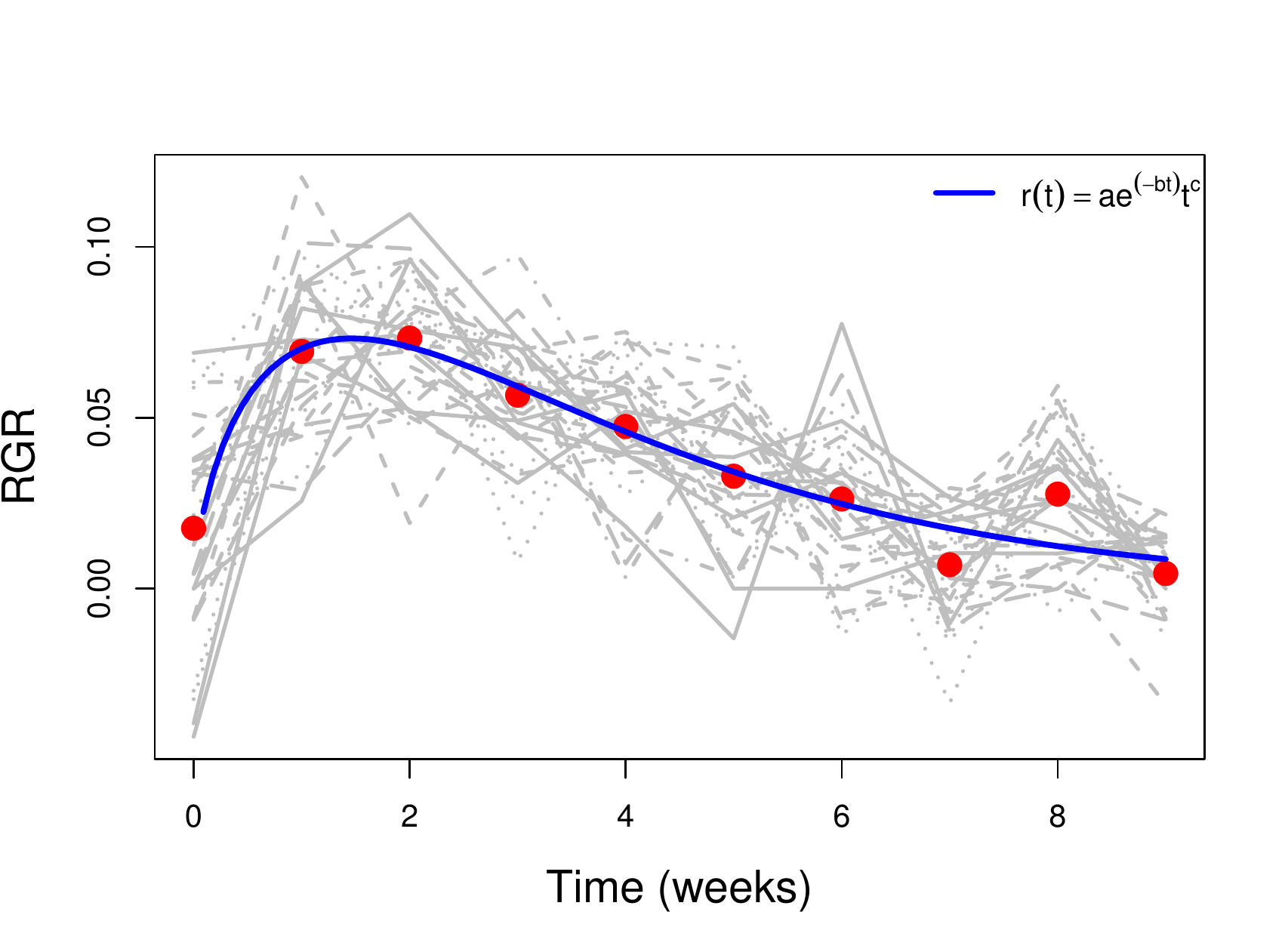}
    \caption{}
    \label{cattle_rgr_profile}
  \end{subfigure}
  \begin{subfigure}{9cm}
\includegraphics[width=9cm]{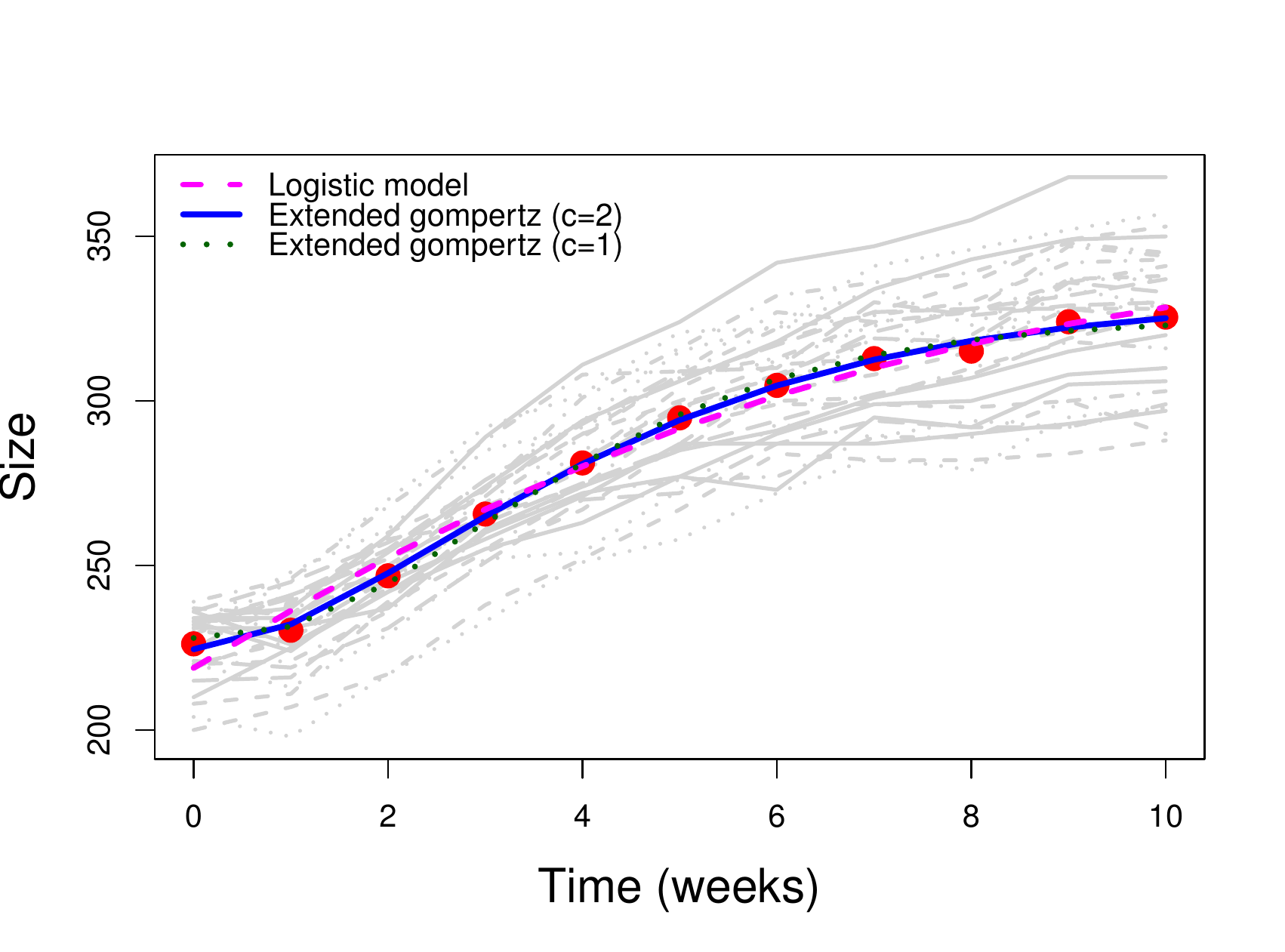}
    \caption{}
    \label{cattle_size_profile}
  \end{subfigure}
  
  \caption{Panel (a): RGR profile of the cattle data. Red dots represent the average RGR at a fixed time point for all animals. The nonlinear least squares fit of the equation $r(t) = ae^{-bt}t^c$ is overlaid on the graph. Estimated parameters are: $\hat{a} = 0.1088834$, $\hat{b} = 0.4375366$ and $\hat{c} =0.6397442$. Panel (b): Size profile of the data and red dots represent the average size. The logistic model and two extended gompertz models are fitted using nonlinear least squares. }
\label{Relation_flowchart_between models0}
    
\end{figure}

\begin{figure}[H]
    \centering
  \begin{subfigure}{9cm}
    \includegraphics[width=9cm]{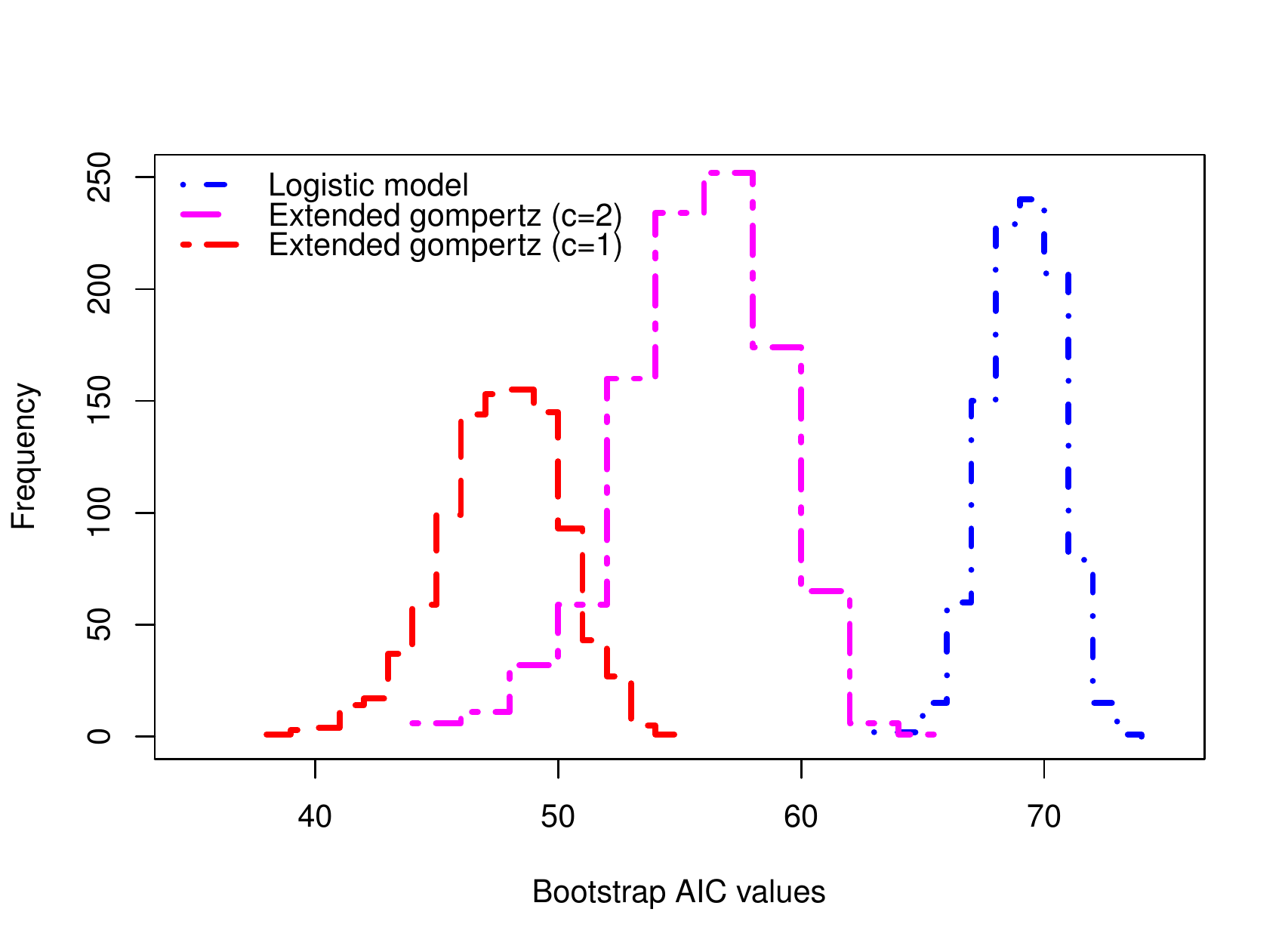}
  \end{subfigure}
 \caption{Frequency histogram of the AIC values obtained from the nonlinear least squares fitting of the models on the bootstrap samples. The number of bootstrap replication is $B=1000$. }
\label{Relation_flowchart_between models1}
\end{figure}

For illustration, we considered the data sets on cattle growth which was analyzed by \citet{Kenward1987}. The cattle data contains the weight of individual cattle at 11 time points over a 133 days period. The animals were given two treatments, namely A and B for intestinal parasites. For the demonstration purpose, we considered the data of animals which were given treatment A which were 30 in total. Here we make an attempt to choose the appropriate mean growth profile based on the data. The measurement schedules are rescaled to $t=0,1,2,\ldots, 10$. The Grey coloured curves in in Fig.~\ref{cattle_rgr_profile} and Fig.~\ref{cattle_size_profile} depicts the RGR and Size profile of individual animal respectively. Given the mean size profile a sigmoid shape growth curve would be an appropriate choice to start the analysis. However, the RGR profile indicates that the logistic growth model is not appropriate for describing the growth pattern. RGR first increases and then decreases. So, according to our proposal in this manuscript, we start with the base model as exponential and the empirical investigation suggests that the growth coefficient $r$ is varying as a function of time as $r(t) = ae^{-bt}t^c$, where $a$, $b$, $c$ are positive. Such a choice of the RGR profile was investigated in \citet{Bhowmick2014} and \citet{BhowmickMathBios2014}. Since for non-integer value of $c$ the exact solution can not be obtained, so we consider the final model for $c=1$ and $c=2$ only. The corresponding models are known as extended gompertz model which has a growth equation described in \citet{Bhowmick2014}. The logistic model, extended gompertz model ($c=1$ and $c=2$) were fitted to the size profiles and the corresponding AIC values are 68.9423, 46.14059 and 55.42874, respectively (Fig.~\ref{cattle_size_profile}). Thus the best model turns out to be the exponential model with continuously varying growth coefficients $r(t) = ae^{-bt}t$. To be more precise with our conclusion, we created $B=1000$ bootstrap data sets by randomly selecting rows of the data with replacement. All three designated models were fitted to each bootstrap sample and corresponding AIC values were recorded. The bootstrap distribution of the AIC for each models gives strong indication for the selection of the extended gompertz model with $c = 1$ (Fig.~\ref{Relation_flowchart_between models1}).

\subsection{Case study - II}
\begin{figure}[H]
    \centering
  \begin{subfigure}{9cm}
    \includegraphics[width=9cm]{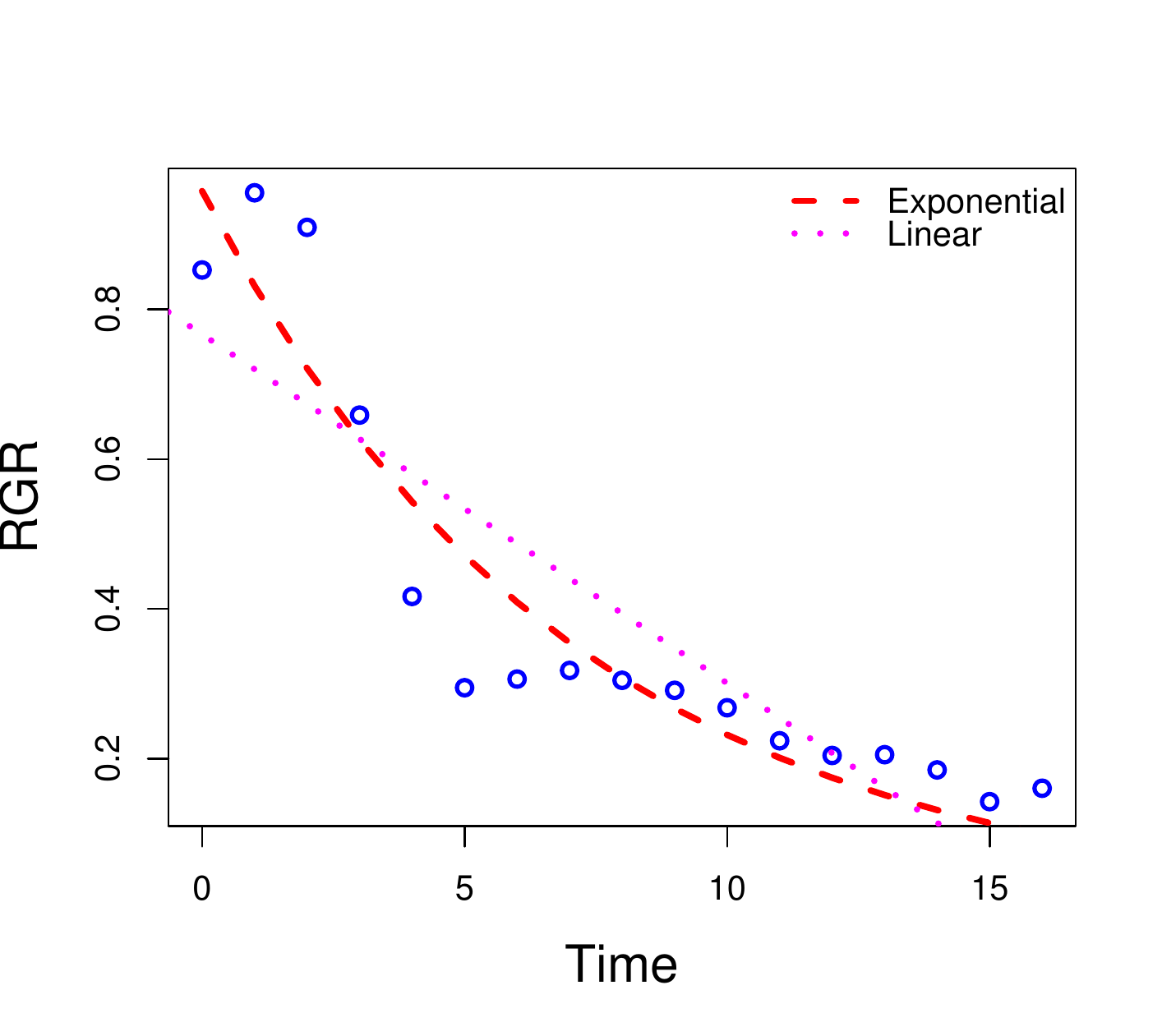}
    \caption{RGR profile of the sales data with respect to time.}
    \label{SalesData_RGR}
  \end{subfigure}
  \begin{subfigure}{9cm}
\includegraphics[width=9cm]{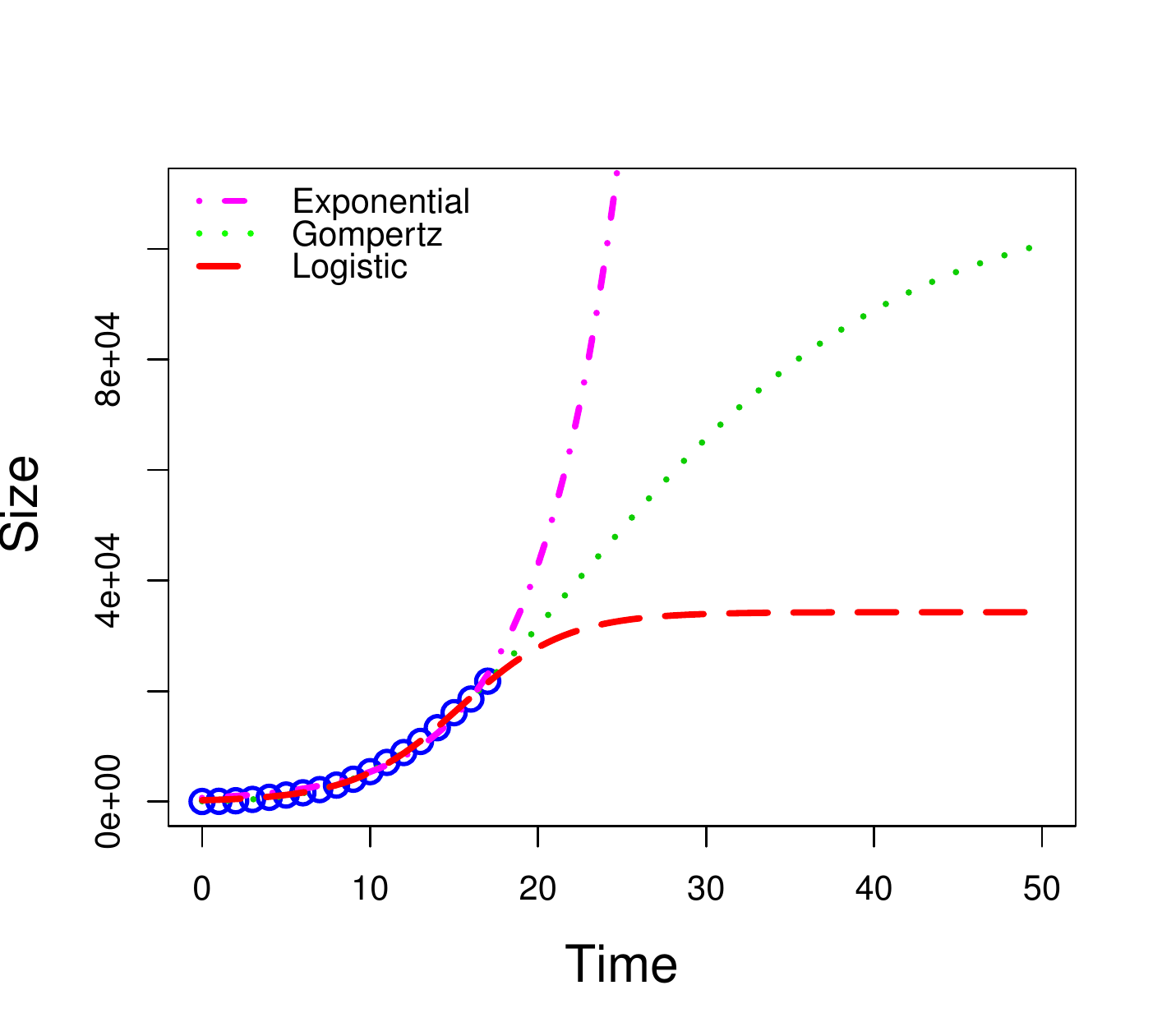}
    \caption{Size profiles of the sales data with respect to time.}
    \label{SalesData_Fitting}
    \end{subfigure}
  
  \caption{RGR and size profiles of the cumulative sales data obtained from \citet{TRAPPEY2008}. (a) Exponential decay of the intrinsic growth rate parameter indicated the choice of Gompertz model to be more appropriate. (b) The size profile data was modelled using three models. Gompertz model gave the best fit.}
\label{Relation_flowchart_between models2}
    
\end{figure}

For illustration of the method, we have considered the data set of cumulative sales of LCD-TV from \citet{TRAPPEY2008}. The data gives the cumulative quarterly sales from 2003 to 2007 (in thousand) which were collected from Market Intelligence Center Taiwan by \citet{TRAPPEY2008}. For simplicity, the measurement schedules are rescaled as $t=0,1,2,\ldots, 18$. The dataset clearly indicates an exponential growth in sales during the period 2003-2007. However, the plot of relative growth rate gives the actual picture (Fig.~\ref{SalesData_RGR}). It is evident from Fig.~\ref{SalesData_RGR}, that the logged difference is a decreasing function with respect to time, which indicates that exponential growth is not an appropriate model to choose. Rather it suggests that, the exponential model with an intrinsic growth rate which is a decreasing function of time may be a better choice. So, for the analysis we fitted the exponential and linear function of intrinsic growth rate. Exponential decay of the form ($r(t) = r_0 e^{-c t}$) has been found to be more appropriate than a linear choice of $r(t)$ based on the Akaike Information Criteria (AIC). AIC of exponential fit is -27.85451 and AIC of linear fit is -14.5925 and the difference in AIC values $\Delta \mbox{AIC} >10$ \citep{burnham2002}. Hence, gompertz model is the appropriate choice for the given data. In Fig.~\ref{SalesData_Fitting}, we have fitted all the three models, viz, exponential, gompertz and logistic and obtained the AIC values as 295.0443, 221.4758 and  247.4407, respectively. Smallest AIC for Gompertz model indicates the preference of this model than other two. The models are also compared using the root mean squared error and the conclusion remained same. The analysis was carried out using nonlinear least squares method using \texttt{nls} function available in \texttt{R}. Complete source codes are provided in the supporting online material.

\subsection{Case study - III}
In the third case study, we have taken the data of cumulative COVID-19 cases in Germany. The data set  being considered from  (\href{https://ourworldindata.org/coronavirus}{https://ourworldindata.org/coronavirus}). The data contains cumulative affected cases of COVID-19 from 31st December 2019 to 8th July 2020. Till 27th Jan 2020, there were no cases reported in Germany.
For the ease in the selection criterion, we have taken 5 days moving average of the data. It has been observed that till day 62, dated 29th Feb 2020, there were minor changes observed in the data, due to which for fitting the model the data is being considered from 1st march 2020 to 8th June 2020, which is 129 days in total. For simplicity, the measurement schedules are rescaled as $t=0,1,2,\ldots, 128$.

\begin{figure}[H]
    \centering
  \begin{subfigure}{9cm}
    \includegraphics[width=9cm]{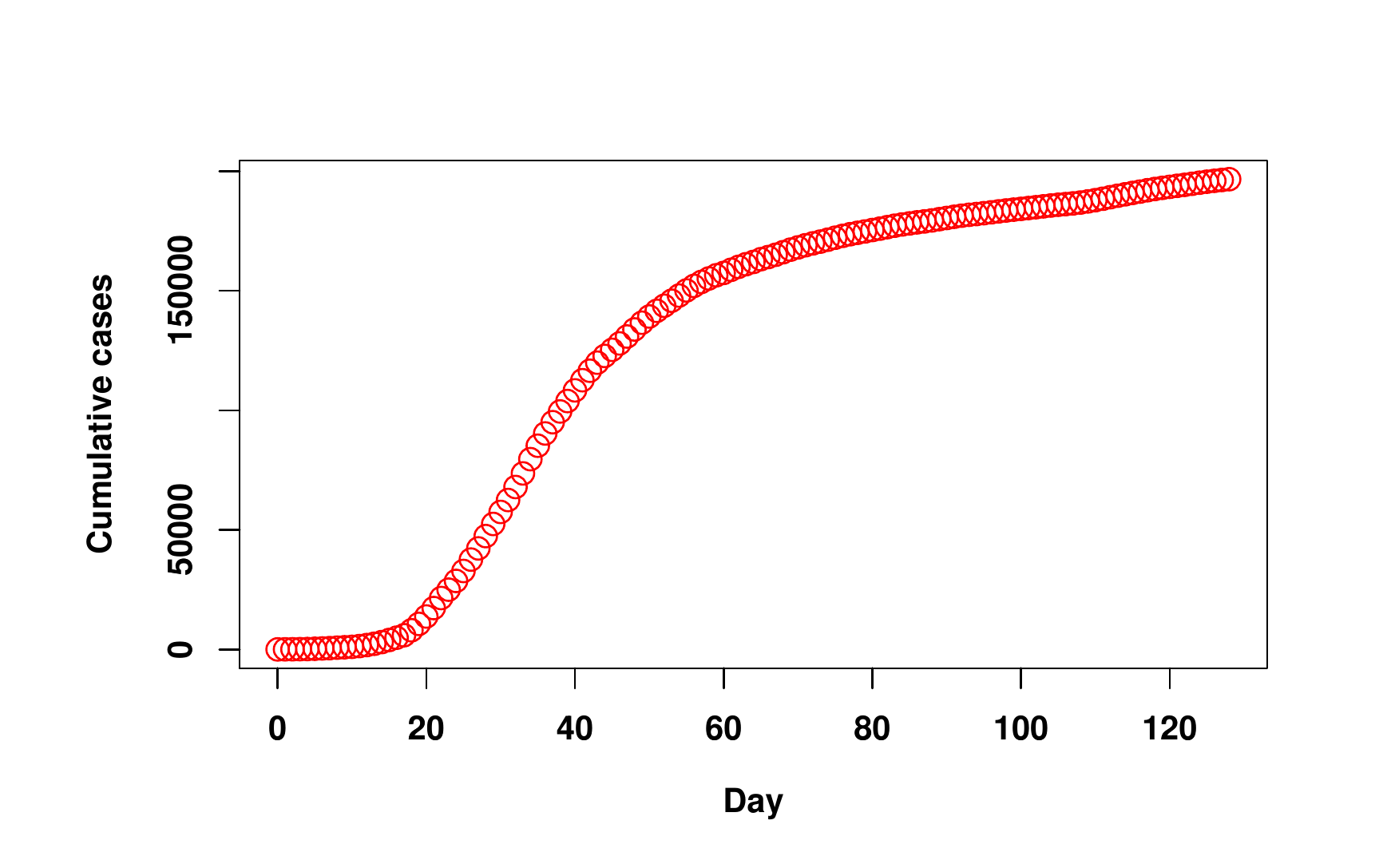}
    \caption{Size profile of Cumulative Case in Germany.}
    \label{COVID_data_plot}
  \end{subfigure}
  \begin{subfigure}{9cm}
\includegraphics[width=9cm]{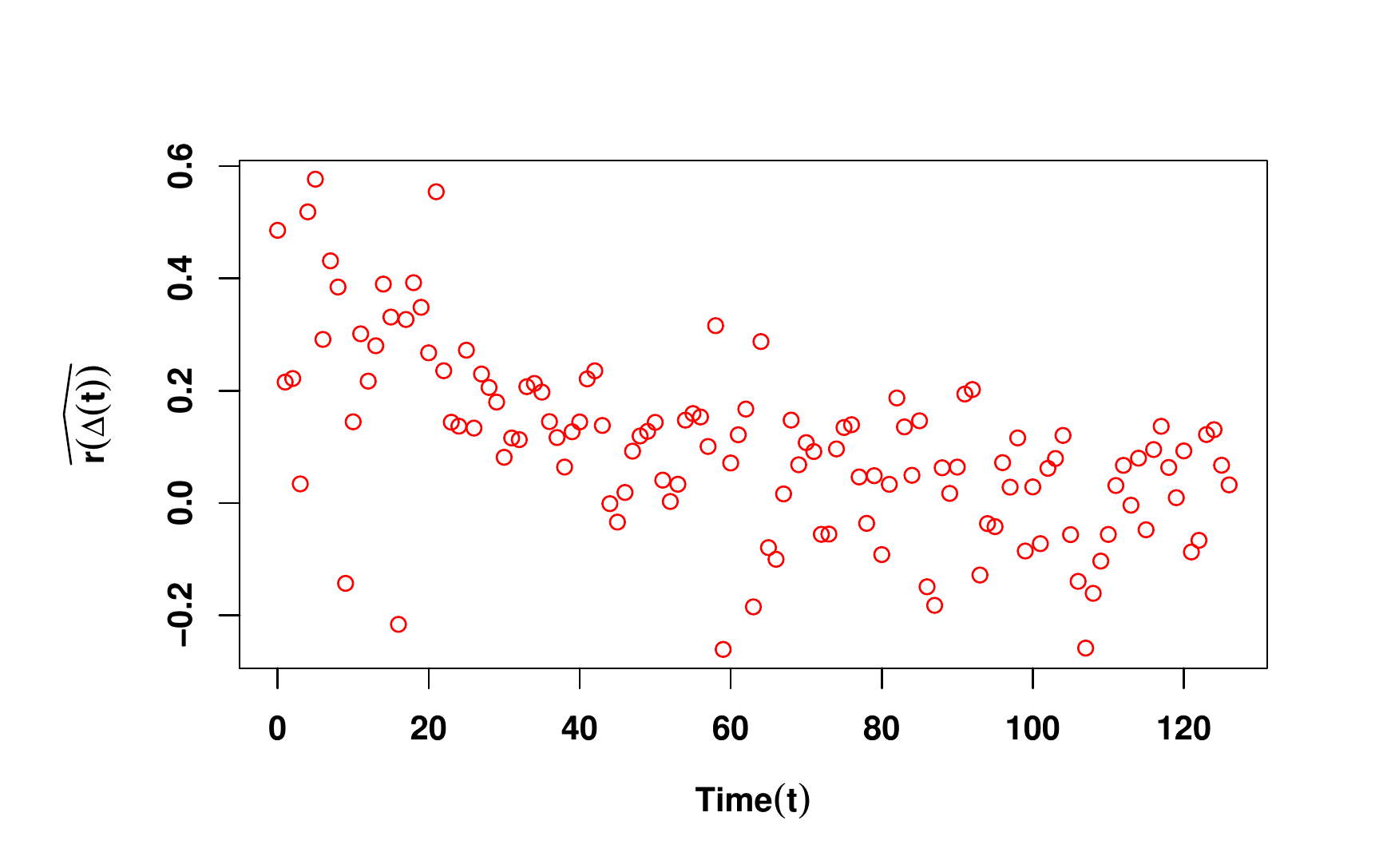}
    \caption{Interval specific estimates of $r$.}
    \label{COVID_data_ISRP_plot}
    \end{subfigure}
  
  \caption{Panel (a): Size profiles of the cumulative number of COVID-19 cases for Germany obtained from the data sets. Panel (b): We plot the ISRP of $r$ with time ($t$) which were computed from the data sets using logistic model.}
\label{COVID_dataset_plot}
\end{figure}

From the size profile (Fig.~\ref{COVID_data_plot}) of the data, it can be clearly seen that logistic growth model seems to be the best choice to start our analysis with. But from the size profile, it is hard to conclude whether or not any variation is present in $r$ (intrinsic growth rate). So for getting insight about any variation present in parameter $r$, we plotted ISRP profile (Fig.~\ref{COVID_data_ISRP_plot}) of $r$ by making use of the formula in \citet{Bhowmick2014}.

\begin{figure}[H]
    \centering
  \begin{subfigure}{9cm}
    \includegraphics[width=9cm]{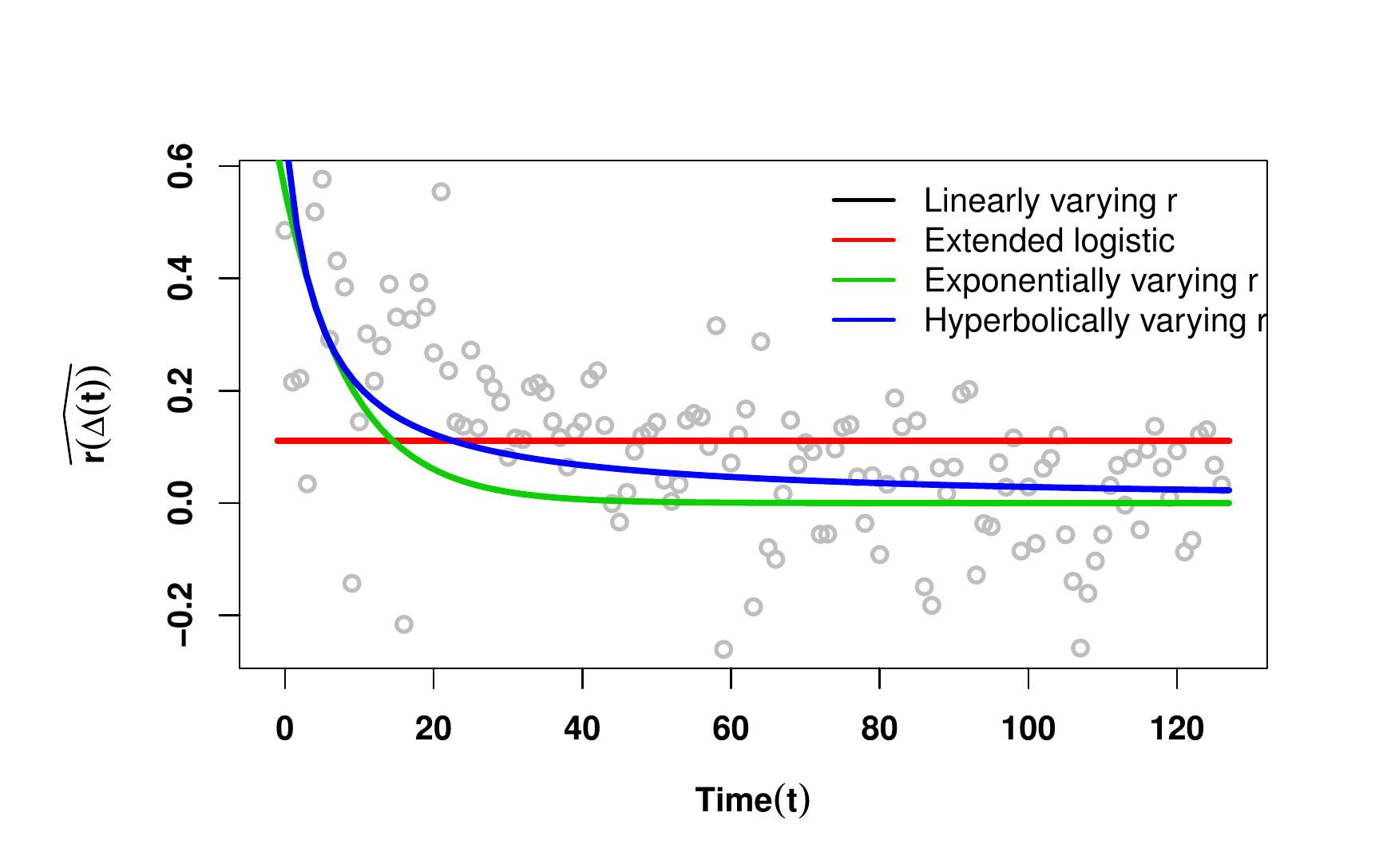}
    \caption{ISRP profile for various continuously varying form of $r$.}
    \label{COVID_data_various_ISRP_plot}
  \end{subfigure}
  \begin{subfigure}{9cm}
\includegraphics[width=9cm]{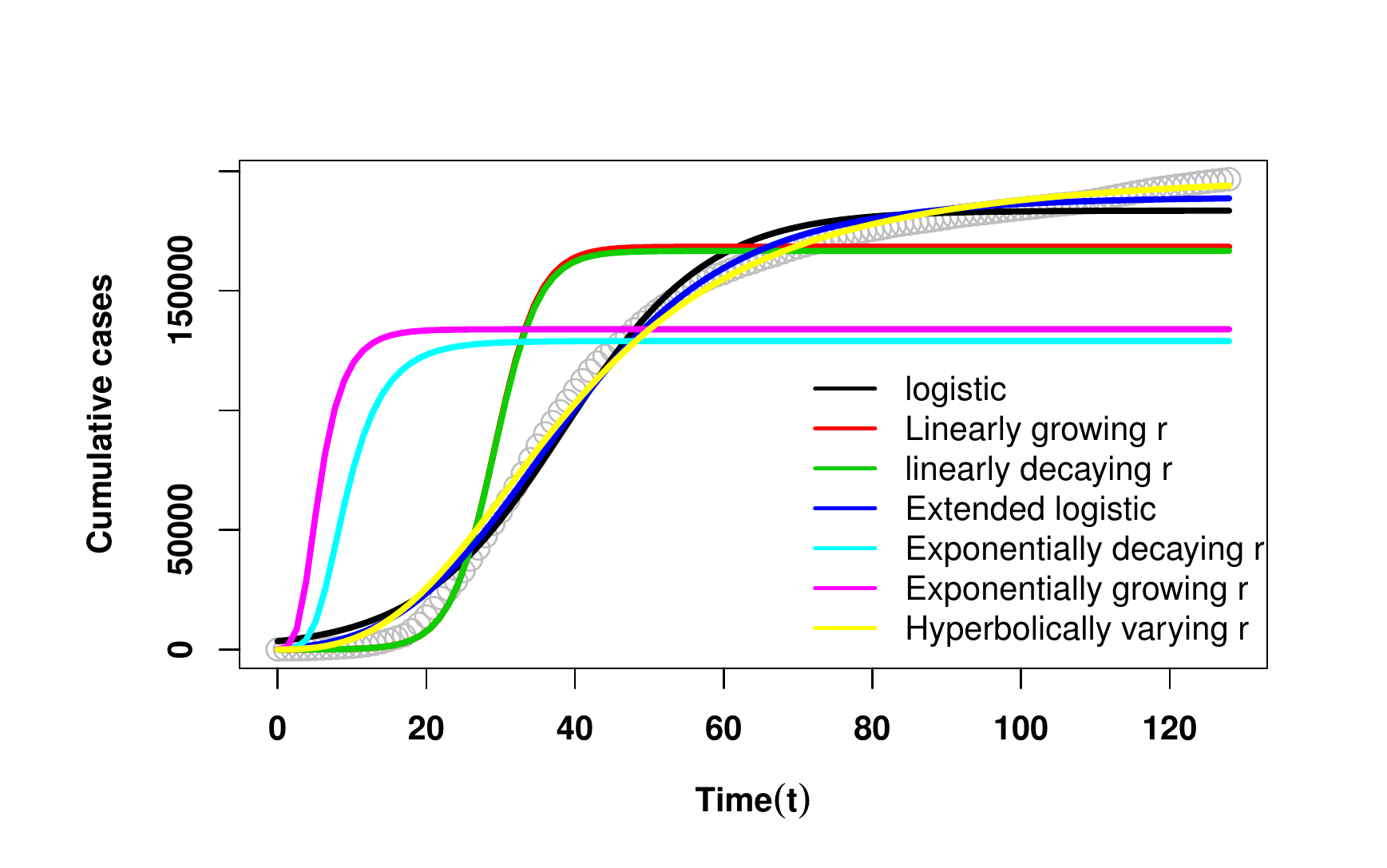}
    \caption{Best fitted model selection.}
    \label{best_fitted_model_selection_for_COVID_data}
    \end{subfigure}
  
  \caption{Panel (a): The continuous variations of parameter $r$ are fitted onto ISRP of $r$ for the data by using \texttt{nls2} function in software \texttt{R}. Since linearly growing and decaying behaves similarly for this data sets as $c=0$ in both cases so, we have considered lineally growing or decaying as linearly varying. For similar reason, we have considered exponentially varying instead of exponentially growing or decaying. Note that extended logistic and linearly varying parameter gives the same fit. Panel (b): The corresponding variations of parameter $r$ are considered in logistic growth model as in Table~\ref{table3}. The analytical solutions in Table~\ref{table3} are fitted onto the data to obtain the best fitted model for the data.}
\label{COVID_data_analysis}
\end{figure}

\begin{table}
\begin{tabular}{|p{0.4cm}|p{3.7cm}|p{1.8cm}|p{1.9cm}|p{4.3cm}|p{4cm}|}
 \hline
 \multicolumn{6}{|c|}{ RMSE and AIC values of continuous variations of parameter$r$ fitted on ISRP of $r$} \\
 \hline
Srl.  & Type of variation & RMSE ~~~values  & AIC ~~~~values & Parameter estimates (with confidence interval) & Confirmed pattern of ISRP\\
 \hline
$1$. & Linearly varying $r$  &  0.1582315 & -101.8885 & $r_0 = 0.111 ~(\pm 0.028 )$, ~~~~ $c = 0 ~(\pm 0.003)$ & No linear variation \\

$2$. & Extended logistic  &  0.1582315 & -101.8885 & $r_0 = 0.111 ~(\pm 1.42\times 10^{-2} )$, ~ $c = 1 ~(\pm 2.15\times 10^{-8})$ & No variation \\

$3$. & Exponentially varying $r$  & 0.1539704 & -108.8223 & $r_0 = 0.556 ~(\pm 0.093 )$, ~~~~ $c = 0.111 ~(\pm 0.028)$ & Exponentially increasing \\

$4$. & \textbf{Hyperbolically varying} \bm{$r$}  & \textbf{0.1385250} & \textbf{-135.6726} & \bm{$r_0 = 0.667~(\pm 0.108 )$}, ~\bm{$c =0.222 ~(\pm 0.068)$} & \textbf{Variation present and best fitted form} \\

\hline
\end{tabular}
\caption{For various continuously varying $r$ in logistic model, we calculate the RMSE and AIC values of ISRP of $r$ to identify the type of variation present in $r$ in the data sets.}
\label{COVID ISRP table}

\bigskip

\begin{tabular}{ |p{1cm}|p{6cm}|p{5cm}|p{5cm}|  }
 \hline
 \multicolumn{4}{|c|}{ RMSE and AIC values of continuously varying model for cumulative COVID data sets  } \\
 \hline
Srl.  & Model  & RMSE values  & AIC values\\
 \hline
$1$. & Logistic model (constant $r$) & 6992.228 & 2658.045 \\

$2$. & Linearly growing $r$ & 24123.600 & 2979.550 \\

$3$. & Linearly decaying $r$ & 24062.943 & 2978.901  \\

$4$. & Extended logistic     &    4793.885 & 2562.661 \\

$5$. & Exponentially decaying $r$ & 57103.655 & 3201.863 \\

$6$. & Exponentially growing $r$ & 61851.387 & 3222.469 \\

$7$. & \textbf{Hyperbolically varying} \bm{$r$} & \textbf{4684.953} & \textbf{2556.731}  \\
\hline
\end{tabular}
\caption{For various continuously varying $r$ in logistic model, we calculate the RMSE and AIC values of the model with continuously varying $r$ to find the best fitted model for the data sets.}
\label{Best fitted model selection table}

\end{table}

We have obtained the empirical estimate of ISRP of $r$ by assuming that the data generating process was logistic. To check whether there is any significant variation in $r$, we fitted the continuous functions given in Table~\ref{table3} and selected the best model based on AIC and RMSE values (Table~\ref{COVID ISRP table}). As there is no sign of presence of periodic variation in data, we have not considered any kind of periodic variation in $r$ as well as as in final fitting. Based on the summary in Table~\ref{COVID ISRP table}, we observed that the parameter $r$ was subjected to hyperbolic variation with respect to time (Fig.~\ref{COVID_data_various_ISRP_plot}). Hence, we suspect that the actual data generation process is not logistic, rather logistic growth with hyperbolically varying $r$. To support our claim, we fitted derived growth equations (in Table~\ref{table3} ) to the cumulative number of cases and found that the logistic model with hyperbolically varying growth coefficient is the best choice amongst all the models (both RMSE and AIC (Table~\ref{Best fitted model selection table}) support the conclusion )(Fig.~\ref{best_fitted_model_selection_for_COVID_data}). The analysis was carried out using nonlinear least squares method using \texttt{nls2} function available in \texttt{R}. Complete source codes are provided in the supporting online material.

Based on these three case studies, we conclude that ISRP profile acts as a key indicator for selecting the best growth model.

\section{Discussion} \label{Discussion}
The main idea of this paper is essentially based on the concept of ISRP developed by \citet{Bhowmick2014}, subsequently studied by \citet{PAL2018}. Their original work  was motivated towards the selection of the best model amongst a class of nonlinear models. In this approach, main criteria is that experimenter should be able to identify the key rate parameter ($b$) for every model. However, if we are in a situation in which the parameter varies continuously, then it puts us into the following problems in applications: (1) One needs to make a very good guess about the final model and the analyst should be able to express the RGR of the final model in the form $bg(t)$, for some real valued function $g$; (2) The final model might turn out to be very complicated in nature so that the computation of ISRP becomes difficult and some parameters may not have an explicit expression of ISRP as well. It can be easily understood that both the tasks are quite difficult and may lead wrong conclusion. Our proposed idea apparently resolved both these issues as discussed below.

Firstly, it reduces the search space of selecting the best model by investigating only a few models. As depicted in Fig.~\ref{Relation_flowchart_between models} and in Table~\ref{table2}, \ref{table3}, \ref{table4}, \ref{table5} and \ref{table6}, most of the models can be obtained from four models (Exponential Model, Logistic Model, Theta-Logistic Model, Confined Exponential Model). So, an experimenter does not need to compute ISRP of all the models which is itself may be a tedious computational process. Only ISRP profile from these four models will give the clue for the selection of the underlying model. Thus, this work further reinforces the use of ISRP and generalizes its application in model selection at a reduced effort.

The other important observation is that ISRP profile of parameter is very informative in a sense that, it depicts homogeneity or heterogeneity across different growth trajectories with respect to the parameter of interest, for example, whether the parameter varies or not, can be traced only in some particular time period. If we observe the ISRP profile of $r$ (Fig.~\ref{r_hat_delta_method_b}),  we notice that till time point $12$, the variation are small, and the same thing happens when it varies continuously (Fig.~\ref{r_hat_extended_quadratic_boxplot}, \ref{r_hat_extended_linear_boxplot}). Hence one could argue that information about the parameter in the data are contained during specific time interval only. Beyond that, data points are not informative as they show very high variability with respect to the parameter of interest.  

Recently, \citet{Chakraborty2019} proposed a unification function to unify a large number of equations. In their unifying function, different choices of parameters lead to different models, thus giving a compact representation of several growth equations. It is worth mentioning that from a mathematical perspective, the unifying function serves a great purpose, but from a statistical point of view it may pose difficulty in dealing with real data sets as the unifying function itself is heavily parametrized by several fixed but unknown real valued parameters. Thus, if we plot a network similar to Fig. \ref{Relation_flowchart_between models}, the network will have only one key node (with unifying function) and all other models will be leaf with no connection between them. Thus in application, essentially an experimenter need to resolve a difficult estimation problem to obtain a simpler equation. On the contrary, our approach in this manuscript starts with the investigation of simpler models and extrapolate to complex models (if required at all). In addition the network type plot, depicted in Fig.~\ref{Relation_flowchart_between models} is more informative than a plot with single key node, because Fig.~\ref{Relation_flowchart_between models} depicts connection between close approximating models as well, hence giving more broader picture. Hence, if we describe the approach in \citet{Chakraborty2019} as a ``complex to simple" strategy, then our approach is ``simple to complex". 

The distribution of the interval specific estimators also depends on the covariance structure of the statistical model. The Koopman covariance structure is homoschedastic which assumes equal population variance for all time points. However, in reality, this may not be always true. For example, \citet{Diniz2012} considered the multiplicative heteroschedasticity and estimated the parameters of Von Bertalanffy model under a fully Bayesian set up. Such instances are abundant in growth studies. \citet{Louzada2014} analyzed the growth data under both normal and skew-normal distribution of the error structure with homoschedastic, homoschedastic of lag 1 (Koopman) and multiplicative heteroschedastic covariance matrix. Till date the distributions of ISRP has only been investigated under the homoschedastic covariance matrix with autocorrelation lag 1. It would be an interesting research avenue to study the distribution of the interval estimators under different assumptions of covariance error structure by means of theoretical interest and broader application of the methods presented in this manuscript and by \citet{Bhowmick2014} and \citet{PAL2018}. 

It is to be noted that in Fig.~\ref{r_hat_extended_quadratic_boxplot}, the values of ISRP ($\widehat{r_j(\Delta t)} $) with respect to time is indicative of deviation from linearity at the initial phase of the growth only. This is quite natural as the impact of the growth coefficient is more prominent at the early stage and as the process evolve, the information about $r$ will be lost. Basically, before the lag phase starts, logistic growth essentially behaves like exponential only. A similar idea has been discussed in the context of estimating the patterns of density dependence for natural populations \citep{clark2010}. The theta-logistic curve was calibrated using a large number of data sets by \citet{sibly2005}. In the model, the per capita growth rate ($pgr$) is modelled by $pgr=r_m \left[1-\left(\frac{N}{K}\right)^{\theta}\right]$, where $r_m$ is the intrinsic growth rate and this model have the maximum $pgr$ when the population size is small. In other words, unless the population sizes are observed at low population densities, estimation of $r_m$ will not be reliable. So, if the population size fluctuates around the carrying capacity, then the estimated model can be one of these types: linear, concave upward or concave downward \citep{clark2010}. Thus, it may lead to dubious inference regarding the fate of the populations. Similar is the idea here, at early stage of growth the parameter $r$ plays crucial role, or in other words, observations taken at the early stage of growth are more informative about $r$ rather than later stage.  

Innovation of interval specific rate parameter is not only useful in better approximation of growth rates, but to understand parameter sensitivity as well in a growth process. If the parameter varies continuously, its identification can be done using interval specific estimates. Similar to the continuously varying parameters, growth models with randomly varying parameters are also significantly available in growth literature. It would be our future endeavor to explore the domain of randomly varying parameters and their impacts in modeling and assessment using real data sets. In addition, it would be our future endeavor to develop general purpose programs that will integrate the connections among growth curve models and guide correct model selection based on the patterns of interval estimates of the model parameters.

\section{Conclusion}
Theory of growth curve modelling is fundamental to understand any biological process. Building a good approximating model for the underlying growth process requires a good identification of the parameters in the model. In this manuscript, we give a detailed historical development of this domain and justified the need of an unifying study specifically targeting this problem. By extending the idea of \citet{Banks1994}, we have identified that several existing growth models in the literature can be connected. We shown here that such connections between growth equations significantly reduce the efforts in choosing an optimal model. Using the idea of Interval Specific Estimator by \citet{Bhowmick2014}, we proposed a statistical method to identify whether the parameter in the model varies with time. Our proposed methodology significantly reduced the efforts involved in model fitting exercises. We believe that this work would be helpful for the practitioners in the field of growth study. The proposed idea is verified by using simulated and real data sets from two different domains (biology and marketing). We believe that this idea is unique and it contains a novel message for the scientific community, in particular for applied researchers.

\section{Acknowledgement}
Karim is thankful to the Council of Scientific \& Industrial Research (CSIR), Government of India for the financial support in the form of junior research fellowship (Grant No. 09/991(0057)/2019-EMR-\Rmnum{1}). Authors are thankful to Ayan Paul for valuable discussions and assisting in computations. 

\bibliographystyle{elsarticle-harv}
\bibliography{main}

\begin{thebibliography}{64}
\expandafter\ifx\csname natexlab\endcsname\relax\def\natexlab#1{#1}\fi
\providecommand{\url}[1]{\texttt{#1}}
\providecommand{\href}[2]{#2}
\providecommand{\path}[1]{#1}
\providecommand{\DOIprefix}{doi:}
\providecommand{\ArXivprefix}{arXiv:}
\providecommand{\URLprefix}{URL: }
\providecommand{\Pubmedprefix}{pmid:}
\providecommand{\doi}[1]{\href{http://dx.doi.org/#1}{\path{#1}}}
\providecommand{\Pubmed}[1]{\href{pmid:#1}{\path{#1}}}
\providecommand{\bibinfo}[2]{#2}
\ifx\xfnm\relax \def\xfnm[#1]{\unskip,\space#1}\fi
\bibitem[{Anderson et~al.(2015)Anderson, Jovanoskia, Towersa and
  Sidhua}]{anderson2015}
\bibinfo{author}{Anderson, C.}, \bibinfo{author}{Jovanoskia, Z.},
  \bibinfo{author}{Towersa, I.}, \bibinfo{author}{Sidhua, H.},
  \bibinfo{year}{2015}.
\newblock \bibinfo{title}{A simple population model with a stochastic carrying
  capacity}.
\newblock \bibinfo{journal}{21st International Congress on Modelling and
  Simulation} .
\bibitem[{Arrigoni and Steiner(1985)}]{arrigoni1985}
\bibinfo{author}{Arrigoni, M.}, \bibinfo{author}{Steiner, A.},
  \bibinfo{year}{1985}.
\newblock \bibinfo{title}{Logistic growth in a fluctuating environment}.
\newblock \bibinfo{journal}{Journal of Mathematical Biology}
  \bibinfo{volume}{21}, \bibinfo{pages}{237--241}.
\bibitem[{Banks(1994)}]{Banks1994}
\bibinfo{author}{Banks, R.B.}, \bibinfo{year}{1994}.
\newblock \bibinfo{title}{Growth and Diffusion Phenomena}.
\newblock \bibinfo{publisher}{Springer}.
\bibitem[{Barker and Sibly(2008)}]{BARKER2008}
\bibinfo{author}{Barker, D.}, \bibinfo{author}{Sibly, R.M.},
  \bibinfo{year}{2008}.
\newblock \bibinfo{title}{The effects of environmental perturbation and
  measurement error on estimates of the shape parameter in the theta-logistic
  model of population regulation}.
\newblock \bibinfo{journal}{Ecological Modelling} \bibinfo{volume}{219},
  \bibinfo{pages}{170 -- 177}.
\bibitem[{Beck(1982)}]{BECK1982}
\bibinfo{author}{Beck, K.}, \bibinfo{year}{1982}.
\newblock \bibinfo{title}{A model of the population genetics of cystic fibrosis
  in the {U}nited {S}tates}.
\newblock \bibinfo{journal}{Mathematical Biosciences} \bibinfo{volume}{58},
  \bibinfo{pages}{243 -- 257}.
\bibitem[{Bhowmick et~al.(2016)Bhowmick, Bandyopadhyay, Rana and
  Bhattacharya}]{BHOWMICK2016}
\bibinfo{author}{Bhowmick, A.R.}, \bibinfo{author}{Bandyopadhyay, S.},
  \bibinfo{author}{Rana, S.}, \bibinfo{author}{Bhattacharya, S.},
  \bibinfo{year}{2016}.
\newblock \bibinfo{title}{A simple approximation of moments of the
  quasi-equilibrium distribution of an extended stochastic theta-logistic model
  with non-integer powers}.
\newblock \bibinfo{journal}{Mathematical Biosciences} \bibinfo{volume}{271},
  \bibinfo{pages}{96 -- 112}.
\bibitem[{Bhowmick and Bhattacharya(2014)}]{BhowmickMathBios2014}
\bibinfo{author}{Bhowmick, A.R.}, \bibinfo{author}{Bhattacharya, S.},
  \bibinfo{year}{2014}.
\newblock \bibinfo{title}{A new growth curve model for biological growth: Some
  inferential studies on the growth of \textit{{C}irrhinus mrigala}}.
\newblock \bibinfo{journal}{Mathematical Biosciences} \bibinfo{volume}{254},
  \bibinfo{pages}{28 -- 41}.
\bibitem[{Bhowmick et~al.(2014)Bhowmick, Chattopadhyay and
  Bhattacharya}]{Bhowmick2014}
\bibinfo{author}{Bhowmick, A.R.}, \bibinfo{author}{Chattopadhyay, G.},
  \bibinfo{author}{Bhattacharya, S.}, \bibinfo{year}{2014}.
\newblock \bibinfo{title}{Simultaneous identification of growth law and
  estimation of its rate parameter for biological growth data: a new approach}.
\newblock \bibinfo{journal}{Journal of Biological Physics}
  \bibinfo{volume}{40}, \bibinfo{pages}{71--95}.
\bibitem[{Bhowmick et~al.(2015)Bhowmick, Saha, Chattopadhyay, Ray and
  Bhattacharya}]{BHOWMICK2015}
\bibinfo{author}{Bhowmick, A.R.}, \bibinfo{author}{Saha, B.},
  \bibinfo{author}{Chattopadhyay, J.}, \bibinfo{author}{Ray, S.},
  \bibinfo{author}{Bhattacharya, S.}, \bibinfo{year}{2015}.
\newblock \bibinfo{title}{Cooperation in species: {I}nterplay of population
  regulation and extinction through global population dynamics database}.
\newblock \bibinfo{journal}{Ecological Modelling} \bibinfo{volume}{312},
  \bibinfo{pages}{150 -- 165}.
\bibitem[{Bridges et~al.(2000)Bridges, Turner, Gates and Smith}]{Bridges2000}
\bibinfo{author}{Bridges, T.C.}, \bibinfo{author}{Turner, L.W.},
  \bibinfo{author}{Gates, R.S.}, \bibinfo{author}{Smith, E.M.},
  \bibinfo{year}{2000}.
\newblock \bibinfo{title}{Relativity of {G}rowth in {L}aboratory {F}arm
  {A}nimals: {I}. {R}epresentation of {P}hysiological {A}ge and the {G}rowth
  {R}ate {T}ime {C}onstant}.
\newblock \bibinfo{journal}{American Society of Agricultural Engineers}
  \bibinfo{volume}{43}, \bibinfo{pages}{1803--1810}.
\bibitem[{Burnham and Anderson(2002)}]{burnham2002}
\bibinfo{author}{Burnham, K.}, \bibinfo{author}{Anderson, D.},
  \bibinfo{year}{2002}.
\newblock \bibinfo{title}{Model selection and multimodel inference: a practical
  information-theoretic approach}.
\newblock \bibinfo{publisher}{Springer Verlag}.
\bibitem[{Casella and Berger(2002)}]{casella2002}
\bibinfo{author}{Casella, G.}, \bibinfo{author}{Berger, R.L.},
  \bibinfo{year}{2002}.
\newblock \bibinfo{title}{Statistical inference}. volume~\bibinfo{volume}{2}.
\newblock \bibinfo{publisher}{Duxbury Pacific Grove, CA}.
\bibitem[{Chakraborty et~al.(2017)Chakraborty, Bhowmick, Chattopadhyay and
  Bhattacharya}]{CHAKRABORTY2017}
\bibinfo{author}{Chakraborty, B.}, \bibinfo{author}{Bhowmick, A.R.},
  \bibinfo{author}{Chattopadhyay, J.}, \bibinfo{author}{Bhattacharya, S.},
  \bibinfo{year}{2017}.
\newblock \bibinfo{title}{Physiological responses of fish under environmental
  stress and extension of growth (curve) models}.
\newblock \bibinfo{journal}{Ecological Modelling} \bibinfo{volume}{363},
  \bibinfo{pages}{172 -- 186}.
\bibitem[{Chakraborty et~al.(2019)Chakraborty, Bhowmick, Chattopadhyay and
  Bhattacharya}]{Chakraborty2019}
\bibinfo{author}{Chakraborty, B.}, \bibinfo{author}{Bhowmick, A.R.},
  \bibinfo{author}{Chattopadhyay, J.}, \bibinfo{author}{Bhattacharya, S.},
  \bibinfo{year}{2019}.
\newblock \bibinfo{title}{A {N}ovel {U}nification {M}ethod to {C}haracterize a
  {B}road {C}lass of {G}rowth {C}urve {M}odels {U}sing {R}elative {G}rowth
  {R}ate}.
\newblock \bibinfo{journal}{Bulletin of Mathematical Biology}
  \bibinfo{volume}{81}, \bibinfo{pages}{2529--2552}.
\bibitem[{Clark et~al.(2010)Clark, Brook, Delean, Re{\c{s}}it~Ak{\c{c}}akaya
  and Bradshaw}]{clark2010}
\bibinfo{author}{Clark, F.}, \bibinfo{author}{Brook, B.W.},
  \bibinfo{author}{Delean, S.}, \bibinfo{author}{Re{\c{s}}it~Ak{\c{c}}akaya,
  H.}, \bibinfo{author}{Bradshaw, C.J.}, \bibinfo{year}{2010}.
\newblock \bibinfo{title}{The theta-logistic is unreliable for modelling most
  census data}.
\newblock \bibinfo{journal}{Methods in Ecology and Evolution}
  \bibinfo{volume}{1}, \bibinfo{pages}{253--262}.
\bibitem[{Coleman(1979)}]{COLEMAN1979}
\bibinfo{author}{Coleman, B.D.}, \bibinfo{year}{1979}.
\newblock \bibinfo{title}{Nonautonomous logistic equations as models of the
  adjustment of populations to environmental change}.
\newblock \bibinfo{journal}{Mathematical Biosciences} \bibinfo{volume}{45},
  \bibinfo{pages}{159 -- 173}.
\bibitem[{Crescenzo and Spina(2016)}]{CRESCENZO2016}
\bibinfo{author}{Crescenzo, A.D.}, \bibinfo{author}{Spina, S.},
  \bibinfo{year}{2016}.
\newblock \bibinfo{title}{Analysis of a growth model inspired by {G}ompertz and
  {K}orf laws, and an analogous birth-death process}.
\newblock \bibinfo{journal}{Mathematical Biosciences} \bibinfo{volume}{282},
  \bibinfo{pages}{121 -- 134}.
\bibitem[{Cushing(1977)}]{Cushing1977}
\bibinfo{author}{Cushing, J.}, \bibinfo{year}{1977}.
\newblock \bibinfo{title}{Periodic {T}ime-{D}ependent {P}redator-{P}rey
  {S}ystems}.
\newblock \bibinfo{journal}{SIAM Journal on Applied Mathematics}
  \bibinfo{volume}{32}, \bibinfo{pages}{82--95}.
\bibitem[{Diniz et~al.(2012)Diniz, Louzada-Neto and Morita}]{Diniz2012}
\bibinfo{author}{Diniz, C.A.R.}, \bibinfo{author}{Louzada-Neto, F.},
  \bibinfo{author}{Morita, L.H.M.}, \bibinfo{year}{2012}.
\newblock \bibinfo{title}{The multiplicative heteroscedastic {V}on
  {B}ertalanffy model}.
\newblock \bibinfo{journal}{Brazilian Journal of Probability and Statistics}
  \bibinfo{volume}{26}, \bibinfo{pages}{71--81}.
\bibitem[{Dubkov and Spagnolo(2008)}]{Dubkov2008}
\bibinfo{author}{Dubkov, A.A.}, \bibinfo{author}{Spagnolo, B.},
  \bibinfo{year}{2008}.
\newblock \bibinfo{title}{Verhulst model with l{\'e}vy white noise excitation}.
\newblock \bibinfo{journal}{The European Physical Journal B}
  \bibinfo{volume}{65}, \bibinfo{pages}{361--367}.
\bibitem[{Ebert and Weisser(1997)}]{Ebart1997}
\bibinfo{author}{Ebert, D.}, \bibinfo{author}{Weisser, W.W.},
  \bibinfo{year}{1997}.
\newblock \bibinfo{title}{Optimal killing for obligate killers: the evolution
  of life histories and virulence of semelparous parasites}.
\newblock \bibinfo{journal}{Proceedings of the Royal Society of London. Series
  B: Biological Sciences} \bibinfo{volume}{264}, \bibinfo{pages}{985--991}.
\bibitem[{Freedman(1980)}]{Freedman}
\bibinfo{author}{Freedman, H.I.}, \bibinfo{year}{1980}.
\newblock \bibinfo{title}{Deterministic mathematical models in population
  ecology}. volume~\bibinfo{volume}{57}.
\newblock \bibinfo{publisher}{Marcel Dekker Incorporated}.
\bibitem[{Gilpin and Ayala(1973)}]{Gilpin3590}
\bibinfo{author}{Gilpin, M.E.}, \bibinfo{author}{Ayala, F.J.},
  \bibinfo{year}{1973}.
\newblock \bibinfo{title}{Global {M}odels of {G}rowth and {C}ompetition}.
\newblock \bibinfo{journal}{Proceedings of the National Academy of Sciences}
  \bibinfo{volume}{70}, \bibinfo{pages}{3590--3593}.
\bibitem[{Gompertz(1825)}]{Gompertz1825}
\bibinfo{author}{Gompertz, B.}, \bibinfo{year}{1825}.
\newblock \bibinfo{title}{Xxiv. on the nature of the function expressive of the
  law of human mortality, and on a new mode of determining the value of life
  contingencies. in a letter to francis baily, esq. f. r. s. \& amp;c}.
\newblock \bibinfo{journal}{Philosophical Transactions of the Royal Society of
  London} \bibinfo{volume}{115}, \bibinfo{pages}{513--583}.
\bibitem[{Hallam and Clark(1981)}]{HALLAM1981}
\bibinfo{author}{Hallam, T.}, \bibinfo{author}{Clark, C.},
  \bibinfo{year}{1981}.
\newblock \bibinfo{title}{Non-autonomous logistic equations as models of
  populations in a deteriorating environment}.
\newblock \bibinfo{journal}{Journal of Theoretical Biology}
  \bibinfo{volume}{93}, \bibinfo{pages}{303 -- 311}.
\bibitem[{Ikeda and Yokoi(1980)}]{IKEDA1980}
\bibinfo{author}{Ikeda, S.}, \bibinfo{author}{Yokoi, T.}, \bibinfo{year}{1980}.
\newblock \bibinfo{title}{Fish population dynamics under nutrient enrichment
  — a case of the {E}ast {S}eto {I}nland {S}ea}.
\newblock \bibinfo{journal}{Ecological Modelling} \bibinfo{volume}{10},
  \bibinfo{pages}{141 -- 165}.
\bibitem[{Kenward(1987)}]{Kenward1987}
\bibinfo{author}{Kenward, M.G.}, \bibinfo{year}{1987}.
\newblock \bibinfo{title}{A method for comparing profiles of repeated
  measurements}.
\newblock \bibinfo{journal}{Journal of the Royal Statistical Society. Series C
  (Applied Statistics)} \bibinfo{volume}{36}, \bibinfo{pages}{296--308}.
\bibitem[{Koopmans(1942)}]{Koopmans}
\bibinfo{author}{Koopmans, T.}, \bibinfo{year}{1942}.
\newblock \bibinfo{title}{Serial {C}orrelation and {Q}uadratic {F}orms in
  {N}ormal {V}ariables}.
\newblock \bibinfo{journal}{The Annals of Mathematical Statistics}
  \bibinfo{volume}{13}, \bibinfo{pages}{14--33}.
\bibitem[{Korf(1939)}]{korf1939}
\bibinfo{author}{Korf, V.}, \bibinfo{year}{1939}.
\newblock \bibinfo{title}{Contribution to mathematical definition of the law of
  stand volume growth}.
\newblock \bibinfo{journal}{Lesnicka Prace} \bibinfo{volume}{18},
  \bibinfo{pages}{339--379}.
\bibitem[{Kot(2001)}]{Kot2001}
\bibinfo{author}{Kot, M.}, \bibinfo{year}{2001}.
\newblock \bibinfo{title}{Elements of Mathematical Ecology.}
\newblock \bibinfo{publisher}{Cambridge Univ. Press},
  \bibinfo{address}{Cambridge, UK}.
\bibitem[{Koya and Goshu(2013)}]{koya2013}
\bibinfo{author}{Koya, P.R.}, \bibinfo{author}{Goshu, A.T.},
  \bibinfo{year}{2013}.
\newblock \bibinfo{title}{Generalized mathematical model for biological
  growths}.
\newblock \bibinfo{journal}{Open Journal of Modelling and Simulation}
  \bibinfo{volume}{1}, \bibinfo{pages}{42}.
\bibitem[{Lakshmi(2003)}]{LAKSHMI2003}
\bibinfo{author}{Lakshmi, B.}, \bibinfo{year}{2003}.
\newblock \bibinfo{title}{Oscillating population models}.
\newblock \bibinfo{journal}{Chaos, Solitons \& Fractals} \bibinfo{volume}{16},
  \bibinfo{pages}{183 -- 186}.
\bibitem[{Lakshmi(2005)}]{LAKSHMI2005}
\bibinfo{author}{Lakshmi, B.}, \bibinfo{year}{2005}.
\newblock \bibinfo{title}{Population models with time dependent parameters}.
\newblock \bibinfo{journal}{Chaos, Solitons \& Fractals} \bibinfo{volume}{26},
  \bibinfo{pages}{719 -- 721}.
\bibitem[{Leach and Andriopoulos(2004)}]{LEACH2004}
\bibinfo{author}{Leach, P.}, \bibinfo{author}{Andriopoulos, K.},
  \bibinfo{year}{2004}.
\newblock \bibinfo{title}{An oscillatory population model}.
\newblock \bibinfo{journal}{Chaos, Solitons \& Fractals} \bibinfo{volume}{22},
  \bibinfo{pages}{1183 -- 1188}.
\bibitem[{L{\'o}pez-Ruiz and Fournier-Prunaret(2005)}]{LOPEZ2004}
\bibinfo{author}{L{\'o}pez-Ruiz, R.}, \bibinfo{author}{Fournier-Prunaret, D.},
  \bibinfo{year}{2005}.
\newblock \bibinfo{title}{Indirect {A}llee effect, bistability and chaotic
  oscillations in a predator–prey discrete model of logistic type}.
\newblock \bibinfo{journal}{Chaos, Solitons \& Fractals} \bibinfo{volume}{24},
  \bibinfo{pages}{85 -- 101}.
\bibitem[{Louzada et~al.(2014)Louzada, Ferreira and Diniz}]{Louzada2014}
\bibinfo{author}{Louzada, F.}, \bibinfo{author}{Ferreira, P.H.},
  \bibinfo{author}{Diniz, C.A.}, \bibinfo{year}{2014}.
\newblock \bibinfo{title}{Skew-normal distribution for growth curve models in
  presence of a heteroscedasticity structure}.
\newblock \bibinfo{journal}{Journal of Applied Statistics}
  \bibinfo{volume}{41}, \bibinfo{pages}{1785--1798}.
\bibitem[{Malthus(1798)}]{malthus1798}
\bibinfo{author}{Malthus, T.R.}, \bibinfo{year}{1798}.
\newblock \bibinfo{title}{An {E}ssey on the {P}rinciple of population, as {I}t
  {A}ffects the future {I}mprovement of {S}ociety with {R}emarks on the
  {S}peculations of {M}r. {G}odwin}.
\newblock \bibinfo{journal}{Condorcet, and Other Writers. J. Johnson in St
  Paul's Churchyard, London} .
\bibitem[{Marusic and Bajzer(1993)}]{MARUSIC1993}
\bibinfo{author}{Marusic, M.}, \bibinfo{author}{Bajzer, Z.},
  \bibinfo{year}{1993}.
\newblock \bibinfo{title}{Generalized {T}wo-{P}arameter {E}quation of
  {G}rowth}.
\newblock \bibinfo{journal}{Journal of Mathematical Analysis and Applications}
  \bibinfo{volume}{179}, \bibinfo{pages}{446 -- 462}.
\bibitem[{M{\'e}ndez et~al.(2010)M{\'e}ndez, Llopis, Campos and
  Horsthemke}]{MENDEZ2010}
\bibinfo{author}{M{\'e}ndez, V.}, \bibinfo{author}{Llopis, I.},
  \bibinfo{author}{Campos, D.}, \bibinfo{author}{Horsthemke, W.},
  \bibinfo{year}{2010}.
\newblock \bibinfo{title}{Extinction conditions for isolated populations
  affected by environmental stochasticity}.
\newblock \bibinfo{journal}{Theoretical Population Biology}
  \bibinfo{volume}{77}, \bibinfo{pages}{250 -- 256}.
\bibitem[{Meyer and Ausubel(1999)}]{MEYER1999}
\bibinfo{author}{Meyer, P.S.}, \bibinfo{author}{Ausubel, J.H.},
  \bibinfo{year}{1999}.
\newblock \bibinfo{title}{Carrying {C}apacity: {A} model with {L}ogistically
  {V}arying {L}imits}.
\newblock \bibinfo{journal}{Technological Forecasting and Social Change}
  \bibinfo{volume}{61}, \bibinfo{pages}{209 -- 214}.
\bibitem[{Nisbet and Gurney(1976)}]{nisbet1976}
\bibinfo{author}{Nisbet, R.}, \bibinfo{author}{Gurney, W.},
  \bibinfo{year}{1976}.
\newblock \bibinfo{title}{Population dynamics in a periodically varying
  environment}.
\newblock \bibinfo{journal}{Journal of Theoretical Biology}
  \bibinfo{volume}{56}, \bibinfo{pages}{459--475}.
\bibitem[{Pal et~al.(2018)Pal, Bhowmick, Yeasmin and Bhattacharya}]{PAL2018}
\bibinfo{author}{Pal, A.}, \bibinfo{author}{Bhowmick, A.R.},
  \bibinfo{author}{Yeasmin, F.}, \bibinfo{author}{Bhattacharya, S.},
  \bibinfo{year}{2018}.
\newblock \bibinfo{title}{Evolution of model specific relative growth rate:
  {I}ts genesis and performance over {F}isher’s growth rates}.
\newblock \bibinfo{journal}{Journal of Theoretical Biology}
  \bibinfo{volume}{444}, \bibinfo{pages}{11 -- 27}.
\bibitem[{Perotto et~al.(1992)Perotto, Cue and Lee}]{perotto1992}
\bibinfo{author}{Perotto, D.}, \bibinfo{author}{Cue, R.}, \bibinfo{author}{Lee,
  A.}, \bibinfo{year}{1992}.
\newblock \bibinfo{title}{Comparison of nonlinear functions for describing the
  growth curve of three genotypes of dairy cattle}.
\newblock \bibinfo{journal}{Canadian Journal of Animal Science}
  \bibinfo{volume}{72}, \bibinfo{pages}{773--782}.
\bibitem[{Richards(1959)}]{Richards1959}
\bibinfo{author}{Richards, F.J.}, \bibinfo{year}{1959}.
\newblock \bibinfo{title}{{A {F}lexible {G}rowth {F}unction for {E}mpirical
  {U}se}}.
\newblock \bibinfo{journal}{Journal of Experimental Botany}
  \bibinfo{volume}{10}, \bibinfo{pages}{290--301}.
\bibitem[{Rogovchenko and Rogovchenko(2009)}]{ROGOVCHENKO2009}
\bibinfo{author}{Rogovchenko, S.P.}, \bibinfo{author}{Rogovchenko, Y.V.},
  \bibinfo{year}{2009}.
\newblock \bibinfo{title}{Effect of periodic environmental fluctuations on the
  {P}earl–{V}erhulst model}.
\newblock \bibinfo{journal}{Chaos, Solitons \& Fractals} \bibinfo{volume}{39},
  \bibinfo{pages}{1169 -- 1181}.
\bibitem[{S{\ae}ther et~al.(1998)S{\ae}ther, Engen, Islam, McCleery and
  Perrins}]{saether1998}
\bibinfo{author}{S{\ae}ther, B.E.}, \bibinfo{author}{Engen, S.},
  \bibinfo{author}{Islam, A.}, \bibinfo{author}{McCleery, R.},
  \bibinfo{author}{Perrins, C.}, \bibinfo{year}{1998}.
\newblock \bibinfo{title}{Environmental stochasticity and extinction risk in a
  population of a small songbird, the great tit}.
\newblock \bibinfo{journal}{The American Naturalist} \bibinfo{volume}{151},
  \bibinfo{pages}{441--450}.
\bibitem[{Safuan et~al.(2011)Safuan, Towers, Jovanoski and Sidhu}]{safuan2011}
\bibinfo{author}{Safuan, H.}, \bibinfo{author}{Towers, I.N.},
  \bibinfo{author}{Jovanoski, Z.}, \bibinfo{author}{Sidhu, H.},
  \bibinfo{year}{2011}.
\newblock \bibinfo{title}{A simple model for the total microbial biomass under
  occlusion of healthy human skin}.
\newblock \bibinfo{journal}{Modelling and Simulation Society of Australia and
  New Zealand} , \bibinfo{pages}{733--739}.
\bibitem[{Safuan et~al.(2013)Safuan, Jovanoski, Towers and Sidhu}]{SAFUAN2013}
\bibinfo{author}{Safuan, H.M.}, \bibinfo{author}{Jovanoski, Z.},
  \bibinfo{author}{Towers, I.N.}, \bibinfo{author}{Sidhu, H.S.},
  \bibinfo{year}{2013}.
\newblock \bibinfo{title}{Exact solution of a non-autonomous logistic
  population model}.
\newblock \bibinfo{journal}{Ecological Modelling} \bibinfo{volume}{251},
  \bibinfo{pages}{99 -- 102}.
\bibitem[{Sharif and Ramanathan(1981)}]{SHARIF1981}
\bibinfo{author}{Sharif, M.}, \bibinfo{author}{Ramanathan, K.},
  \bibinfo{year}{1981}.
\newblock \bibinfo{title}{Binomial innovation diffusion models with dynamic
  potential adopter population}.
\newblock \bibinfo{journal}{Technological Forecasting and Social Change}
  \bibinfo{volume}{20}, \bibinfo{pages}{63 -- 87}.
\bibitem[{Shepherd and Litvak(2004)}]{shepard2004}
\bibinfo{author}{Shepherd, T.D.}, \bibinfo{author}{Litvak, M.K.},
  \bibinfo{year}{2004}.
\newblock \bibinfo{title}{Density-dependent habitat selection and the ideal
  free distribution in marine fish spatial dynamics: considerations and
  cautions}.
\newblock \bibinfo{journal}{Fish and Fisheries} \bibinfo{volume}{5},
  \bibinfo{pages}{141--152}.
\bibitem[{Sibly et~al.(2005)Sibly, Barker, Denham, Hone and Pagel}]{sibly2005}
\bibinfo{author}{Sibly, R.M.}, \bibinfo{author}{Barker, D.},
  \bibinfo{author}{Denham, M.C.}, \bibinfo{author}{Hone, J.},
  \bibinfo{author}{Pagel, M.}, \bibinfo{year}{2005}.
\newblock \bibinfo{title}{On the regulation of populations of mammals, birds,
  fish, and insects}.
\newblock \bibinfo{journal}{Science} \bibinfo{volume}{309},
  \bibinfo{pages}{607--610}.
\bibitem[{Timm(2002)}]{Timm2002}
\bibinfo{author}{Timm, N.}, \bibinfo{year}{2002}.
\newblock \bibinfo{title}{Applied Multivariate Analysis: Springer Texts in
  Statistics}.
\newblock \bibinfo{publisher}{Springer-Verlag New York Incorporated}.
\bibitem[{Trappey and Wu(2008)}]{TRAPPEY2008}
\bibinfo{author}{Trappey, C.V.}, \bibinfo{author}{Wu, H.Y.},
  \bibinfo{year}{2008}.
\newblock \bibinfo{title}{An evaluation of the time-varying extended logistic,
  simple logistic, and gompertz models for forecasting short product
  lifecycles}.
\newblock \bibinfo{journal}{Advanced Engineering Informatics}
  \bibinfo{volume}{22}, \bibinfo{pages}{421 -- 430}.
\newblock \bibinfo{note}{PLM Challenges}.
\bibitem[{Tsoularis and Wallace(2002)}]{TSOULARIS2002}
\bibinfo{author}{Tsoularis, A.}, \bibinfo{author}{Wallace, J.},
  \bibinfo{year}{2002}.
\newblock \bibinfo{title}{Analysis of logistic growth models}.
\newblock \bibinfo{journal}{Mathematical Biosciences} \bibinfo{volume}{179},
  \bibinfo{pages}{21 -- 55}.
\bibitem[{Turner et~al.(1969)Turner, Blumenstein and Sebaugh}]{Turner1969}
\bibinfo{author}{Turner, M.E.}, \bibinfo{author}{Blumenstein, B.A.},
  \bibinfo{author}{Sebaugh, J.L.}, \bibinfo{year}{1969}.
\newblock \bibinfo{title}{265 note: A generalization of the logistic law of
  growth}.
\newblock \bibinfo{journal}{Biometrics} \bibinfo{volume}{25},
  \bibinfo{pages}{577--580}.
\bibitem[{Turner et~al.(1976)Turner, Bradley, Kirk and Pruitt}]{TURNER1976}
\bibinfo{author}{Turner, M.E.}, \bibinfo{author}{Bradley, E.L.},
  \bibinfo{author}{Kirk, K.A.}, \bibinfo{author}{Pruitt, K.M.},
  \bibinfo{year}{1976}.
\newblock \bibinfo{title}{A theory of growth}.
\newblock \bibinfo{journal}{Mathematical Biosciences} \bibinfo{volume}{29},
  \bibinfo{pages}{367 -- 373}.
\bibitem[{Utida(1957)}]{Utida1957}
\bibinfo{author}{Utida, S.}, \bibinfo{year}{1957}.
\newblock \bibinfo{title}{Cyclic {F}luctuations of {P}opulation {D}ensity
  {I}ntrinsic to the {H}ost-{P}arasite {S}ystem}.
\newblock \bibinfo{journal}{Ecology} \bibinfo{volume}{38},
  \bibinfo{pages}{442--449}.
\bibitem[{Verhulst(1838)}]{Verhulst1838}
\bibinfo{author}{Verhulst, P.}, \bibinfo{year}{1838}.
\newblock \bibinfo{title}{Notice sur la loi que la population suit dans son
  accroissement.}
\newblock \bibinfo{journal}{Correspondances Mathématiques et Physiques.}
  \bibinfo{volume}{10}, \bibinfo{pages}{113--121}.
\bibitem[{Von~Bertalanffy(1949)}]{Von1949}
\bibinfo{author}{Von~Bertalanffy, L.}, \bibinfo{year}{1949}.
\newblock \bibinfo{title}{Problems of {O}rganic {G} rowth}.
\newblock \bibinfo{journal}{Nature} \bibinfo{volume}{163}.
\bibitem[{Von~Bertalanffy(1960)}]{von1960}
\bibinfo{author}{Von~Bertalanffy, L.}, \bibinfo{year}{1960}.
\newblock \bibinfo{title}{In fundamental aspects of normal and malignant
  growth}.
\newblock \bibinfo{journal}{Elsevier, Amsterdam} \bibinfo{volume}{35},
  \bibinfo{pages}{137--295}.
\bibitem[{Wasserman(2004)}]{wasserman2004}
\bibinfo{author}{Wasserman, L.}, \bibinfo{year}{2004}.
\newblock \bibinfo{title}{All of statistics: {A} concise course in statistical
  inference brief contents}.
\newblock \bibinfo{journal}{Simulation} \bibinfo{volume}{100},
  \bibinfo{pages}{461}.
\bibitem[{Weibull et~al.(1951)}]{weibull1951}
\bibinfo{author}{Weibull, W.}, et~al., \bibinfo{year}{1951}.
\newblock \bibinfo{title}{A statistical distribution function of wide
  applicability}.
\newblock \bibinfo{journal}{Journal of applied mechanics} \bibinfo{volume}{18},
  \bibinfo{pages}{293--297}.
\bibitem[{Yoshioka(2019)}]{Yoshioka2019}
\bibinfo{author}{Yoshioka, H.}, \bibinfo{year}{2019}.
\newblock \bibinfo{title}{A simplified stochastic optimization model for
  logistic dynamics with control-dependent carrying capacity}.
\newblock \bibinfo{journal}{Journal of Biological Dynamics}
  \bibinfo{volume}{13}, \bibinfo{pages}{148--176}.
\bibitem[{Zhao and Tang(2011)}]{ZHAO2011}
\bibinfo{author}{Zhao, T.}, \bibinfo{author}{Tang, S.}, \bibinfo{year}{2011}.
\newblock \bibinfo{title}{Impulsive harvesting and by-catch mortality for the
  theta logistic model}.
\newblock \bibinfo{journal}{Applied Mathematics and Computation}
  \bibinfo{volume}{217}, \bibinfo{pages}{9412 -- 9423}.

\end{thebibliography}

\newpage

\appendix
\section{Expression of the required partial derivatives for computation of variance of interval estimators by Delta Method}\label{app:partial}
\subsection{Logistic model}
\begin{enumerate}
\item \underline{\textbf{Variance of $\mathbf{\widehat{r_j(\Delta t)}}$}}:\\
For compact representation, we shall use matrix notations. From eqn.(\ref{eqn:r_hat_delta_t}), we obtain that
\begin{eqnarray} \nonumber
&&\frac{\partial \phi}{\partial x}\bigg|_{\bm{\mu}} =-\left[\frac{1}{h}\frac{y}{x(y-x)}\right]\bigg|_{\bm{\mu}}\\ \nonumber
&&\frac{\partial \phi}{\partial y}\bigg|_{\bm{\mu}} =\left[\frac{1}{h}\frac{z-x}{(y-x)(z-y)}\right]\bigg|_{\bm{\mu}} \\ \nonumber
&&\frac{\partial \phi}{\partial z}\bigg|_{\bm{\mu}} =\left[-\frac{1}{h}\frac{y}{z(z-y)}\right]\bigg|_{\bm{\mu}}
\end{eqnarray}

In matrix notation, we have the following expression for $\nabla \phi$:
\[ \nabla \phi \big|_{\bm{\mu}}= \begin{pmatrix} 
\frac{\partial \phi}{\partial x}\\
\frac{\partial \phi}{\partial y}\\
\frac{\partial \phi}{\partial z}
\end{pmatrix}\bigg|_{\bm{\mu}}
=\begin{pmatrix}
  -\frac{1}{h}\frac{y}{x(y-x)}  \\
 \frac{1}{h}\frac{z-x}{(y-x)(z-y)}  \\
   -\frac{1}{h}\frac{y}{z(z-y)}
\end{pmatrix}\bigg|_{\bm{\mu}}\]
The expression of the vector of partial derivatives after evaluating at $\widehat{\bm{\mu}}$ is given in the main text.

\item \underline{\textbf{Variance of $\mathbf{\widehat{K_j(\Delta t)}}$}}:\\
In real variable, eqn.(\ref{eqn:K_hat_delta_t}) is written as:
\begin{equation*}
\psi(x,y,z)=\left[\frac{1}{c}-\frac{(y-x)^{\frac{t_j}{h}+2}z^{\frac{t_j}{h}+1}}{(z-y)^{\frac{t_j}{h}}yx^{\frac{t_j}{h}+1}[z(y-x)-x(z-y)]}\right]^{-1} , 
\end{equation*}
where $c=\overline{X}_{0}$ is constant. We define $\zeta=\frac{(y-x)^{\frac{t_j}{h}+2}z^{\frac{t_j}{h}+1}}{(z-y)^{\frac{t_j}{h}}yx^{\frac{t_j}{h}+1}[z(y-x)-x(z-y)]}$ and $\eta=\frac{1}{c}-\zeta$.
Taking logarithm on $\zeta$, we obtain
\begin{equation*}
\ln{\zeta}=\left(\frac{t_j}{h}+2\right) \ln{(y-x)} +\left(\frac{t_j}{h}+1\right)\ln{z}-\frac{t_j}{h}\ln{(z-y)}-\ln{y}-\left(\frac{t_j}{h}+1\right)\ln{x}-\ln{[z(y-x)-x(z-y)]}.\end{equation*}
Taking partial derivative with respect to $x$, we obtain: \begin{equation*} 
\frac{\partial \psi}{\partial x}= -\frac{1}{\eta^2}\frac{\partial \eta}{\partial x},~~~~~~~~~\frac{\partial \eta}{\partial x}= -\frac{\partial \zeta}{\partial x},~~\mbox{and}~~~~~\frac{\partial \zeta}{\partial x}= \zeta\frac{\partial \ln{\zeta}}{\partial x}
\end{equation*}
So, finally we obtain that,  \begin{equation*}
 \frac{\partial \psi}{\partial x}= \frac{\zeta}{\eta^2}\frac{\partial \ln{\zeta}}{\partial x},   
\end{equation*}
Final expressions of all the required partial derivatives are as follows:
\begin{eqnarray} \nonumber
&&\frac{\partial \ln{\zeta}}{\partial x} =\left[-\frac{\frac{t_j}{h}+2}{y-x}-\frac{\frac{t_j}{h}+1}{x}+ \frac{2z-y}{ z(y-x)-x(z-y)}\right] \\ \nonumber
&&\frac{\partial \ln{\zeta}}{\partial y} =\left[\frac{\frac{t_j}{h}+2}{y-x}+\frac{\frac{t_j}{h}}{z-y}-\frac{1}{y}-\frac{z+x}{z(y-x)-x(z-y)}\right] \\ \nonumber
&&\frac{\partial \ln{\zeta}}{\partial z}
=\left[\frac{\frac{t_j}{h}+1}{z}-\frac{\frac{t_j}{h}}{z-y}+\frac{2x-y}{z(y-x)-x(z-y)}\right],
\end{eqnarray}
In matrix notation, we have the following expression for $\nabla \psi$:
\[ \nabla \psi \big|_{\bm{\mu}}= \left(\frac{\zeta}{\eta^2}\right)\bigg|_{\bm{\mu}} \begin{pmatrix} 
\frac{\partial \ln{\zeta}}{\partial x}\\
\frac{\partial \ln{\zeta}}{\partial y}\\
\frac{\partial \ln{\zeta}}{\partial z}
\end{pmatrix}\bigg|_{\bm{\mu}}
=\left(\frac{\zeta}{\eta^2}\right)\bigg|_{\bm{\mu}}\begin{pmatrix}
 -\frac{\frac{t_j}{h}+2}{y-x}-\frac{\frac{t_j}{h}+1}{x}+ \frac{2z-y}{ z(y-x)-x(z-y)}\\
 \frac{\frac{t_j}{h}+2}{y-z}+\frac{\frac{t_j}{h}}{z-y}-\frac{1}{y}-\frac{z+x}{z(y-x)-x(z-y)} \\
 \frac{\frac{t_j}{h}+1}{z}-\frac{\frac{t_j}{h}}{z-y}+\frac{2x-y}{z(y-x)-x(z-y)} 
\end{pmatrix}\bigg|_{\bm{\mu}}.\]

\end{enumerate}

\subsection{Exponential model}

The exponential model \citep{malthus1798} is given by  \begin{equation}\label{eqn:exponential_model}
\frac{dX(t)}{dt}= rX; X(0) = X_0  \end{equation} and the solution $X_t$ is given by
\begin{equation*}
X_t=X_0e^{rt}.
\end{equation*}
In this model, only one variable $r$ is present and the ISRP of $r$ is given by \citep{PAL2018} \begin{equation*}
\widehat{r_{j}(\Delta t)}=\frac{1}{h}\ln{\left(\frac{\overline{X}_{j+1}}{\overline{X}_{j}}\right)}=\phi\left(\overline{X}_{j},\overline{X}_{j+1}\right)~~~(\mbox{say}).
\end{equation*}
In real variable $\phi$ is written as:
\begin{equation*}
\phi(x,y)=\frac{1}{h}\ln{\left(\frac{y}{x}\right)} .   
\end{equation*}
Now taking partial derivative of $\phi$ with respect to $x$ and $y$, we obtain 
\begin{eqnarray} \nonumber
&& \frac{\partial \phi}{\partial x}=-\frac{1}{hx}\\ \nonumber
&& \frac{\partial \phi}{\partial y}=\frac{1}{hy},
\end{eqnarray}
Using eqn. (\ref{delta_metod}), where $\mu_{t}=\mu_0e^{rt}$, the distribution of $\widehat{r_{j}(\Delta t)}$ is given as:
 \[ \sqrt{n} \left[\phi \begin{pmatrix}
   \overline{X}_{j} \\
    \overline{X}_{j+1}
\end{pmatrix}
- \phi \begin{pmatrix}
   \mu_{j} \\
     \mu_{j+1}
\end{pmatrix}\right]
\xrightarrow{\text{d}}\mathcal{\bm{N}}\left(0,\bm{\nabla'}\phi|_{\bm{\mu}} \Sigma \bm{\nabla}\phi|_{\bm{\mu}}\right), \]
where \[ \nabla \phi \big|_{\bm{\mu}}= \begin{pmatrix} 
\frac{\partial \phi}{\partial x}\\
\frac{\partial \phi}{\partial y}
\end{pmatrix}\bigg|_{\bm{\mu}}
=\begin{pmatrix}
  -\frac{1}{hx}\\
 \frac{1}{hy}
\end{pmatrix}\bigg|_{\bm{\mu}}.\]

\subsection{Theta-logistic model}
The theta-logistic model is given by \begin{equation} \label{eqn:theta_logistic_model}
\frac{dX(t)}{dt}= rX\left[1-\left(\frac{X}{K}\right)^{\theta}\right],  X(0) = X_0 \end{equation} and the solution $X_t$ is given by
\begin{equation*}
X_t=X_0K\left[X_0^{\theta}+e^{-\theta rt}\left(K^{\theta}-X_0^{\theta}\right)\right]^{-\frac{1}{\theta}}.
\end{equation*}
In this model, two variables $r$ and $K$ and one limiting constant $\theta$ are present. Here, we only calculate the variance of $r$ and $K$ only.

\begin{enumerate}

\item \underline{\textbf{Variance of $\mathbf{\widehat{r_j(\Delta t)}}$}}:\\
The ISRP of $r$ is given by \citep{PAL2018}, \begin{equation*}
    \widehat{{r}_{j}(\Delta t)} =\frac{1}{h}\ln{\left[\frac{\frac{1}{\overline{X}_j^{\theta}}-\frac{1}{\overline{X}_{j+1}^{\theta}}}{\frac{1}{\overline{X}_{j+1}^{\theta}}-\frac{1}{\overline{X}_{j+2}^{\theta}}}\right]}= \phi (\overline{X}_{j},\overline{X}_{j+1},\overline{X}_{j+2}) ~~~(\mbox{say}),
\end{equation*}
In terms of real variable the function $\phi$ is given as 
\begin{equation*}
\phi(x,y,z)=\frac{1}{h}\ln{\left[\frac{\frac{1}{x^{\theta}}-\frac{1}{y^{\theta}}}{\frac{1}{y^{\theta}}-\frac{1}{z^{\theta}}}\right]}.
\end{equation*}
Now taking partial derivative of $\phi$ with respect to $x$, $y$ and $z$, we obtain 
\begin{eqnarray} \nonumber
&& \frac{\partial \phi}{\partial x}= -\frac{1}{h}\frac{\theta y^{\theta}}{x\left(y^{\theta}-x^{\theta}\right)}\\ \nonumber
&& \frac{\partial \phi}{\partial y}=\frac{1}{h}\frac{y^{\theta -1}\left(z^{\theta}-x^{\theta}\right)}{(y-x)(z-y)}\\ \nonumber
&& \frac{\partial \phi}{\partial z}=-\frac{1}{h}\frac{\theta y^{\theta}}{z\left(z^{\theta}-y^{\theta}\right)},
\end{eqnarray}
Using eqn. (\ref{delta_metod}), where $\mu_{t}=\mu_0 K\left[\mu_0^{\theta}+e^{-\theta rt}\left(K^{\theta}-\mu_0^{\theta}\right)\right]^{-\frac{1}{\theta}}$, the distribution of $\widehat{r_{j}(\Delta t)}$ is given as:
 \[ \sqrt{n} \left[\phi \begin{pmatrix}
   \overline{X}_{j} \\
    \overline{X}_{j+1}\\
    \overline{X}_{j+2}
\end{pmatrix}
- \phi \begin{pmatrix}
   \mu_{j} \\
     \mu_{j+1}\\
     \mu_{j+2}
\end{pmatrix}\right]
\xrightarrow{\text{d}}\mathcal{\bm{N}}\left(0,\bm{\nabla'}\phi|_{\bm{\mu}} \Sigma \bm{\nabla}\phi|_{\bm{\mu}}\right), \]
where  \[ \nabla \phi \big|_{\bm{\mu}}= \begin{pmatrix} 
\frac{\partial \phi}{\partial x}\\
\frac{\partial \phi}{\partial y}\\
\frac{\partial \phi}{\partial z}
\end{pmatrix}\bigg|_{\bm{\mu}}
=\begin{pmatrix}
  -\frac{1}{h}\frac{\theta y^{\theta}}{x\left(y^{\theta}-x^{\theta}\right)}  \\
 \frac{1}{h}\frac{y^{\theta -1}\left(z^{\theta}-x^{\theta}\right)}{(y-x)(z-y)}  \\
   -\frac{1}{h}\frac{\theta y^{\theta}}{z\left(z^{\theta}-y^{\theta}\right)}
\end{pmatrix}\bigg|_{\bm{\mu}}.\]

\item \underline{\textbf{Variance of $\mathbf{\widehat{K_j(\Delta t)}}$}}:\\
The ISRP of $K$ is given by \citep{PAL2018},
\begin{equation*}
    \widehat{{K}_{j}(\Delta t)} =\left[\frac{1}{\overline{X}_0^{\theta}}-\frac{\frac{1}{\overline{X}_j^{\theta}}-\frac{1}{\overline{X}_{j+1}^{\theta}}}{\exp\left({-\widehat{{r}_{j}(\Delta t)}}t_j\right)\left[1-\exp\left({-\widehat{{r}_{j}(\Delta t)}}h\right)\right]}\right]^{-\frac{1}{\theta}}=\psi (\overline{X}_{j},\overline{X}_{j+1},\overline{X}_{j+2}) ~~~(\mbox{say}),
\end{equation*} 
 After simplification $\psi$ is given as :
\begin{equation*}
\psi(\overline{X}_{j},\overline{X}_{j+1},\overline{X}_{j+2})=\left[\frac{1}{c}-\frac{\left(\overline{X}_{j+1}^{\theta}-\overline{X}_{j}^{\theta}\right)^{\frac{t_j}{h}+2}~\left(\overline{X}_{j+2}^{\theta}\right)^{\frac{t_j}{h}+1}}{\left(\overline{X}_{j+2}^{\theta}-\overline{X}_{j+1}^{\theta}\right)^{\frac{t_j}{h}}~\overline{X}_{j+1}^{\theta}~\left(\overline{X}_{j}^{\theta}\right)^{\frac{t_j}{h}+1}~\left[\overline{X}_{j+2}^{\theta}\left(\overline{X}_{j+1}^{\theta}-\overline{X}_{j}^{\theta}\right)-\overline{X}_{j}^{\theta}\left(\overline{X}_{j+2}^{\theta}-\overline{X}_{j+1}^{\theta}\right)\right]}\right]^{-\frac{1}{\theta}}, 
\end{equation*}
where $c=\overline{X}_{0}^{\theta}$ is constant.
In terms of real variable the function $\psi$ is written as:
\begin{equation*}
\psi(x,y,z)=\left[\frac{1}{c}-\frac{\left(y^{\theta}-x^{\theta}\right)^{\frac{t_j}{h}+2}\left(z^{\theta}\right)^{\frac{t_j}{h}+1}}{\left(z^{\theta}-y^{\theta}\right)^{\frac{t_j}{h}}\left(y^{\theta}\right)\left(x^{\theta}\right)^{\frac{t_j}{h}+1}\left[z^{\theta}(y^{\theta}-x^{\theta})-x^{\theta}(z^{\theta}-y^{\theta})\right]}\right]^{-\frac{1}{\theta}}.
\end{equation*}
Now, we define $\zeta=\frac{\left(y^{\theta}-x^{\theta}\right)^{\frac{t_j}{h}+2}\left(z^{\theta}\right)^{\frac{t_j}{h}+1}}{\left(z^{\theta}-y^{\theta}\right)^{\frac{t_j}{h}}\left(y^{\theta}\right)\left(x^{\theta}\right)^{\frac{t_j}{h}+1}\left[z^{\theta}(y^{\theta}-x^{\theta})-x^{\theta}(z^{\theta}-y^{\theta})\right]}$ and $\eta=\frac{1}{c}-\zeta$.
Taking logarithm on $\zeta$, we obtain
\begin{eqnarray*}
\ln{\zeta}&=&\left(\frac{t_j}{h}+2\right) \ln{\left(y^{\theta}-x^{\theta}\right)} +\left(\frac{t_j}{h}+1\right)\theta~ \ln{z} -\frac{t_j}{h}\ln{\left(z^{\theta}-y^{\theta}\right)}-\theta \ln{y}\\
&-&\left(\frac{t_j}{h}+1\right)\theta \ln{x} - \ln{\left[z^{\theta}\left(y^{\theta}-x^{\theta}\right)-x^{\theta}\left(z^{\theta}-y^{\theta}\right)\right]}.\end{eqnarray*}
Taking partial derivative with respect to $x$, we obtain: \begin{equation*} 
\frac{\partial \psi}{\partial x}= -\frac{1}{\theta \eta^{\frac{\theta+1}{\theta}}}\frac{\partial \eta}{\partial x},~~~~~~~~~\frac{\partial \eta}{\partial x}= -\frac{\partial \zeta}{\partial x},~~\mbox{and}~~~~~\frac{\partial \zeta}{\partial x}= \zeta\frac{\partial \ln{\zeta}}{\partial x}
\end{equation*}
So, finally we obtain that,  \begin{equation*}
 \frac{\partial \psi}{\partial x}= \frac{\zeta}{\theta \eta^{\frac{\theta+1}{\theta}}}\frac{\partial \ln{\zeta}}{\partial x},   
\end{equation*}
Final expressions of all the required partial derivatives are as follows:
\begin{eqnarray} \nonumber
&&\frac{\partial \ln{\zeta}}{\partial x} =\left[-\frac{\left(\frac{t_j}{h}+2\right)\theta x^{\theta-1}}{y^{\theta}-x^{\theta}}-\frac{\left(\frac{t_j}{h}+1\right)\theta}{x}+ \frac{\theta x^{\theta} \left(2z^{\theta}-y^{\theta}\right)}
{z^{\theta}\left(y^{\theta}-x^{\theta}\right)-x^{\theta}\left(z^{\theta}-y^{\theta}\right)}\right] \\ \nonumber
&&\frac{\partial \ln{\zeta}}{\partial y} =\left[\frac{\theta \left(\frac{t_j}{h}+2\right)\theta y^{\theta-1}}{y^{\theta}-x^{\theta}}+ \frac{\frac{t_j}{h}\theta y^{\theta-1}}{z^{\theta}-y^{\theta}}- \frac{\theta}{y}-\frac{\theta \left(z^{\theta}+x^{\theta}\right)y^{\theta-1}}{z^{\theta}\left(y^{\theta}-x^{\theta}\right)-x^{\theta}\left(z^{\theta}-y^{\theta}\right)}\right] \\ \nonumber
&&\frac{\partial \ln{\zeta}}{\partial z}
=\left[\frac{\left(\frac{t_j}{h}+1\right)\theta}{z}-\frac{\theta \frac{t_j}{h}z^{\theta -1}}{z^{\theta}-y^{\theta}}+\frac{\theta z^{\theta-1} \left(2x^{\theta}-y^{\theta}\right)}{ z^{\theta}\left(y^{\theta}-x^{\theta}\right)-x^{\theta}\left(z^{\theta}-y^{\theta}\right)}\right],
\end{eqnarray}
Using eqn. (\ref{delta_metod}), where $\mu_{t}=\mu_0 K\left[\mu_0^{\theta}+e^{-\theta rt}\left(K^{\theta}-\mu_0^{\theta}\right)\right]^{-\frac{1}{\theta}}$, the distribution of $\widehat{K_{j}(\Delta t)}$ is given as:
 \[ \sqrt{n} \left[\psi \begin{pmatrix}
   \overline{X}_{j} \\
    \overline{X}_{j+1}\\
    \overline{X}_{j+2}
\end{pmatrix}
- \phi \begin{pmatrix}
   \mu_{j} \\
     \mu_{j+1}\\
     \mu_{j+2}
\end{pmatrix}\right]
\xrightarrow{\text{d}}\mathcal{\bm{N}}\left(0,\bm{\nabla'}\psi|_{\bm{\mu}} \Sigma \bm{\nabla}\psi|_{\bm{\mu}}\right), \]

where  \[ \nabla \psi \big|_{\bm{\mu}}=\left(\frac{\zeta} {\theta \eta^{\frac{\theta+1}{\theta}}}\right)\bigg|_{\bm{\mu}} 
\begin{pmatrix} 
\frac{\partial \ln{\zeta}}{\partial x}\\
\frac{\partial \ln{\zeta}}{\partial y}\\
\frac{\partial \ln{\zeta}}{\partial z}
\end{pmatrix}\bigg|_{\bm{\mu}}
=\left(\frac{\zeta} {\theta \eta^{\frac{\theta+1}{\theta}}}\right)\bigg|_{\bm{\mu}}\begin{pmatrix}
 -\frac{\left(\frac{t_j}{h}+2\right)\theta x^{\theta-1}}{y^{\theta}-x^{\theta}}-\frac{\left(\frac{t_j}{h}+1\right)\theta}{x}+ \frac{\theta x^{\theta} \left(2z^{\theta}-y^{\theta}\right)}
{z^{\theta}\left(y^{\theta}-x^{\theta}\right)-x^{\theta}\left(z^{\theta}-y^{\theta}\right)}\\
 \frac{\theta \left(\frac{t_j}{h}+2\right)\theta y^{\theta-1}}{y^{\theta}-x^{\theta}}+ \frac{\frac{t_j}{h}\theta y^{\theta-1}}{z^{\theta}-y^{\theta}}- \frac{\theta}{y}-\frac{\theta \left(z^{\theta}+x^{\theta}\right)y^{\theta-1}}{z^{\theta}\left(y^{\theta}-x^{\theta}\right)-x^{\theta}\left(z^{\theta}-y^{\theta}\right)} \\
\frac{\left(\frac{t_j}{h}+1\right)\theta}{z}-\frac{\theta \frac{t_j}{h}z^{\theta -1}}{z^{\theta}-y^{\theta}}+\frac{\theta z^{\theta-1} \left(2x^{\theta}-y^{\theta}\right)}{ z^{\theta}\left(y^{\theta}-x^{\theta}\right)-x^{\theta}\left(z^{\theta}-y^{\theta}\right)} 
\end{pmatrix}\bigg|_{\bm{\mu}}.\]

\end{enumerate}

\subsection{Confined exponential model}
The confined exponential model is given by \begin{equation} \label{confined_exponential_model}
\frac{dX(t)}{dt}=r(K-X); X(0) = X_0
\end{equation}
where $X(t)$ is the population size at time $t$; $r$ and $K$ are the intrinsic growth rate and carrying capacity (asymptotic size) respectively,) and the solution $X_t$ is given by
\begin{equation*}
X_t=K-(K-X_0)e^{-rt}.
\end{equation*}
In this model, two variables $r$ and $K$ are present.

\begin{enumerate}

\item \underline{\textbf{Variance of $\mathbf{\widehat{r_j(\Delta t)}}$}}: \\
The ISRP of $r$ is given by \citep{PAL2018}, \begin{equation*}
\widehat{r_{j}(\Delta t)}=\frac{1}{h}\ln{\left(\frac{\overline{X}_{j+1}-\overline{X}_{j}}{\overline{X}_{j+2}-\overline{X}_{j+1}}\right)}=\phi\left(\overline{X}_{j}, \overline{X}_{j+1},\overline{X}_{j+2} \right)~~~(\mbox{say}),
\end{equation*}
In real variable $\phi$ is written as:
\begin{equation*}
\phi(x,y,z)=\frac{1}{h}\ln{\left(\frac{y-x}{z-y}\right)}.    
\end{equation*}
Now taking partial derivative of $\phi$ with respect to $x$, $y$ and $z$, we obtain 
\begin{eqnarray} \nonumber
&& \frac{\partial \phi}{\partial x}=-\frac{1}{h}\frac{1}{(y-x)}\\ \nonumber
&& \frac{\partial \phi}{\partial y}=\frac{1}{h}\frac{z-x}{(y-x)(z-y)}\\ \nonumber
&& \frac{\partial \phi}{\partial z}=-\frac{1}{h}\frac{1}{(z-y)},
\end{eqnarray}
Using eqn. (\ref{delta_metod}), where $\mu_{t}=K-(K-\mu_0)e^{-rt}$, the distribution of $\widehat{r_{j}(\Delta t)}$ is given as:
 \[ \sqrt{n} \left[\phi \begin{pmatrix}
   \overline{X}_{j} \\
    \overline{X}_{j+1}\\
    \overline{X}_{j+2}
\end{pmatrix}
- \phi \begin{pmatrix}
   \mu_{j} \\
     \mu_{j+1}\\
     \mu_{j+2}
\end{pmatrix}\right]
\xrightarrow{\text{d}}\mathcal{\bm{N}}\left(0,\bm{\nabla'}\phi|_{\bm{\mu}} \Sigma \bm{\nabla}\phi|_{\bm{\mu}}\right), \]
where  \[ \nabla \phi \big|_{\bm{\mu}}= \begin{pmatrix} 
\frac{\partial \phi}{\partial x}\\
\frac{\partial \phi}{\partial y}\\
\frac{\partial \phi}{\partial z}
\end{pmatrix}\bigg|_{\bm{\mu}}
=\begin{pmatrix}
  -\frac{1}{h}\frac{1}{(y-x)}  \\
 \frac{1}{h}\frac{z-x}{(y-x)(z-y)}  \\
   -\frac{1}{h}\frac{1}{(z-y)}
\end{pmatrix}\bigg|_{\bm{\mu}}.\]

\item \underline{\textbf{Variance of $\mathbf{\widehat{K_j(\Delta t)}}$}}:\\
The ISRP of $K$ is given by \citep{PAL2018}, \begin{equation*}
\widehat{K_{j}(\Delta t)}=\left[\overline{X}_0+\frac{\overline{X}_{j+1}-\overline{X}_j}{\exp\left({-\widehat{{r}_{j}(\Delta t)}}t_j\right)\left[1-\exp\left({-\widehat{{r}_{j}(\Delta t)}}h\right)\right]}\right]= \psi (\overline{X}_{j},\overline{X}_{j+1},\overline{X}_{j+2}) ~~~(\mbox{say}).
\end{equation*} 
After simplification $\psi$ is given as:
\begin{equation*}
\psi(\overline{X}_{j},\overline{X}_{j+1},\overline{X}_{j+2})= \left[\overline{X}_{0}+\frac{\left(\overline{X}_{j+1}-\overline{X}_{j}\right)^{\frac{t_j}{h}+2}}{\left(\overline{X}_{j+2}-\overline{X}_{j+1}\right)^{\frac{t_j}{h}}\left(2\overline{X}_{j+1}-\overline{X}_{j+2}-\overline{X}_{j}\right)}\right].
\end{equation*}

In real variable, $\psi$ is written as:
\begin{equation*}
\psi(x,y,z)=\left[c+\frac{(y-x)^{\frac{t_j}{h}+2}}{(z-y)^{\frac{t_j}{h}}(2y-z-x)}\right] , 
\end{equation*}
where $c=\overline{X}_{0}$ is constant. We define $\zeta=\frac{(y-x)^{\frac{t_j}{h}+2}}{(z-y)^{\frac{t_j}{h}}(2y-z-x)}$.
Taking logarithm on $\zeta$, we obtain
\begin{equation*}
\ln{\zeta}=\left(\frac{t_j}{h}+2\right)\ln{(y-x)} -\frac{t_j}{h}\ln{(z-y)}-\ln{[2y-z-x]}.\end{equation*}

Taking partial derivative with respect to $x$, we obtain: \begin{equation*} 
\frac{\partial \psi}{\partial x}=\frac{\partial \zeta}{\partial x} ~~\mbox{and}~~~~~\frac{\partial \zeta}{\partial x}= \zeta\frac{\partial \ln{\zeta}}{\partial x},
\end{equation*}
So, finally we obtain that,  \begin{equation*}
 \frac{\partial \psi}{\partial x}= \zeta \frac{\partial \ln{\zeta}}{\partial x},   
\end{equation*}
Final expressions of all the required partial derivatives are as follows:
\begin{eqnarray} \nonumber
&&\frac{\partial \ln{\zeta}}{\partial x} =\left[-\frac{\frac{t_j}{h}+2}{y-x}+\frac{1}{2y-z-x}\right] \\ \nonumber
&&\frac{\partial \ln{\zeta}}{\partial y} =\left[\frac{\frac{t_j}{h}+2}{y-z}+\frac{\frac{t_j}{h}}{z-y}-\frac{2}{2y-z-x}\right] \\ \nonumber
&&\frac{\partial \ln{\zeta}}{\partial z}
=\left[-\frac{\frac{t_j}{h}}{z-y}+\frac{1}{2y-z-x}\right],
\end{eqnarray}

Using eqn. (\ref{delta_metod}), where $\mu_{t}=K-(K-\mu_0)e^{-rt}$, the distribution of $\widehat{K_{j}(\Delta t)}$ is given as:

 \[ \sqrt{n} \left[\psi \begin{pmatrix}
   \overline{X}_{j} \\
    \overline{X}_{j+1}\\
    \overline{X}_{j+2}
\end{pmatrix}
- \psi \begin{pmatrix}
   \mu_{j} \\
     \mu_{j+1}\\
     \mu_{j+2}
\end{pmatrix}\right]
\xrightarrow{\text{d}}\mathcal{\bm{N}}\left(0,\bm{\nabla'}\psi|_{\bm{\mu}} \Sigma \bm{\nabla}\psi|_{\bm{\mu}}\right), \]
where  \[ \nabla \psi \big|_{\bm{\mu}}= \zeta|_{\bm{\mu}} \begin{pmatrix} 
\frac{\partial \ln{\zeta}}{\partial x}\\
\frac{\partial \ln{\zeta}}{\partial y}\\
\frac{\partial \ln{\zeta}}{\partial z}
\end{pmatrix}\bigg|_{\bm{\mu}}
=\zeta|_{\bm{\mu}}\begin{pmatrix}
 -\frac{\frac{t_j}{h}+2}{y-x}+\frac{1}{2y-z-x}\\
 \frac{\frac{t_j}{h}+2}{y-z}+\frac{\frac{t_j}{h}}{z-y}-\frac{2}{2y-z-x} \\
 -\frac{\frac{t_j}{h}}{z-y}+\frac{1}{2y-z-x} 
\end{pmatrix}\bigg|_{\bm{\mu}}.\]

\end{enumerate}

\newpage
\nolinenumbers
\beginsupplement
\section{Supporting online information}

\subsection{Analytical expression exist}
We first discuss about the cases where after varying the parameter analytical expression of the solution of the model exists.

\subsubsection{Variation in exponential growth model}
We start our discussion by the oldest candidate in the growth curve literature, the exponential model \citep{malthus1798}. \citet{Banks1994} showed that after varying the parameter ($r$) in the exponential model (eqn.~\ref{eqn:exponential_model}), one can build connections with the normal distribution, gompertz growth, linear growth, hyperbolic growth and power law exponential model. So, we do not discuss them here. We will concentrate on the other type of variation which are not available in existing literature.

\begin{figure}[H]
\begin{subfigure}[t]{0.33\textwidth}
  \includegraphics[width=\linewidth]{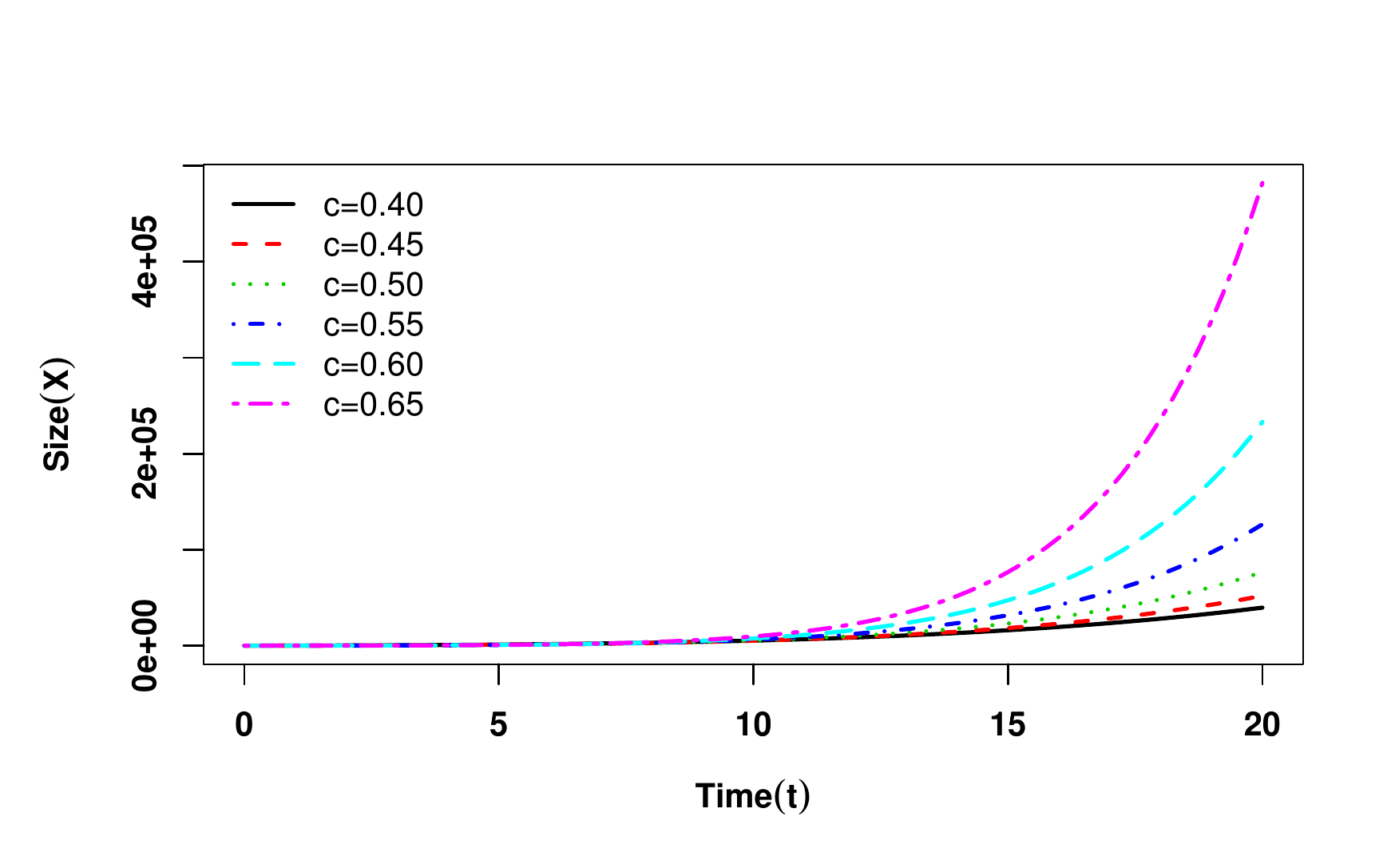}
\caption{$r=r_0t^{c-1}$, ($r_0=1$)}
    \label{subim1}
\end{subfigure}\hfill
\begin{subfigure}[t]{0.33\textwidth}
    \includegraphics[width=\linewidth]{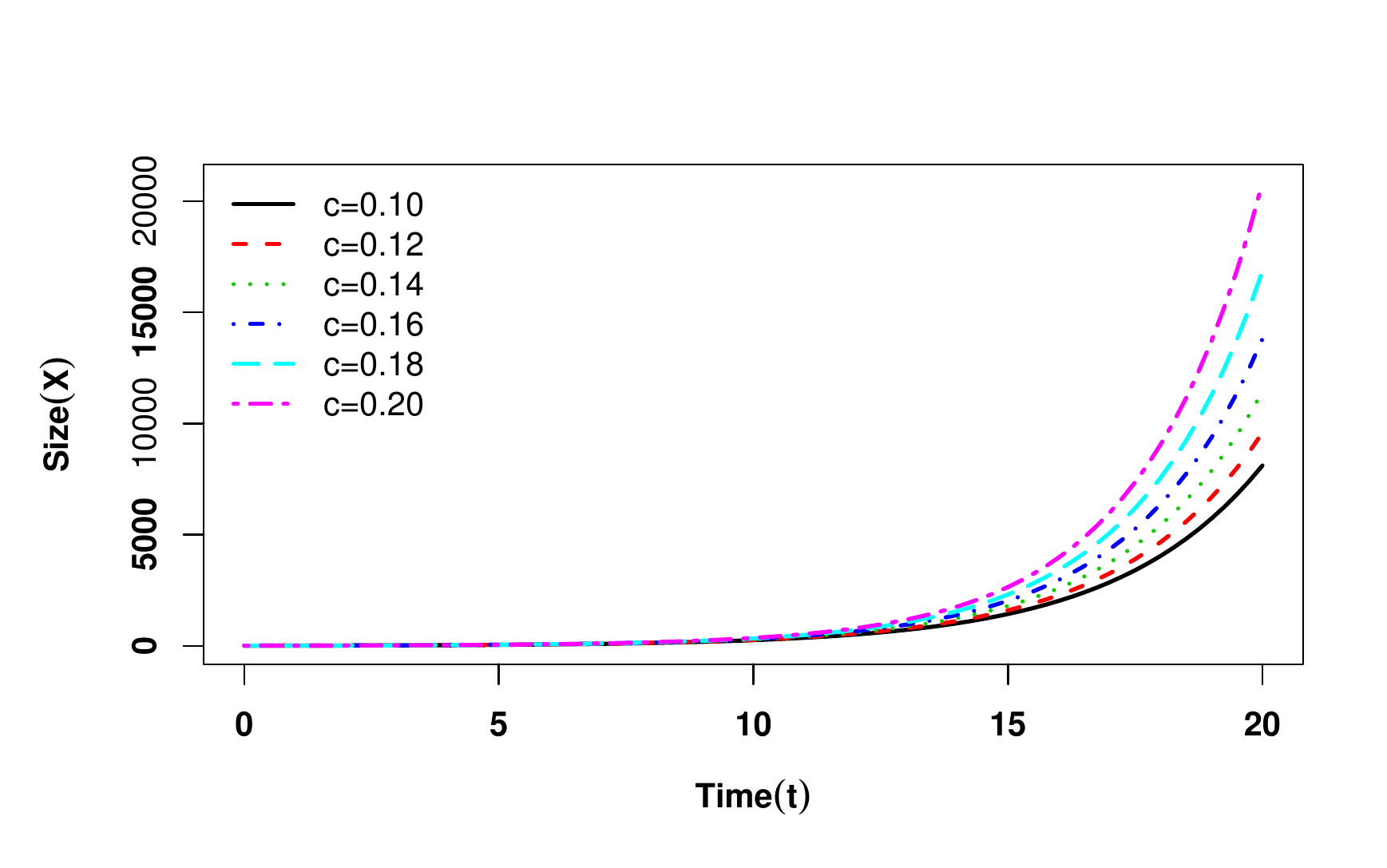}
\caption{$r=r_0(1+ct)$, ($r_0=0.2$)}
    \label{subim2}
\end{subfigure}
\begin{subfigure}[t]{0.33\textwidth}
    \includegraphics[width=\linewidth]{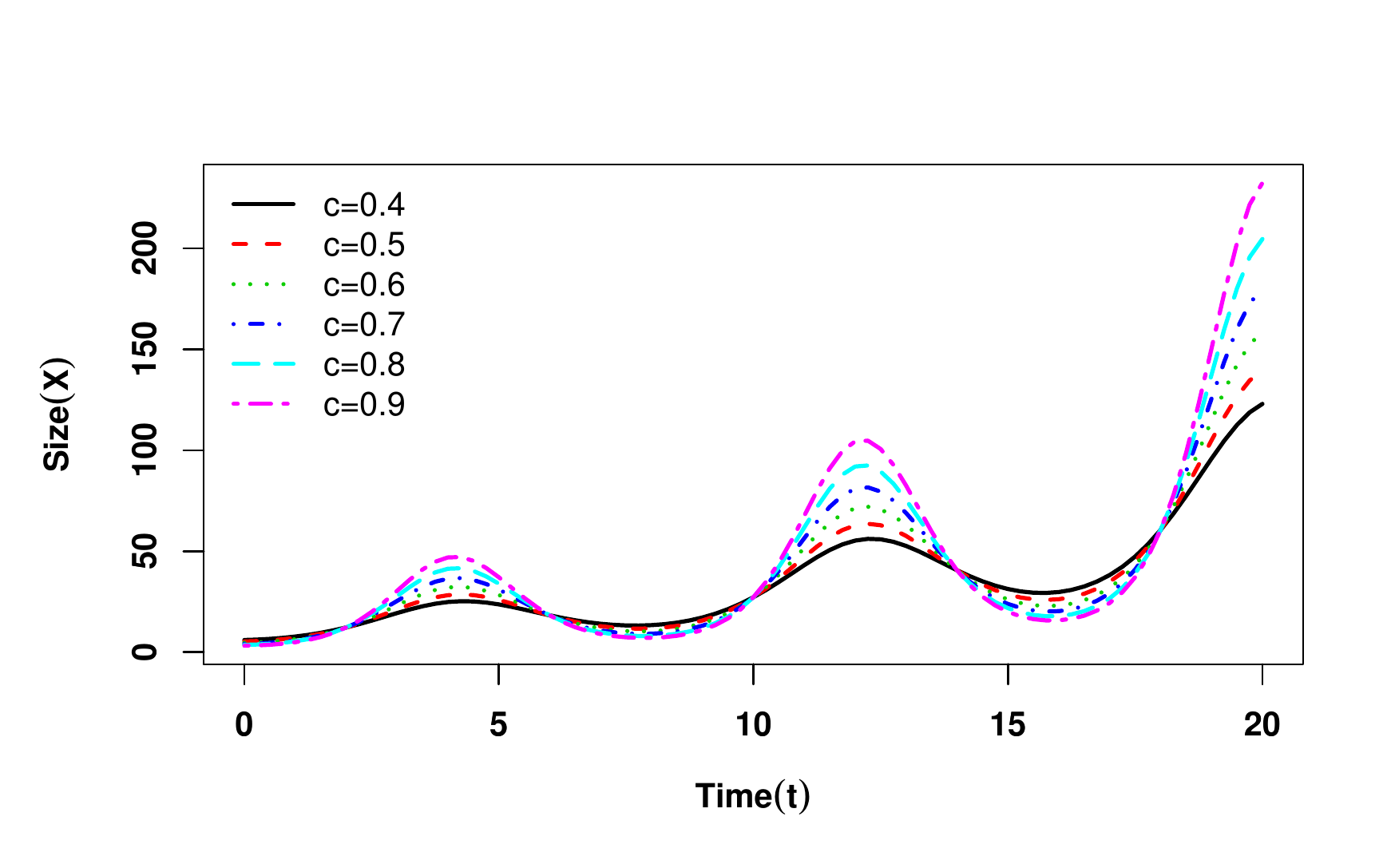}
 \caption{$r=r_0+c\sin({\omega t})$, ($r_0=0.1$)}
    \label{subim3}
\end{subfigure}\hfill

\begin{subfigure}[t]{0.33\textwidth}
    \includegraphics[width=\linewidth]{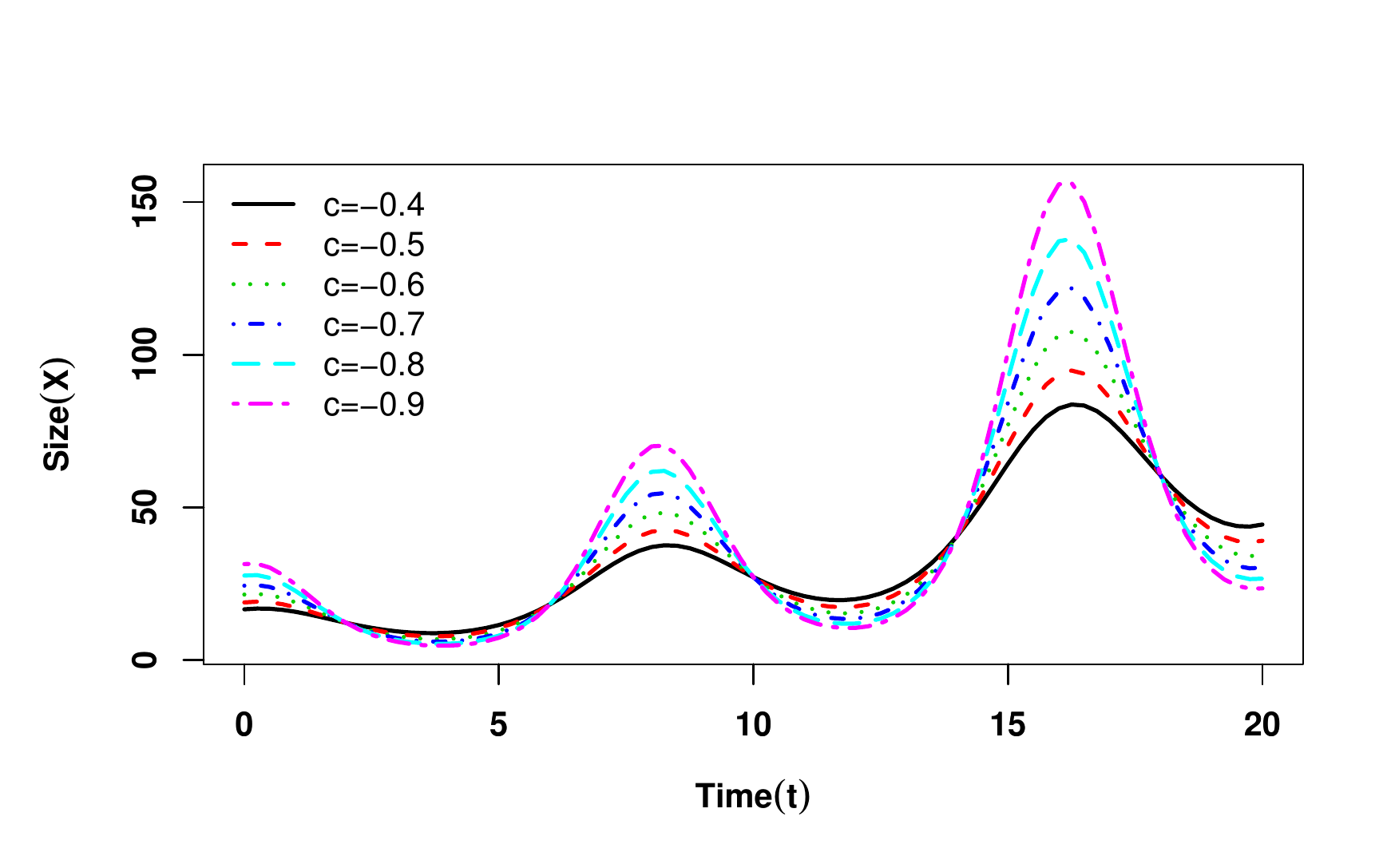}
 \caption{$r=r_0+c\sin({\omega t})$, ($r_0=0.1$)}
    \label{subim4}
\end{subfigure}\hfill
\begin{subfigure}[t]{0.33\textwidth}
    \includegraphics[width=\textwidth]{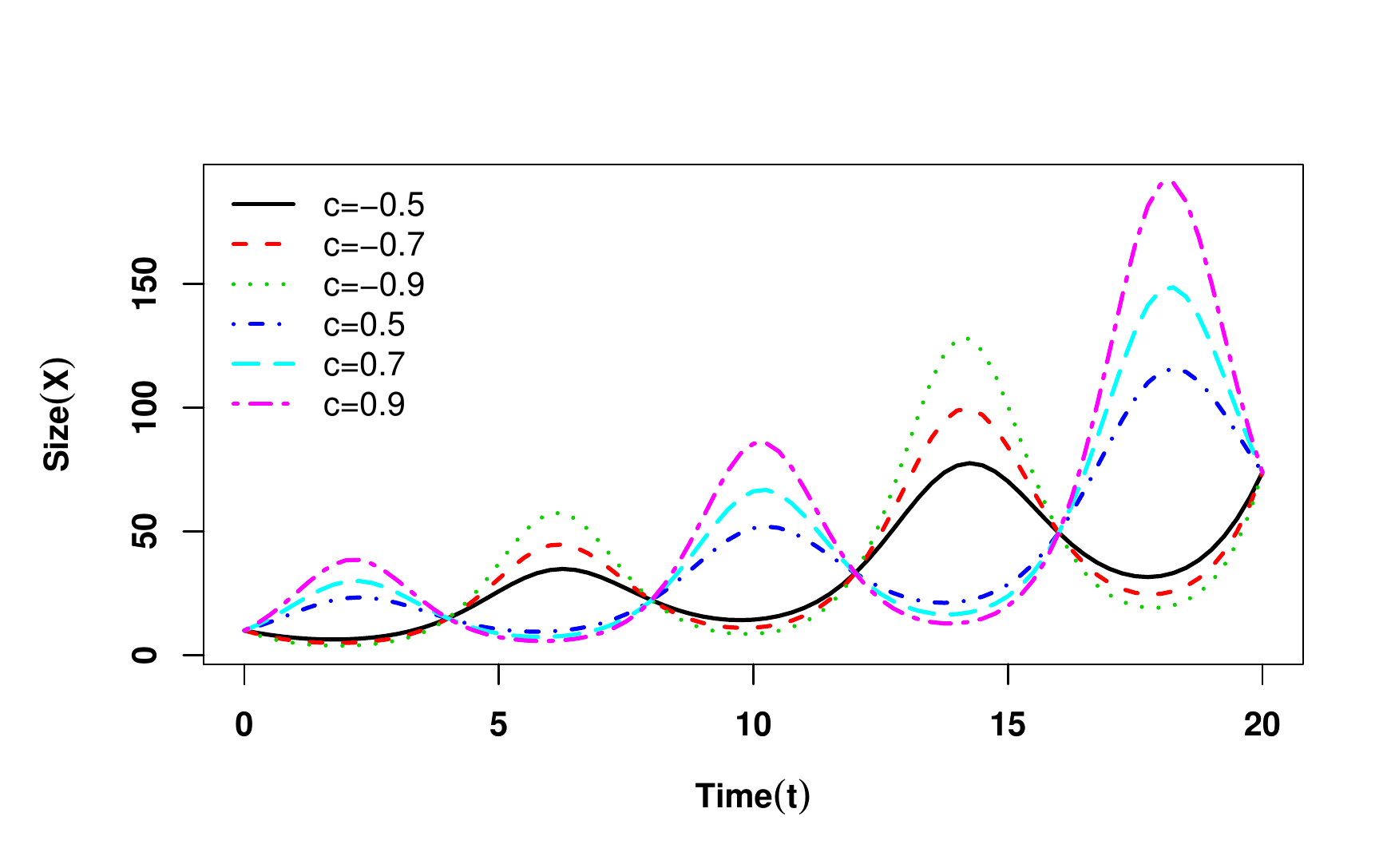}
 \caption{$r=r_0+c\cos({\omega t})$, ($r_0=0.1$)}
    \label{subim5}
\end{subfigure}

 \caption{Time ($t$) vs size ($X(t)$) plot for continuously varying $r$ in Exponential Model. In the upper panel, in Fig. (a) , we consider $r$ as a polynomial function of time and in Fig. (b), we consider $r$ as a linearly increasing function of time. In lower panel (Fig. (c), (d) and (e)), $r$ varies periodically. In all cases, $r_0$ be the initial values of the parameter $r$. $X_0$ is kept fixed at 10 and for Fig. (c), (d) and (e),we consider $\omega=\frac{\pi}{4}$.}
 \label{Exponential_plot}
\end{figure}

\begin{enumerate}
\item $r = r_0t^{c-1}$; $c > 0$ :\\
If we take this type of variation in the parameter $r$ then the eqn. (\ref{eqn:exponential_model}) turns out to Korf model \citep{korf1939} (Table~\ref{table2}; srl. $6$)  for which the asymptotic size tends towards $\infty$ and behaves like exponential model (Fig.~\ref{subim1}) and as $c$ increases $X(t)$ increases and goes towards asymptotic size at a much faster rate.

\item $r=r_0(1+ct)$; ($c>0$) :\\
For this variation in $r$, eqn. (\ref{eqn:exponential_model}) turns out to a new model (Table~\ref{table2}; srl. $7$) whose asymptotic size is also $\infty$ like exponential model but $X(t)$ tends towards it at a much faster rate than the exponential model (Fig.~\ref{subim2}).

\item $r= r_0 + c\sin({\omega t})$; $\omega >  0$ :\\ 
For periodically varying function of $r$, eqn. (\ref{eqn:exponential_model}) turns into a new model (Table~\ref{table2}; srl. $8$) whose asymptotic size remains $\infty$ but $X(t)$ goes towards it periodically. If we keep increasing the value of  $r$ we can see much bigger period in  population size ($X$) and also $X(t)$ goes towards the asymptotic size at a much faster rate (Fig.~\ref{subim3}, \ref{subim4}). If we take cosine function instead of sine function then also we get a new model (Table~\ref{table2}; srl. $9$) having similar behaviour (Fig. \ref{subim5}).
\end{enumerate}

\subsubsection{Variation in logistic growth model}
The logistic model is given by \citep{Verhulst1838}
\begin{equation} \label{eqn:logistic_model}
\frac{dX(t)}{dt}= rX\left(1-\frac{X}{K}\right); X(0) = X_0 , \end{equation}
where $X(t)$ be the population size at time $t$, $r$ and $K$ be the intrinsic growth rate and carrying capacity (asymptotic size). \citet{Banks1994} discussed the dynamics of the logistic equation by varying $K$ as function of time. Here, we discuss the growth behaviour of logistic equation by varying the parameter $r$. Apart from the connection to different existing growth equations some new models are also obtained. In the following, we categorically discuss different cases.

\begin{enumerate}

\item $r=r_0(1+ct)$; ($c>0$) :\\
The resulting equation (Table~\ref{table3}; srl.$3$) has the same asymptotic size $K$ like logistic model (eqn.~\ref{eqn:logistic_model}). An increase in the value of $r_0$ (or $c$), $X(t)$ tends to $K$ at a faster rate, however, the point of inflection remains same as logistic equation (Fig.~\ref{subim6}).

\item $r=r_0(1-ct)$; ($c>0$) :\\
Under this transformation, a new growth equation (Table~\ref{table3}; srl.$4$) is obtained. In this case growth is not monotonic. For small $c$, $X(t)$ first increases and then decreases to zero. As $r_0$ (or $c$) increases  $X(t) \to 0$ at a faster rate (Fig.~\ref{subim7}).

\begin{figure}[H]
\begin{subfigure}[t]{0.33\textwidth}
    \includegraphics[width=\linewidth]{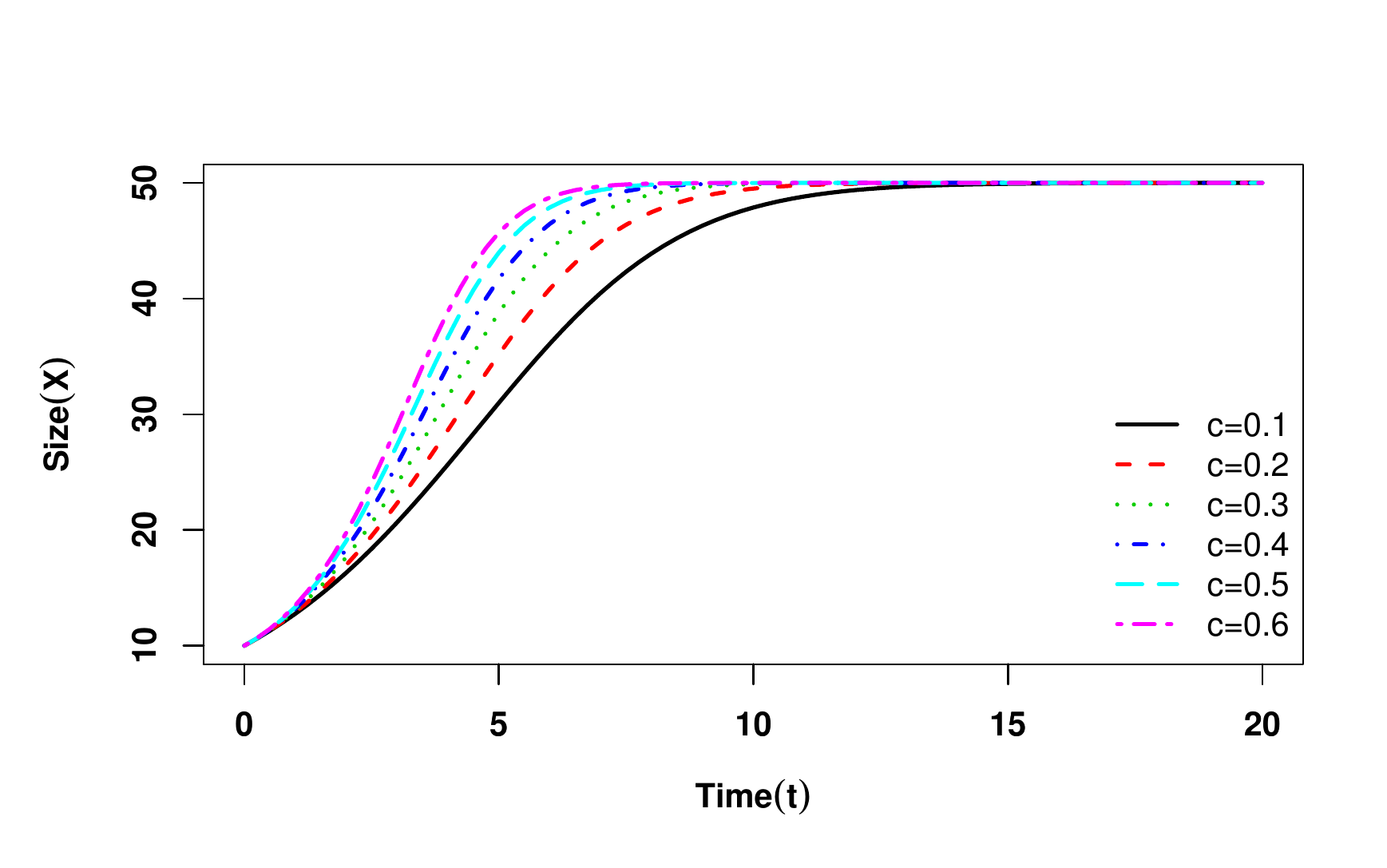}
\caption{$r=r_0(1+ct)$, ($r_0=0.3$)}
\label{subim6}
\end{subfigure}\hfill
\begin{subfigure}[t]{0.33\textwidth}
  \includegraphics[width=\linewidth]{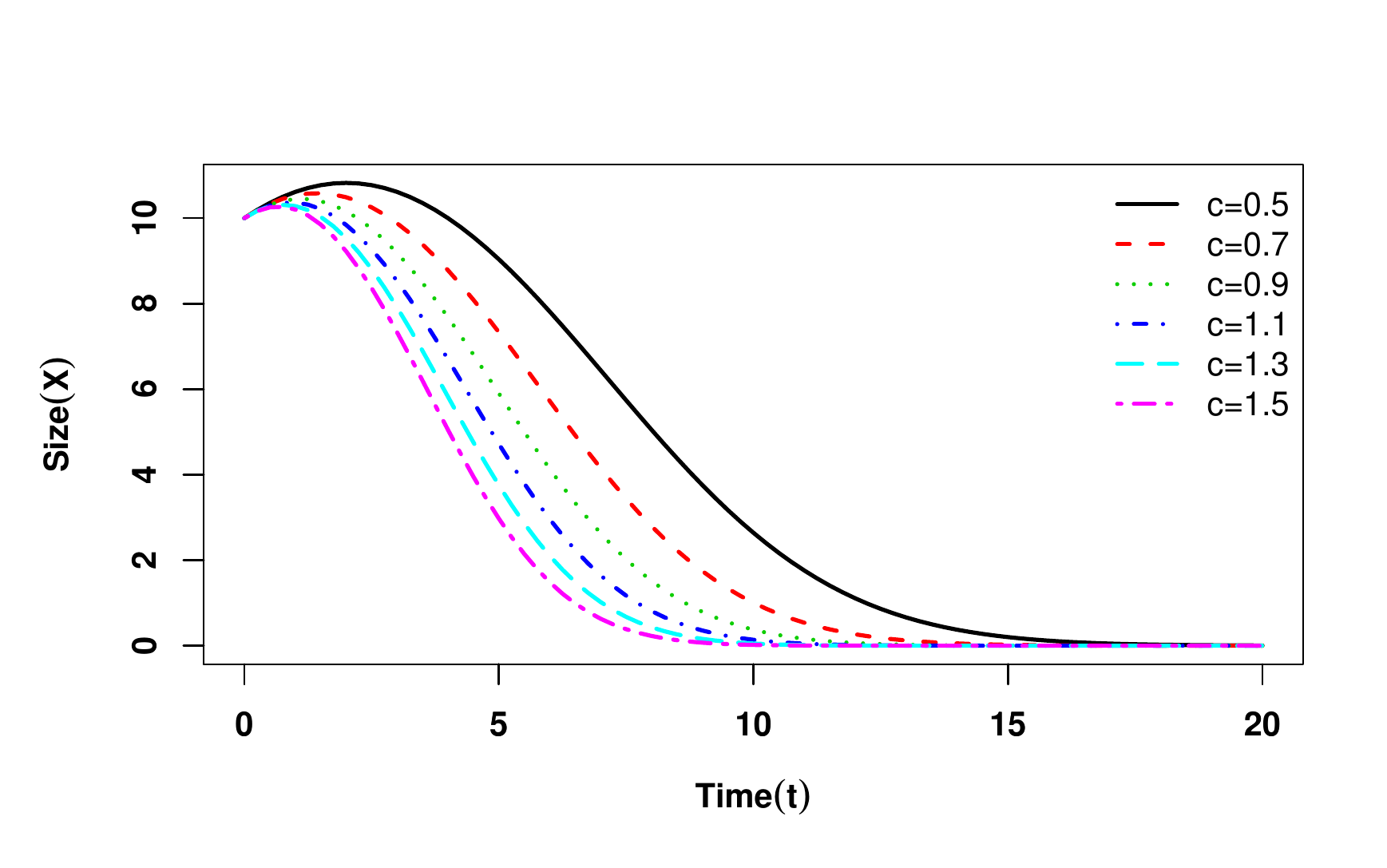}
\caption{$r=r_0(1-ct)$, ($r_0=0.1$)}
\label{subim7}
\end{subfigure}\hfill
\begin{subfigure}[t]{0.33\textwidth}
    \includegraphics[width=\linewidth]{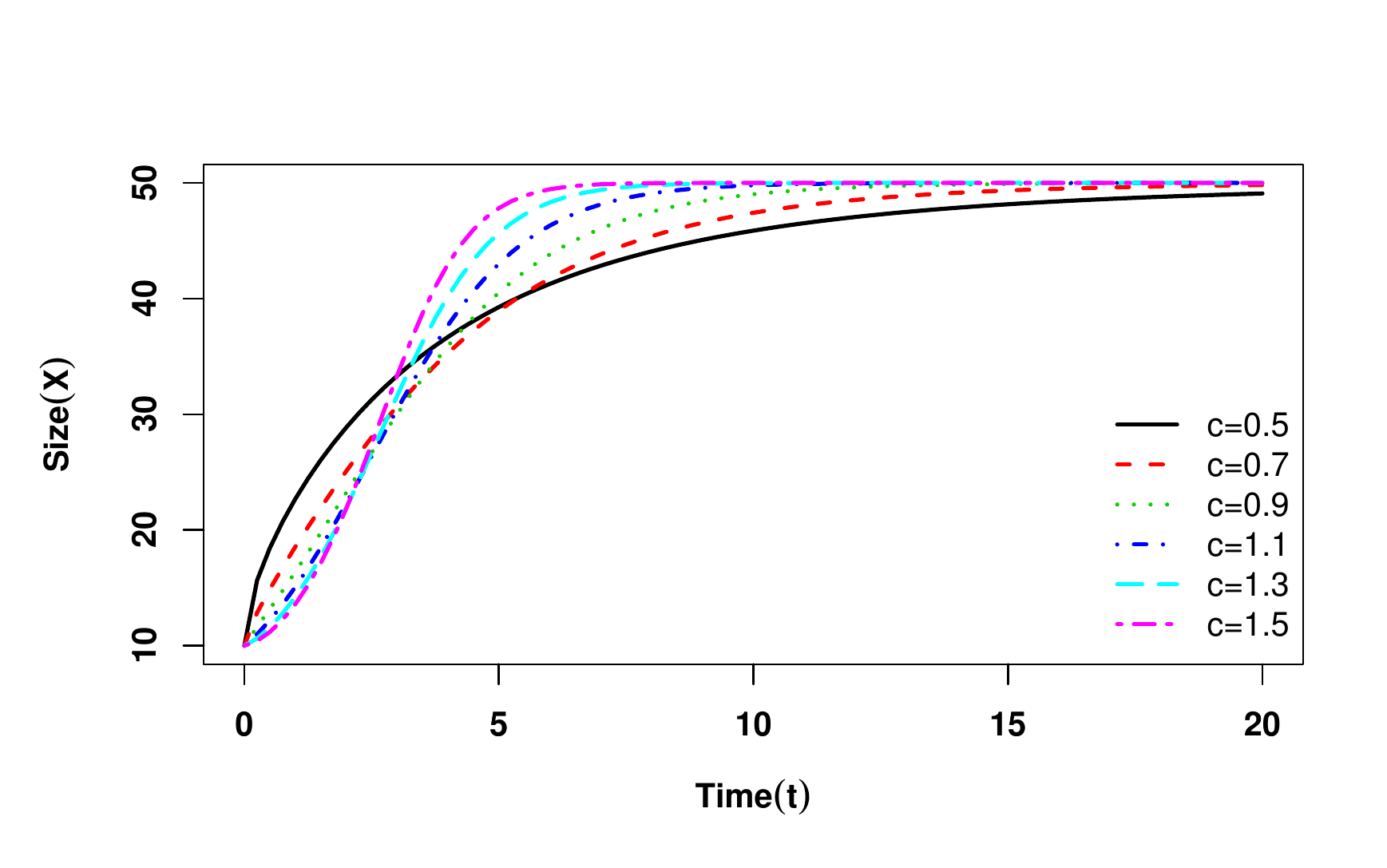}
\caption{$r=r_0t^{c-1}$, ($r_0=0.6$)}
\label{subim8}
\end{subfigure}

\begin{subfigure}[t]{0.33\textwidth}
    \includegraphics[width=\linewidth]{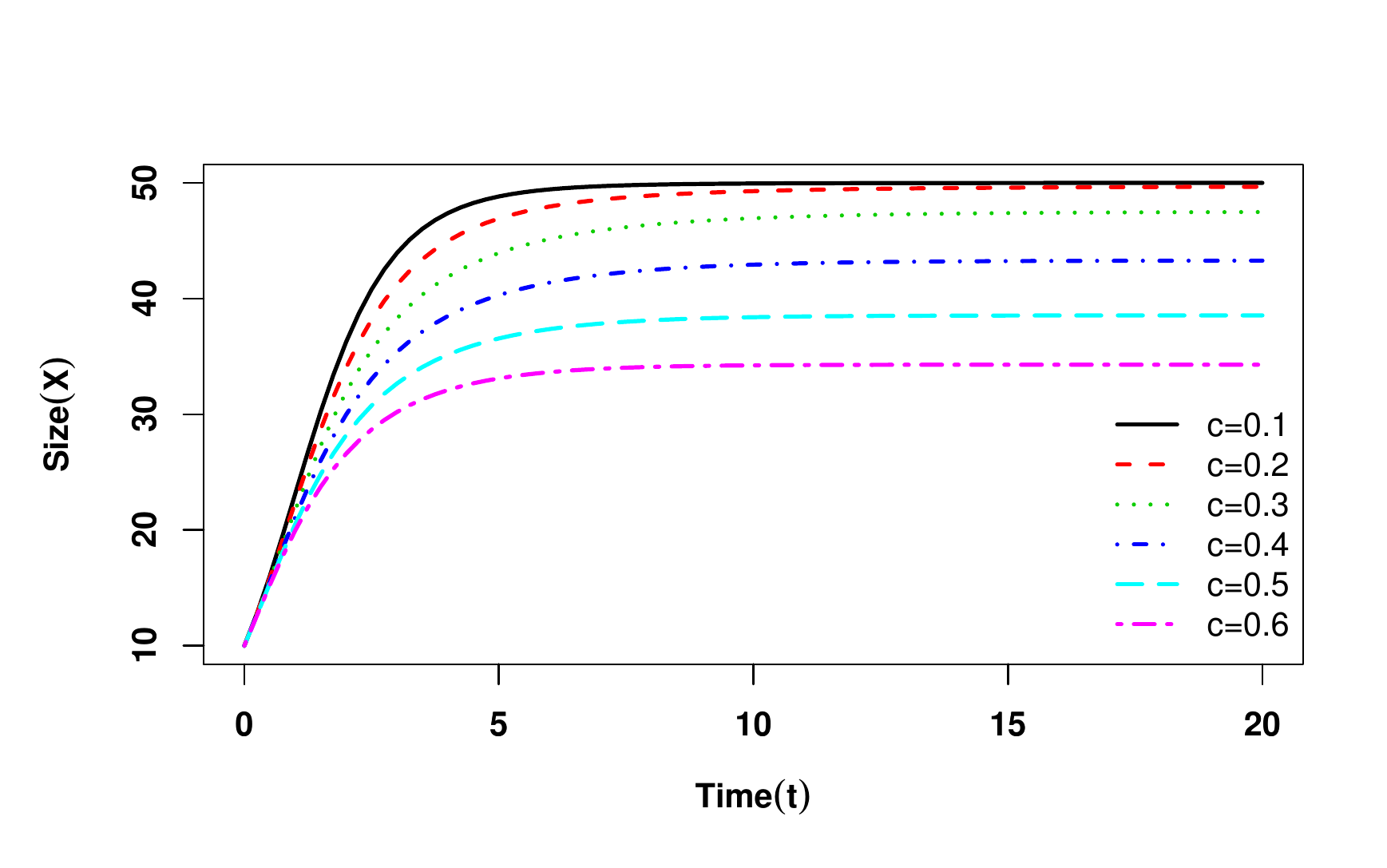}
\caption{$r=r_0e^{-ct}$, ($r_0=1.3$)}
\label{subim9}
\end{subfigure}\hfill
\begin{subfigure}[t]{0.33\textwidth}
    \includegraphics[width=\linewidth]{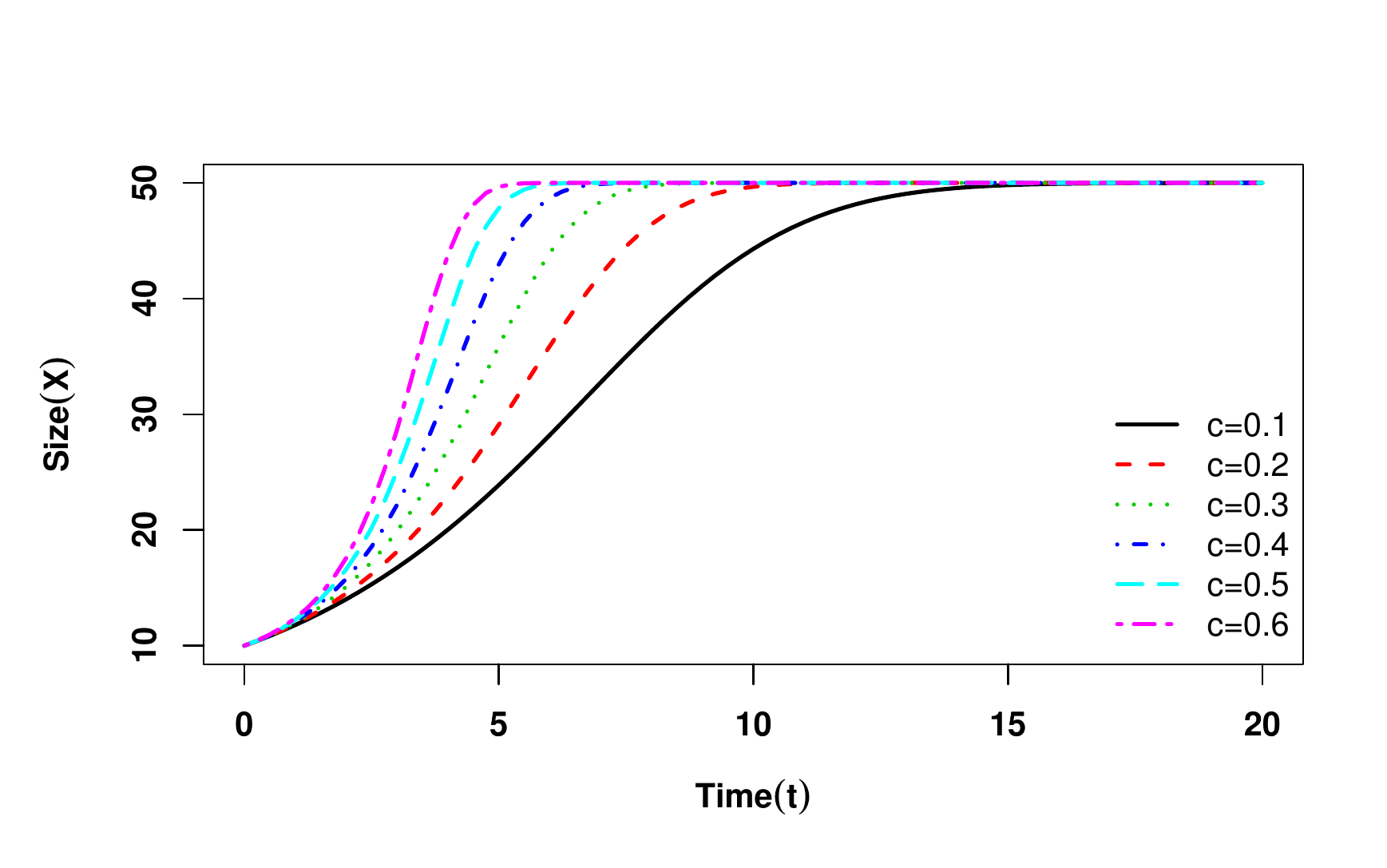}
\caption{$r=r_0e^{ct}$, ($r_0=0.2$)}
\label{subim10}
\end{subfigure}\hfill
\begin{subfigure}[t]{0.33\textwidth}
    \includegraphics[width=\textwidth]{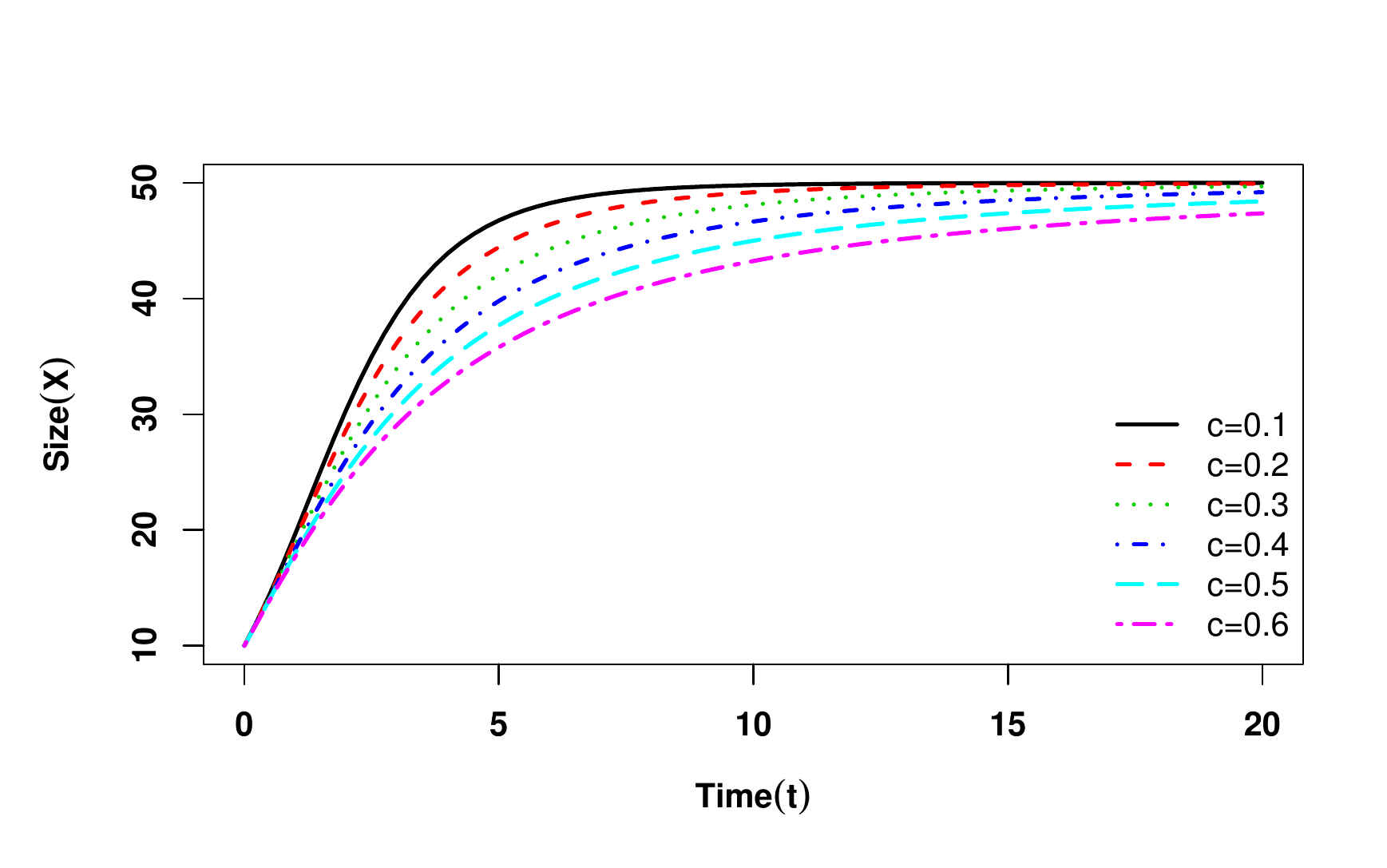}
\caption{$r=\frac{r_0}{1+ct}$, ($r_0=1$)}
\label{subim11}
\end{subfigure}

\begin{subfigure}[t]{0.33\textwidth}
    \includegraphics[width=\linewidth]{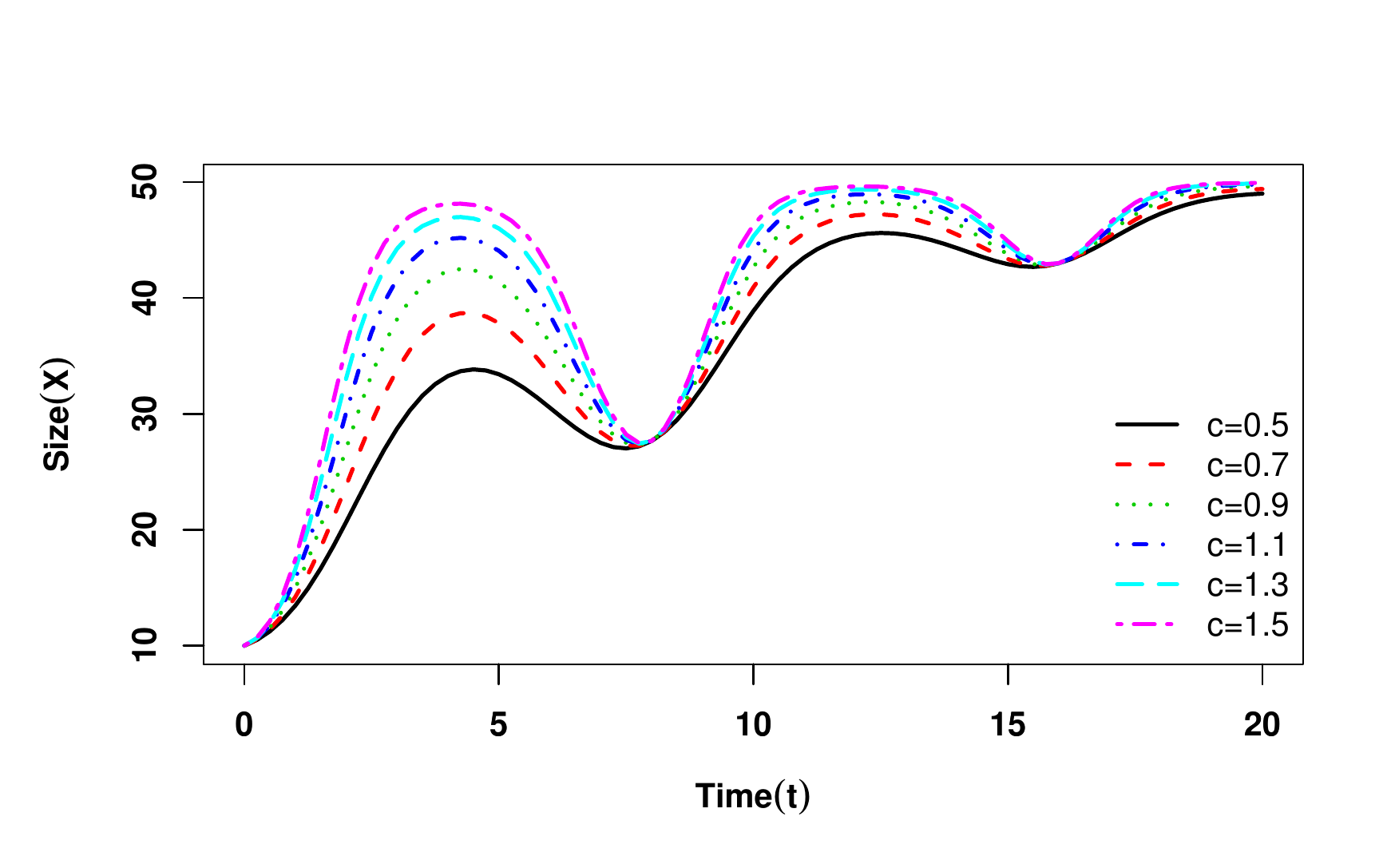}
\caption{$r=r_0+c\sin({\omega t})$, ($r_0=0.2, \omega=\frac{\pi}{4}$)}
\label{subim12}
\end{subfigure}\hfill
\begin{subfigure}[t]{0.33\textwidth}
    \includegraphics[width=\linewidth]{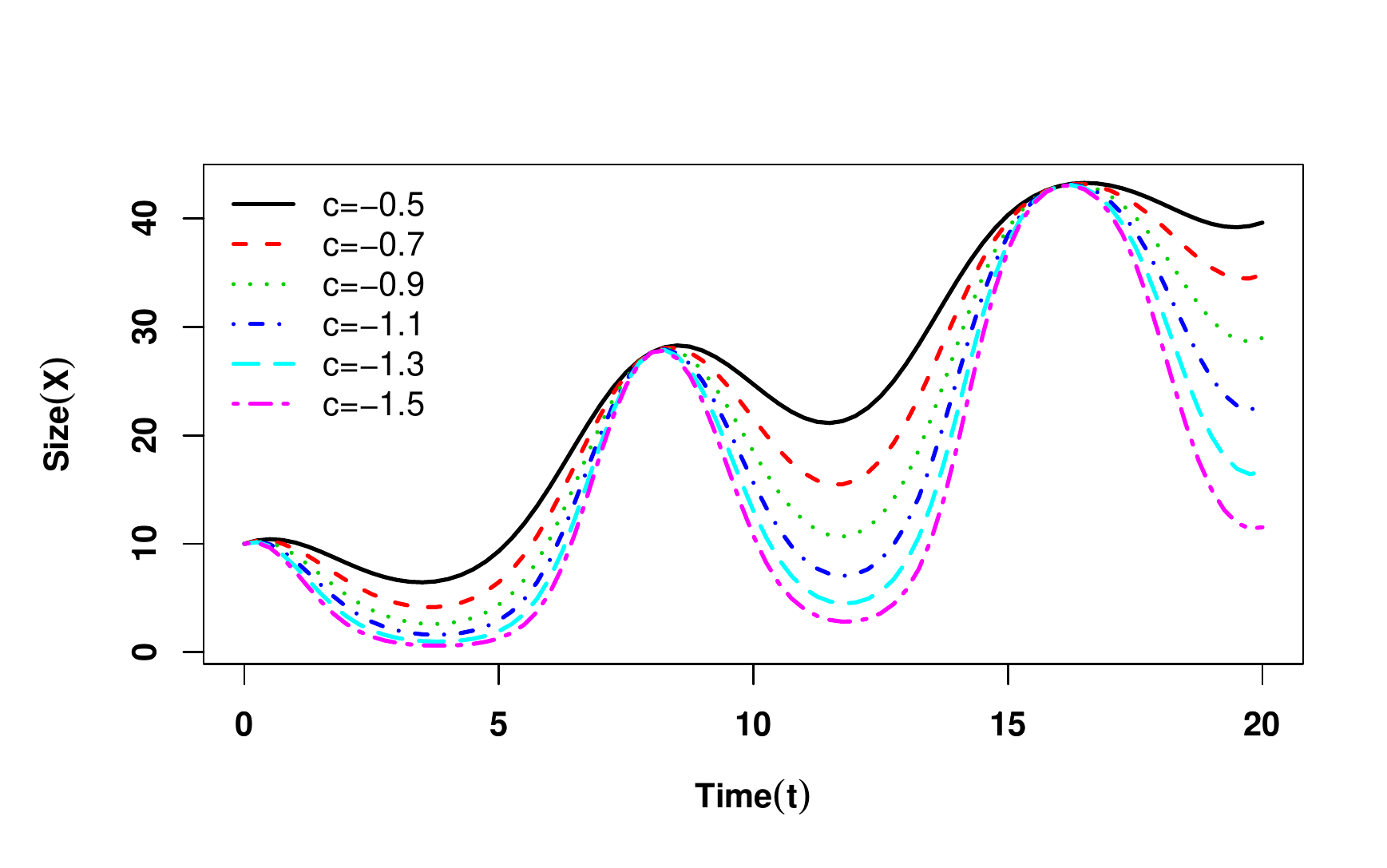}
\caption{$r=r_0+c\sin({\omega t})$, ($r_0=0.2, \omega=\frac{\pi}{4}$)}
\label{subim13}
\end{subfigure}
\begin{subfigure}[t]{0.33\textwidth}
    \includegraphics[width=\linewidth]{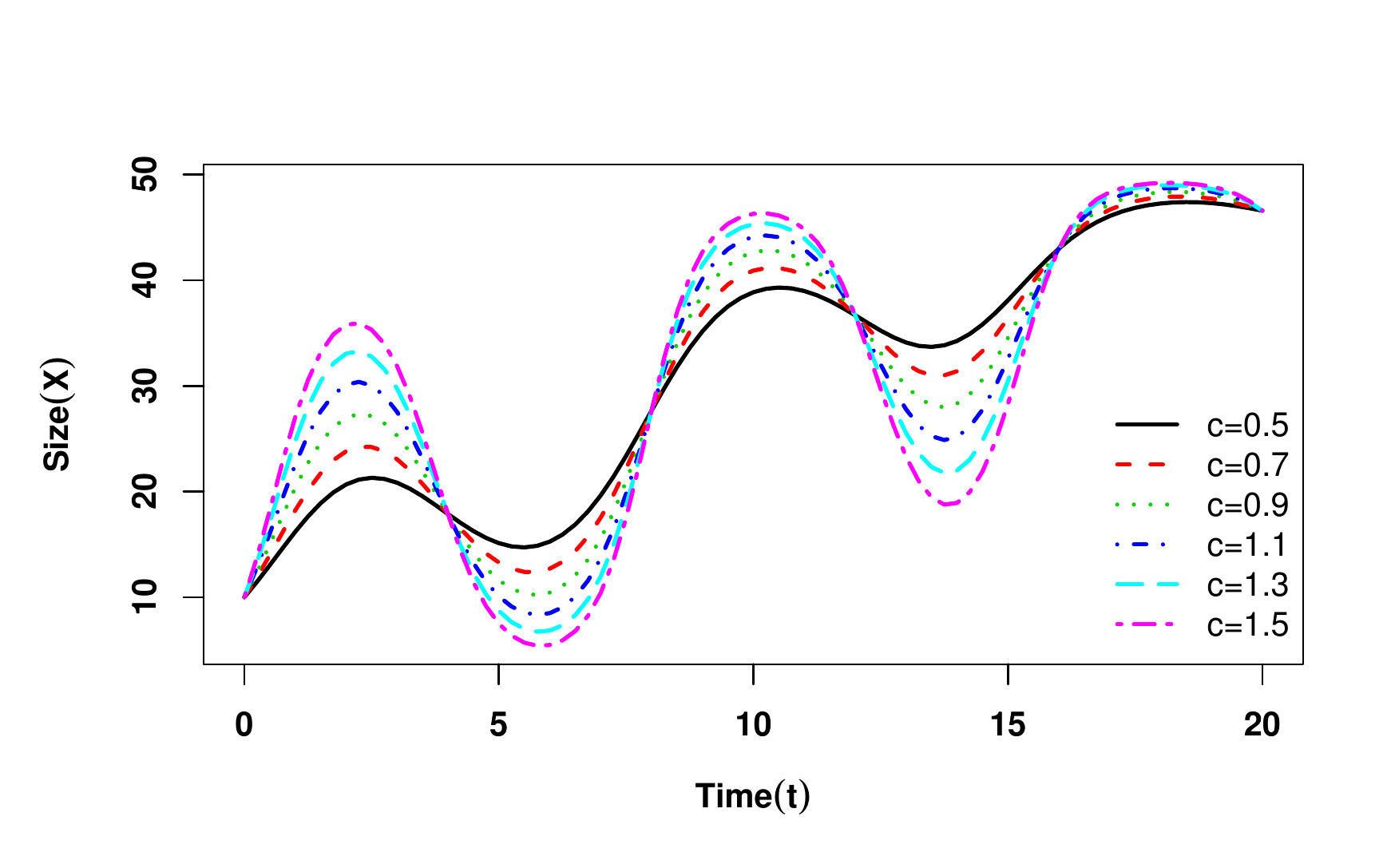}
\caption{$r=r_0+c\cos({\omega t})$, ($r_0=0.2, \omega=\frac{\pi}{4}$)}
\label{subim14}
\end{subfigure}\hfill

\begin{subfigure}[t]{0.33\textwidth}
    \includegraphics[width=\linewidth]{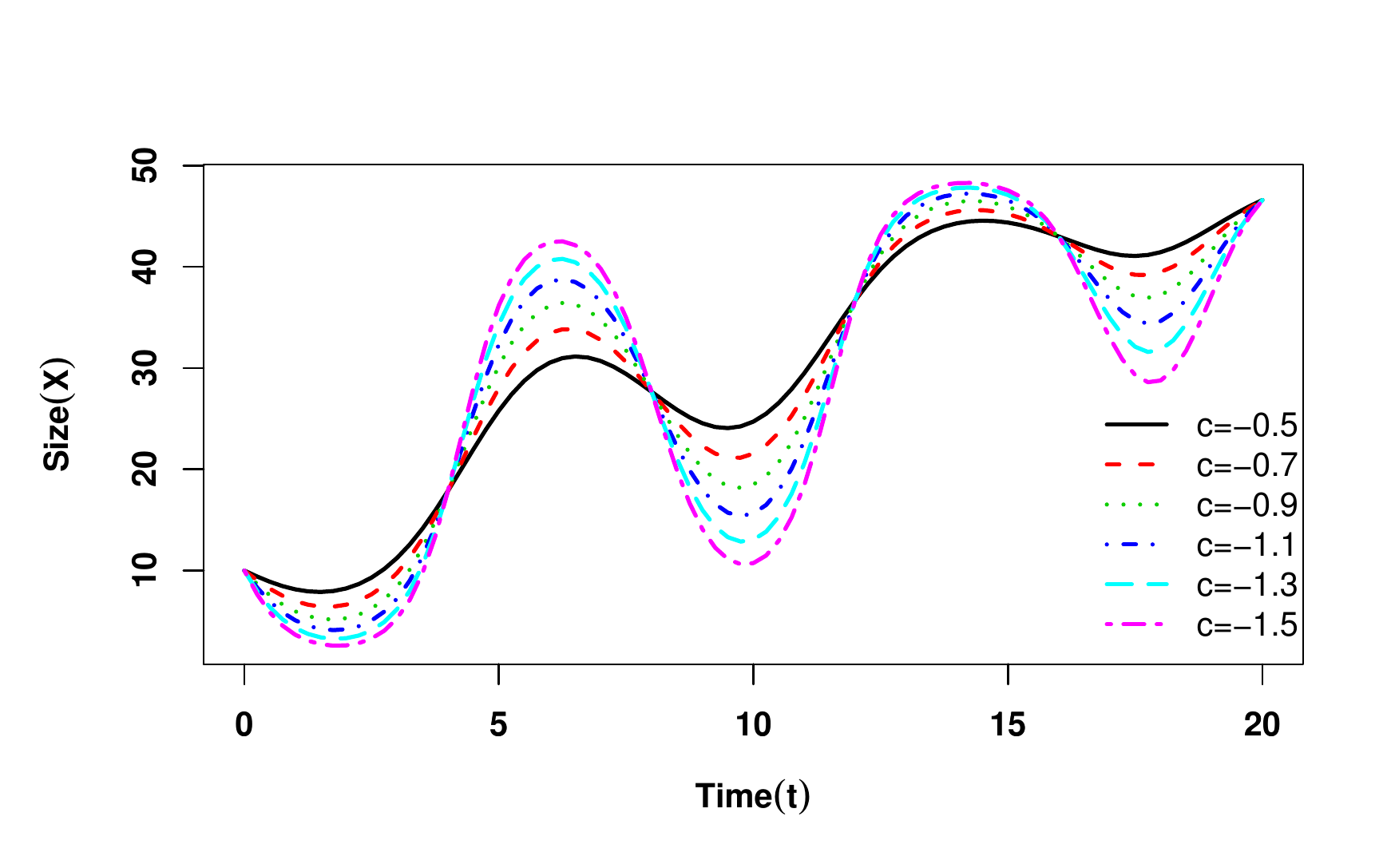}
\caption{$r=r_0+c\cos({\omega t})$, ($r_0=0.2, \omega=\frac{\pi}{4}$)}
\label{subim15}
\end{subfigure}

\caption{Time ($t$) vs size ($X$) plot for continuously varying parameter $r$ in Logistic Model. In the first panel (Fig. (a), (b) and (c)), we consider $r$ as a linearly increasing, decreasing and polynomial function of time, respectively. In the second  panel, in Fig. (d) and (e), we consider $r$ as a exponentially decreasing and increasing function of time, respectively and in Fig. (f), we consider $r$ as inverse function of time. In the third panel, in Fig. (g) and (h) $r$ varies periodically (sine function) with time and in Fig. (i) and  In the forth panel (in Fig. (j)), $r$ varies periodically (cosine function) with time. In all the cases, $r_0$ be the initial values of the parameter $r$. $X_0 = 10$ and $K=50$ are kept fixed.}
\label{Logistic_plot}
\end{figure}

\item $r = r_0t^{c-1}$; ($c>0$) :\\
If we take $r$ as a polynomial function of time in eqn. (\ref{eqn:logistic_model}), then we get extended logistic growth model \citep{CHAKRABORTY2017} (Table~\ref{table3}; srl.$5$). In this case, $X(t)$ converges to $K$ as $t \to \infty$ at faster rate than in the logistic growth model (Fig.~\ref{subim8}). The point of inflection remains same as logistic model.

\item $r = r_0e^{-ct}$; ($c>0$):\\
For exponentially decaying $r$, eqn. (\ref{eqn:logistic_model}) turns into a new model (Table~\ref{table3}; srl.$6$)  in which $\displaystyle \lim_{t\to \infty} X(t)$ depends on $c$ and $r_0$. As $c$ (or $r_0$) increases the asymptotic size decreases (Fig.~\ref{subim9}).

\item $r = r_0e^{ct}$; ($c>0$):\\
For exponentially growing $r$, eqn. (\ref{eqn:logistic_model}) turns into a new model (Table~\ref{table3}; srl.$7$) in which the asymptotic size remains $K$. As $c$ increases $X(t)$ converges to $K$  faster than logistic model (Fig.~\ref{subim10}). The point of inflection also remains the same.

\item $r=\frac{r_0}{1+ct}$; ($c>0$):\\
For this type of variation in $r$, eqn. (\ref{eqn:logistic_model}) turns into a new growth model (Table~\ref{table3}; srl.$8$), whose asymptotic size remains $K$ but $X(t)$ tends towards $K$ very slowly (Fig.~\ref{subim11}) compared to logistic growth model. For large value of $c$, the tendency to go towards $K$ becomes slower.

\item $r=r_0+c\sin({\omega t})$, ( $\omega >0$):\\
In this case, eqn.(\ref{eqn:logistic_model}) turns into a new model (Table~\ref{table3}; srl.$9$). For $c>0$, $X(t) \to K$ in a damped oscillation manner (Fig.~\ref{subim12}) and also for $c<0$, a similar behaviour is observed (Fig.~\ref{subim13}). If we take cosine function instead of sine function in eqn.(\ref{eqn:logistic_model}), then also we get a new model (Table~\ref{table3}; srl.$10$) with similar kind of behaviour (Fig.~\ref{subim14}, \ref{subim15})).
\end{enumerate}

\subsubsection{Variation in theta-logistic growth model}
The theta-logistic model (eqn.~\ref{eqn:theta_logistic_model}) (also referred as generalized logistic) is one of the most widely used growth equations in ecological literature. The model was first proposed by \citet{Gilpin3590} in the concept of competitive interactive systems. Several studies are available in the literature based on this model (\citet{sibly2005, BARKER2008, clark2010, ZHAO2011, BHOWMICK2016} and other references there in). We consider both time and density dependent variation in $r$ and $K$ is kept as constant since almost for all the cases, the resulting growth equations do not have analytical solution. They must be solved numerically. We investigate the equations for different range of values of $\theta$. In the following we categorically discuss each cases.

\begin{enumerate}
\item $\theta = 1$ :\\
This is essentially the logistic model whose dynamics are well studied \citep{Kot2001} (Table~\ref{table4}; srl.$1$)

\item $r=\frac{r_0}{\theta}$, $\theta \to 0$ :\\
If $r$ takes this type of form and $\theta$ tends to zero in eqn. (\ref{eqn:theta_logistic_model}), then the equation turns out into gompertz model \citep{Gompertz1825}. For gompertz model the carrying capacity remains $K$ (for fixed initial size $X_0$) but point of inflection changes into $\frac{K}{e}$ (Fig.~\ref{subim16}) (Table~\ref{table4}; srl.$2$).

\item $\theta \geq -1$ :\\
If we consider this limitation in $\theta$ in eqn.(\ref{eqn:theta_logistic_model}), then it turns into Richard's growth law \citep{Richards1959}. For negative value of $\theta$,  $\displaystyle \lim_{t\to \infty} X(t)= 0$ but for positive value of $\theta$,  $\displaystyle \lim_{t\to \infty} X(t)= K$ (Fig.~\ref{subim17}) (Table~\ref{table4}; srl.$3$). If we take $\theta$ equal to zero then $X(t)$ is constant. However in application one should evaluate limiting case $\theta \to 0$ as demonstrated in \citet{saether1998}. 

\item $r = r_0t^{c-1}$ ($c > 0$), $\theta>0$ :\\
For such variation, the eqn. (\ref{eqn:theta_logistic_model}) turns into Koya-Goshu Model \citep{koya2013}. In the revised equation the asymptotic size remains $K$. For $c < 1$, there is no point of inflection i.e. no lag phase or log phase, but for $c > 1$, point of inflection is present in Koya-Goshu model (Fig.~\ref{subim18}) (Table~\ref{table4}; srl.$4$).

\item $r = r_0(1+ct)$ ($c>0$):\\
If we take linearly increasing time dependent functional form of $r$ in eqn. (\ref{eqn:theta_logistic_model}), then a new growth equation is obtained whose asymptotic size remains $K$ (Fig.~\ref{subim19}) (Table~\ref{table4}; srl.$5$). As $c$ increases, $X(t)$ goes towards $K$ at faster rate.

\item $r = r_0(1-ct)$ ($c>0$):\\
Like the previous case, here also a new growth equation is obtained whose asymptotic size is zero (instead of $K$). If we increase the value of $c$, $X(t)$ goes towards zero at faster rate (Fig.~\ref{subim20}) (Table~\ref{table4}; srl.$6$).

\item $r = \frac{r_0}{\theta}t^{c-1}$, ($c > 0$), $\theta \to 0$ :\\
If we take this type of variation of $r$ and limitation of $\theta$ in eqn. (\ref{eqn:theta_logistic_model}), then it turns into extended gompertz model \citep{Bhowmick2014} for which $K$ remains the asymptotic size and for large $c$, $X(t)$ goes towards the asymptotic size at a much faster rate (Fig.~\ref{subim21}) (Table~\ref{table4}; srl.$7$).

\begin{figure}[H]
\begin{subfigure}[t]{0.33\textwidth}
    \includegraphics[width=\linewidth]{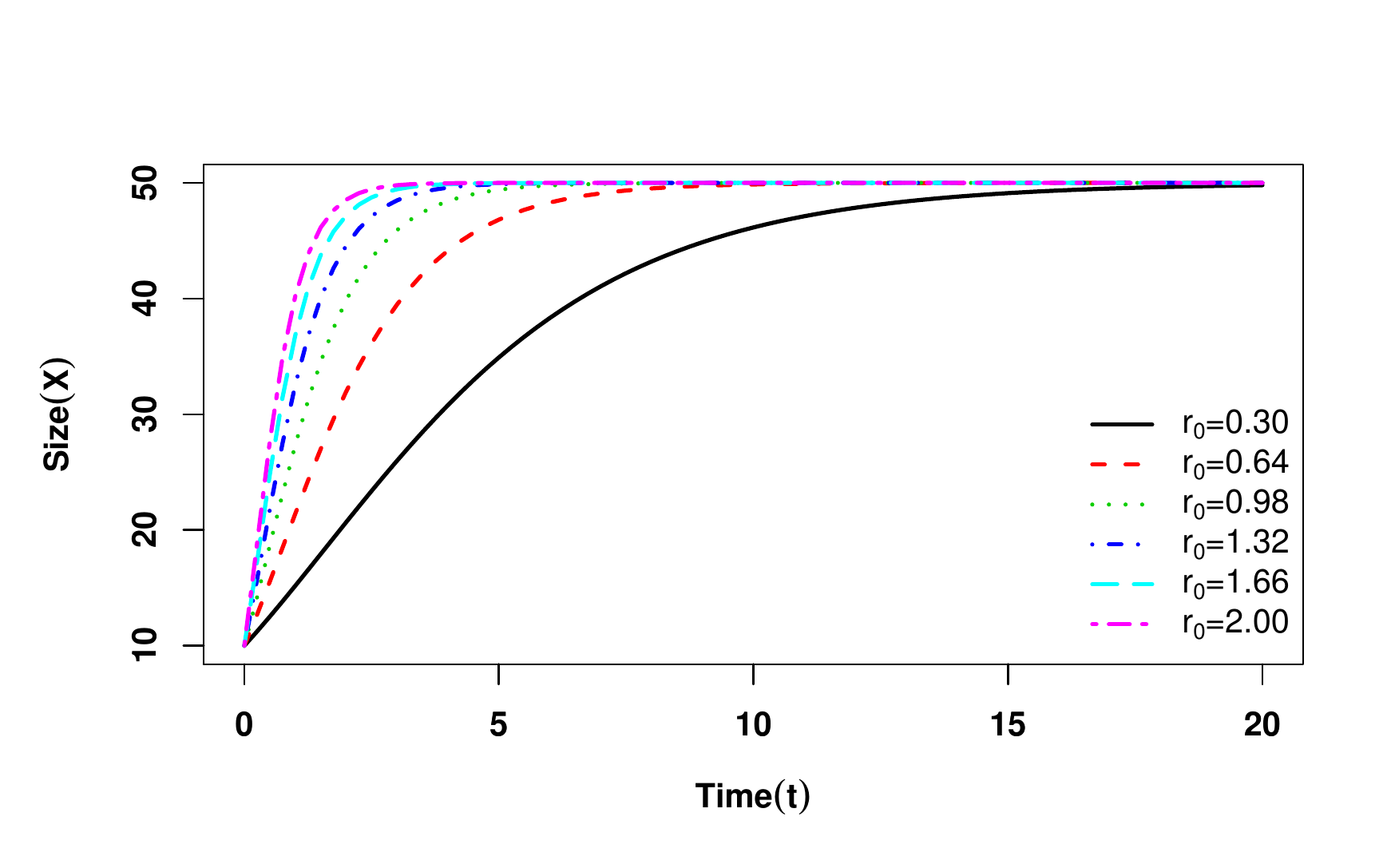}
\caption{$r=\frac{r_0}{\theta}$, $\theta \to 0$}
\label{subim16}
\end{subfigure}\hfill
\begin{subfigure}[t]{0.33\textwidth}
  \includegraphics[width=\linewidth]{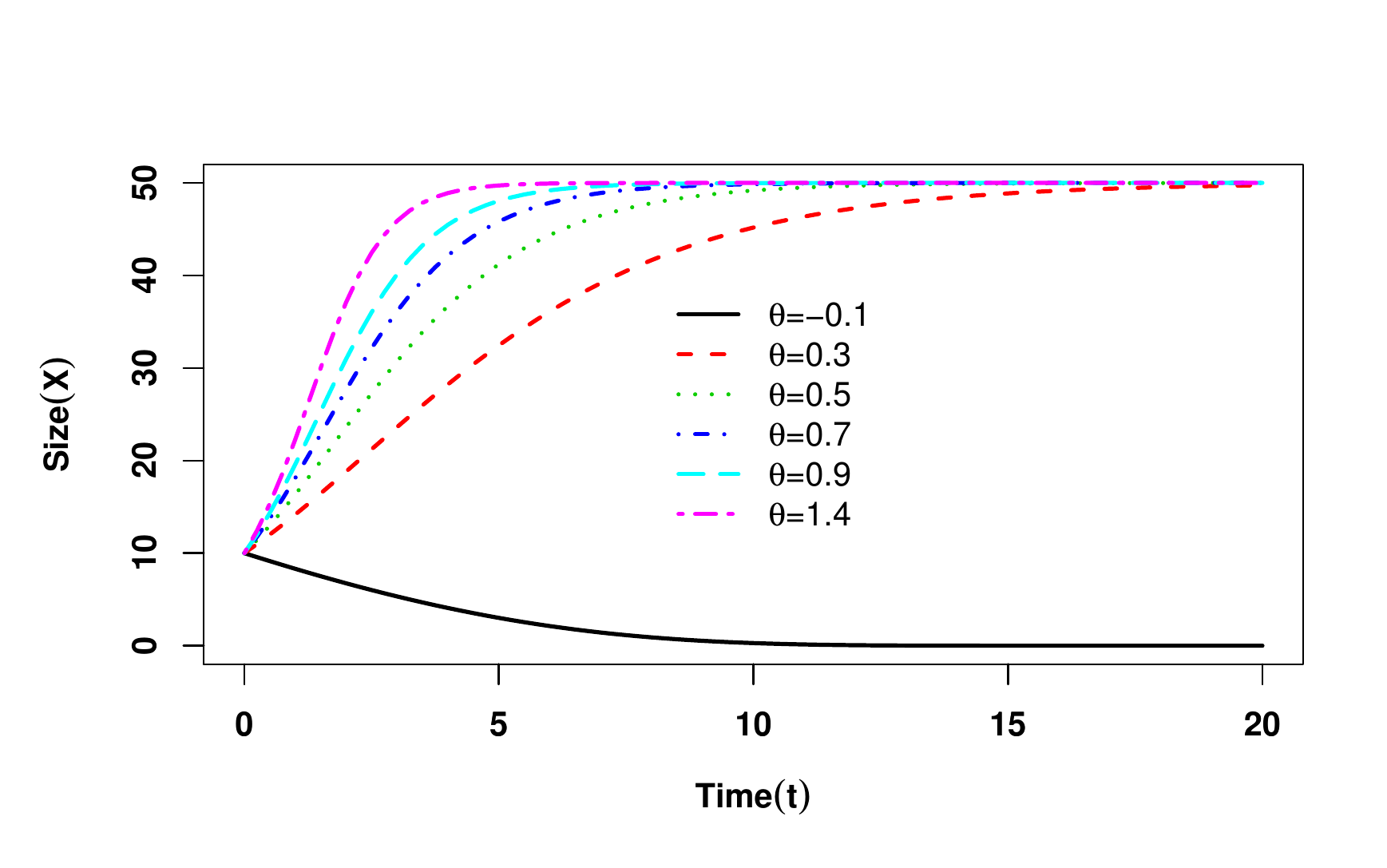}
\caption{$\theta \geq -1$, $r_0=1$}
\label{subim17}
\end{subfigure}\hfill
\begin{subfigure}[t]{0.33\textwidth}
    \includegraphics[width=\linewidth]{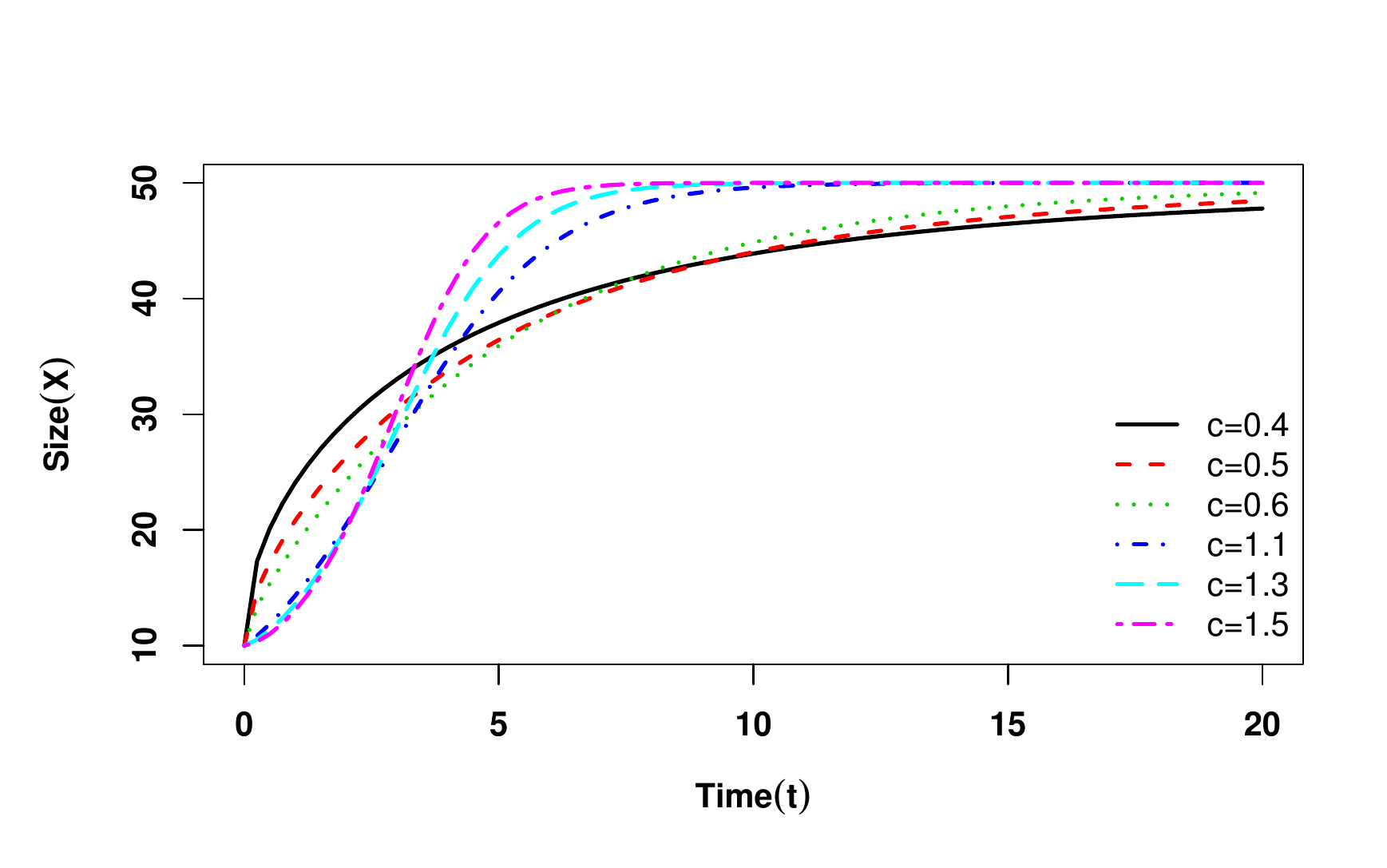}
\caption{$r=r_0t^{c-1}$, ($r_0=0.5$, $\theta=1.1$)}
\label{subim18}
\end{subfigure}

\begin{subfigure}[t]{0.33\textwidth}
    \includegraphics[width=\linewidth]{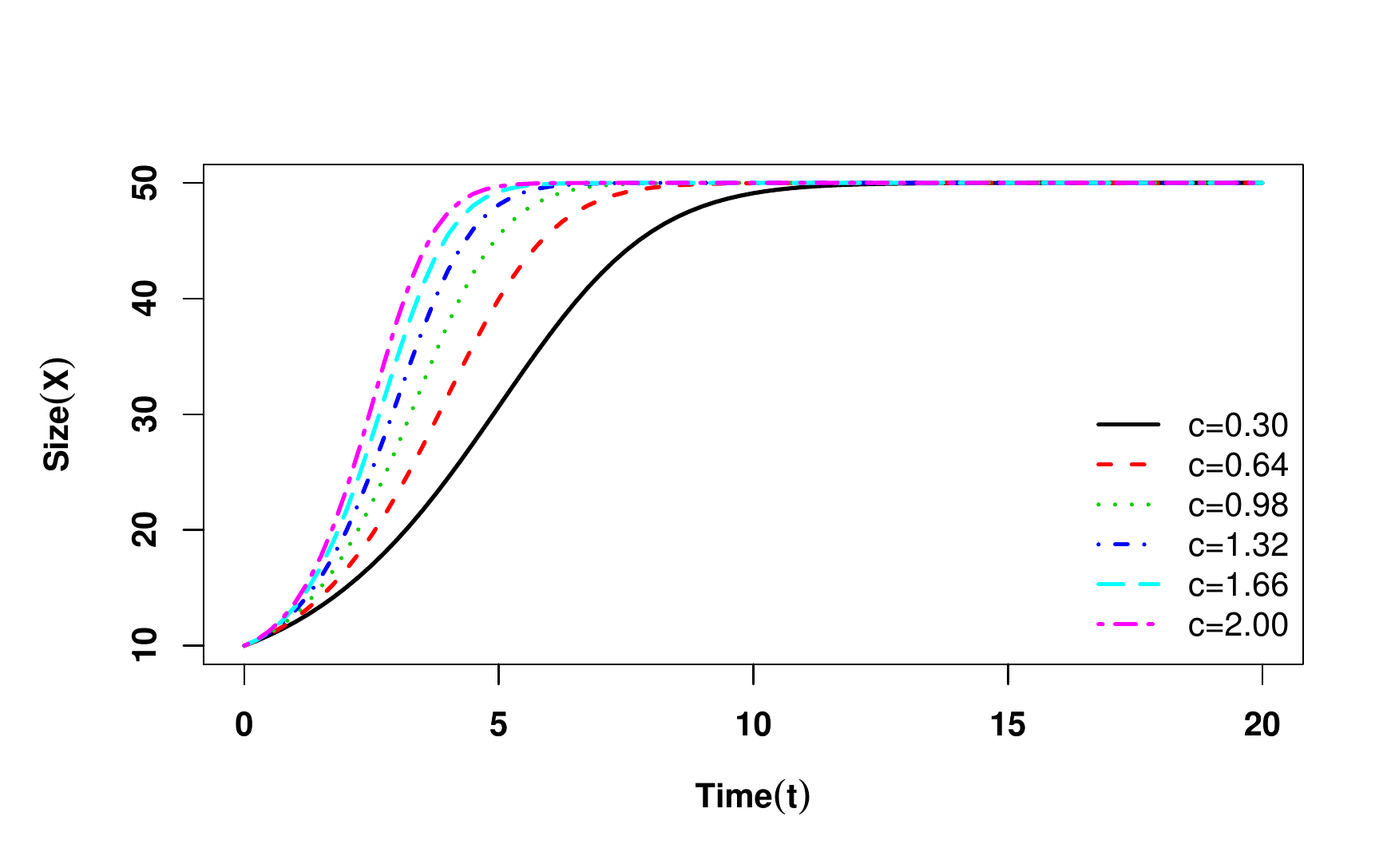}
\caption{$r=r_0(1+ct)$, ($r_0=0.2$, $\theta=1.1$)}
\label{subim19}
\end{subfigure}\hfill
\begin{subfigure}[t]{0.33\textwidth}
    \includegraphics[width=\linewidth]{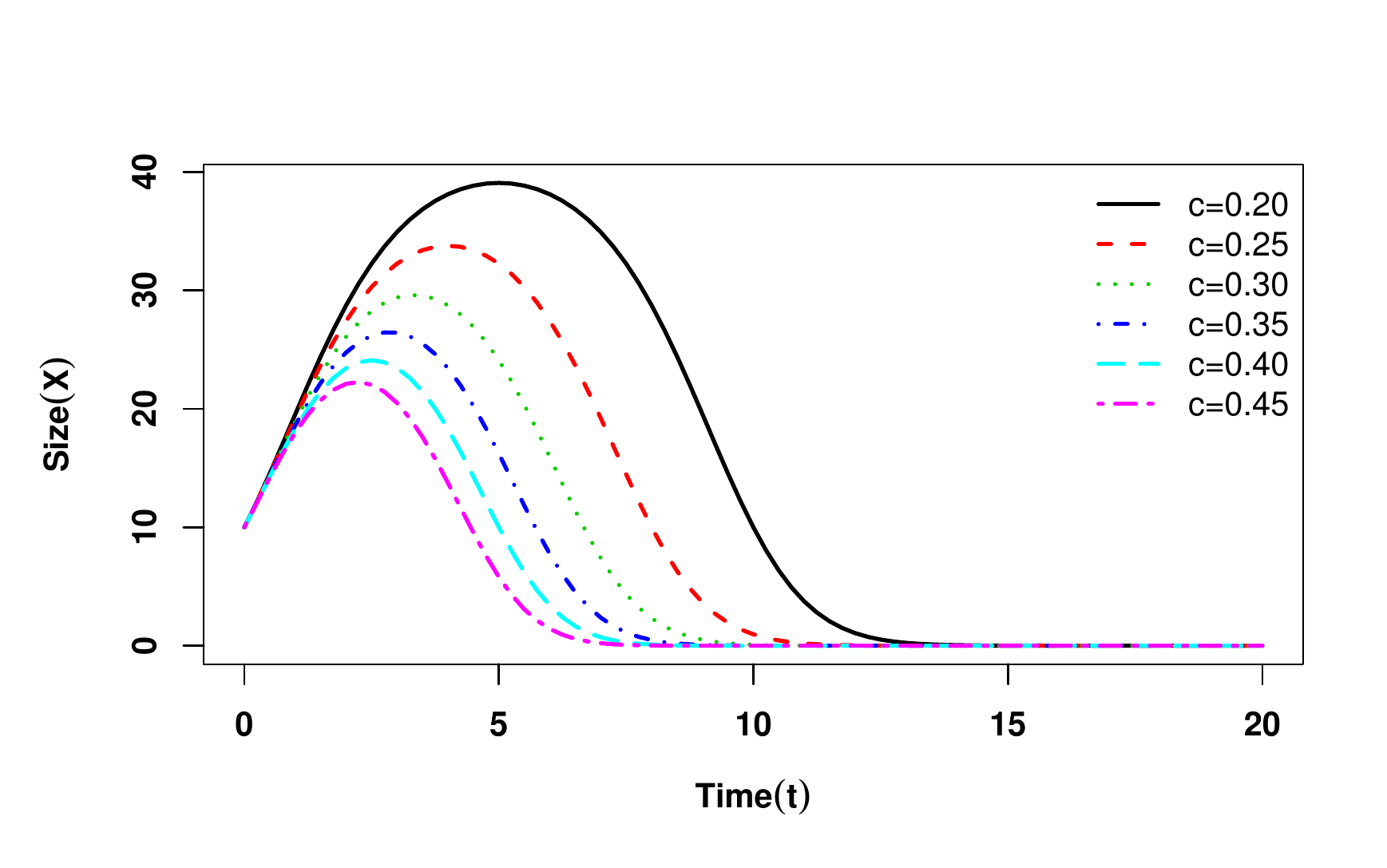}
\caption{$r=r_0(1-ct)$, ($r_0=1$, $\theta=1.1$)}
\label{subim20}
\end{subfigure}\hfill
\begin{subfigure}[t]{0.33\textwidth}
    \includegraphics[width=\textwidth]{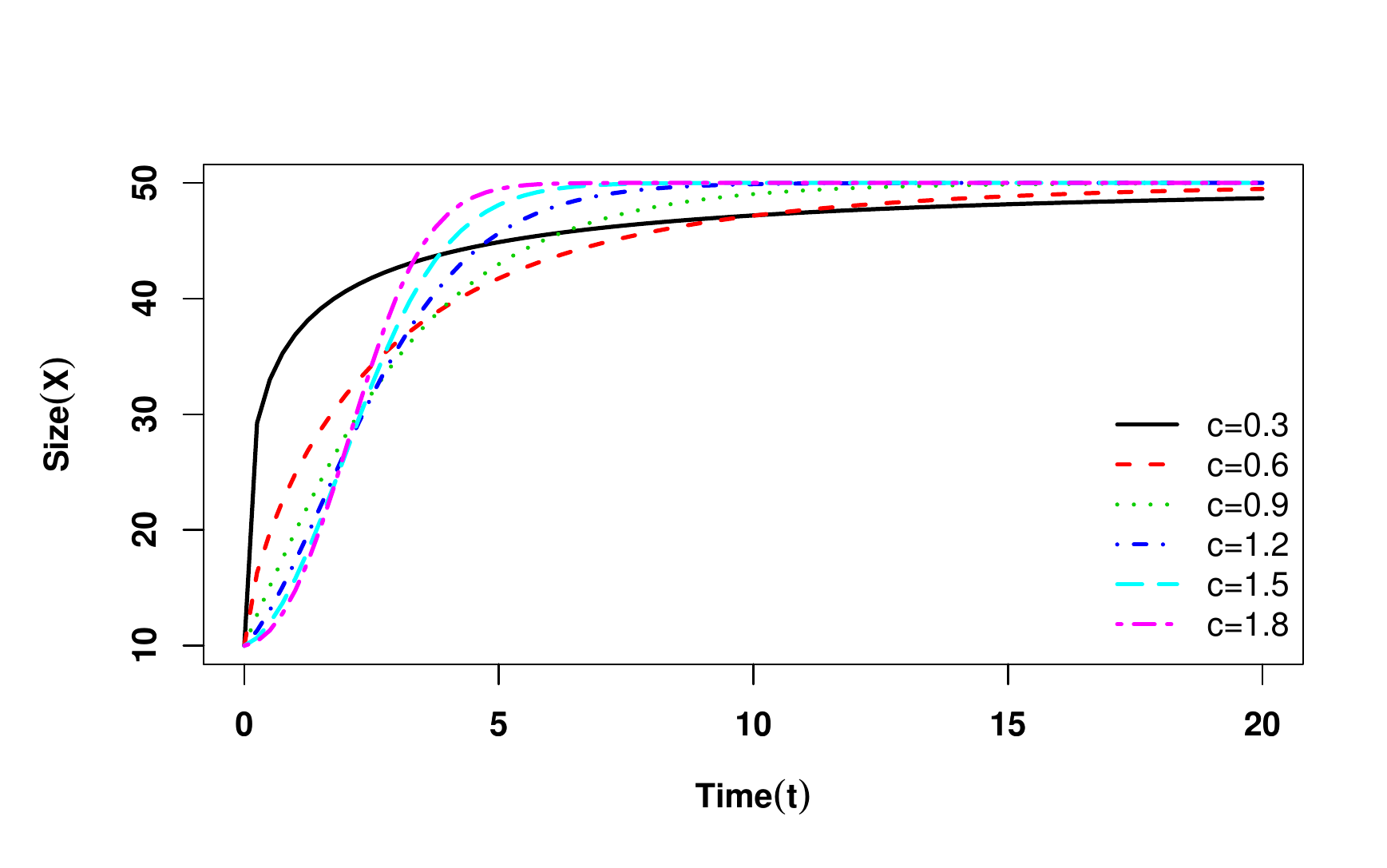}
\caption{$r=\frac{r_0}{\theta}t^{c-1}$, ($r_0=0.5$, $\theta \to 0$)}
\label{subim21}
\end{subfigure}

\begin{subfigure}[t]{0.33\textwidth}
    \includegraphics[width=\linewidth]{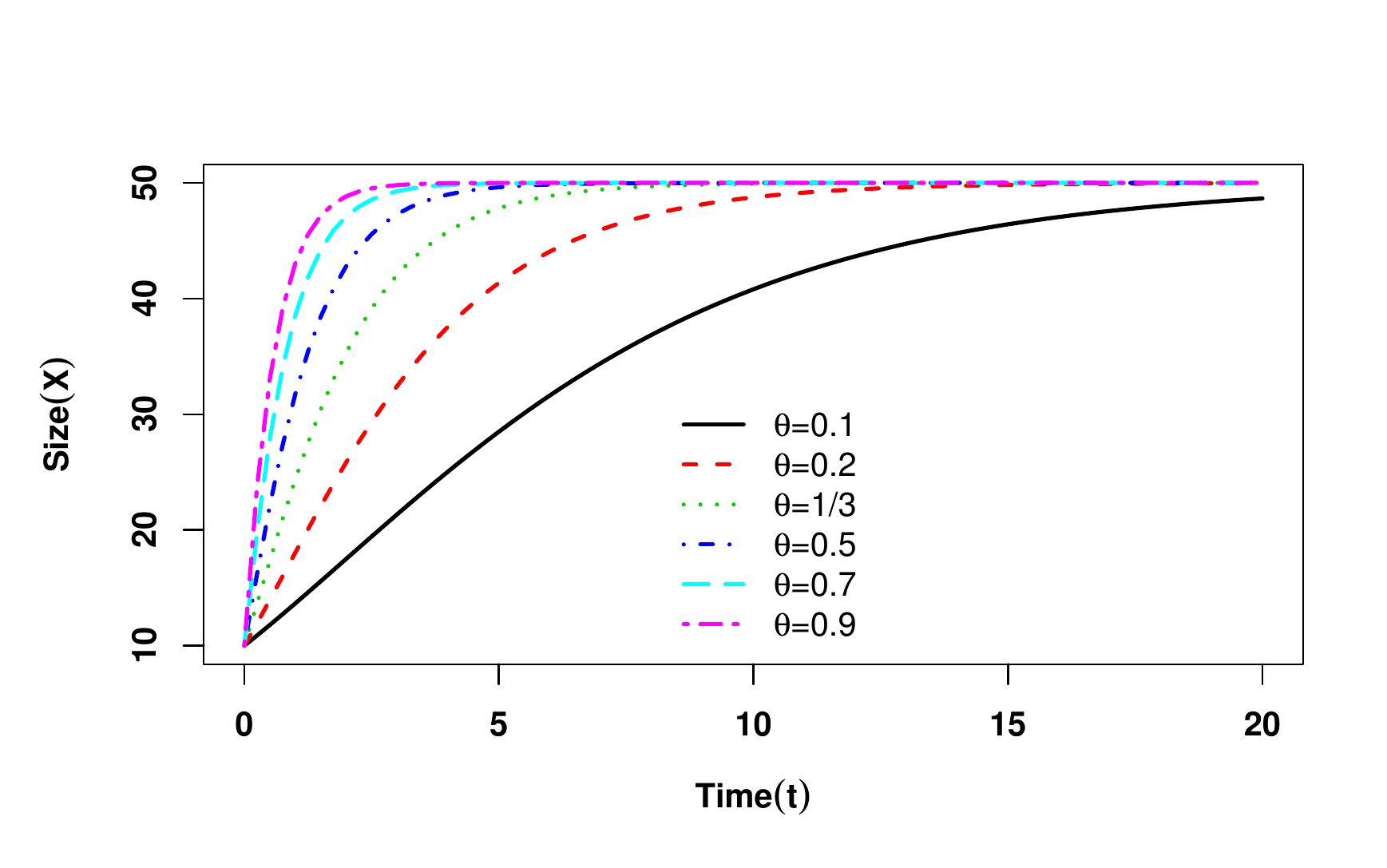}
\caption{$r=r_0\left(\frac{K}{X}\right)^{\theta}$, ($r_0=2$)}
\label{subim22}
\end{subfigure}\hfill
\begin{subfigure}[t]{0.33\textwidth}
    \includegraphics[width=\linewidth]{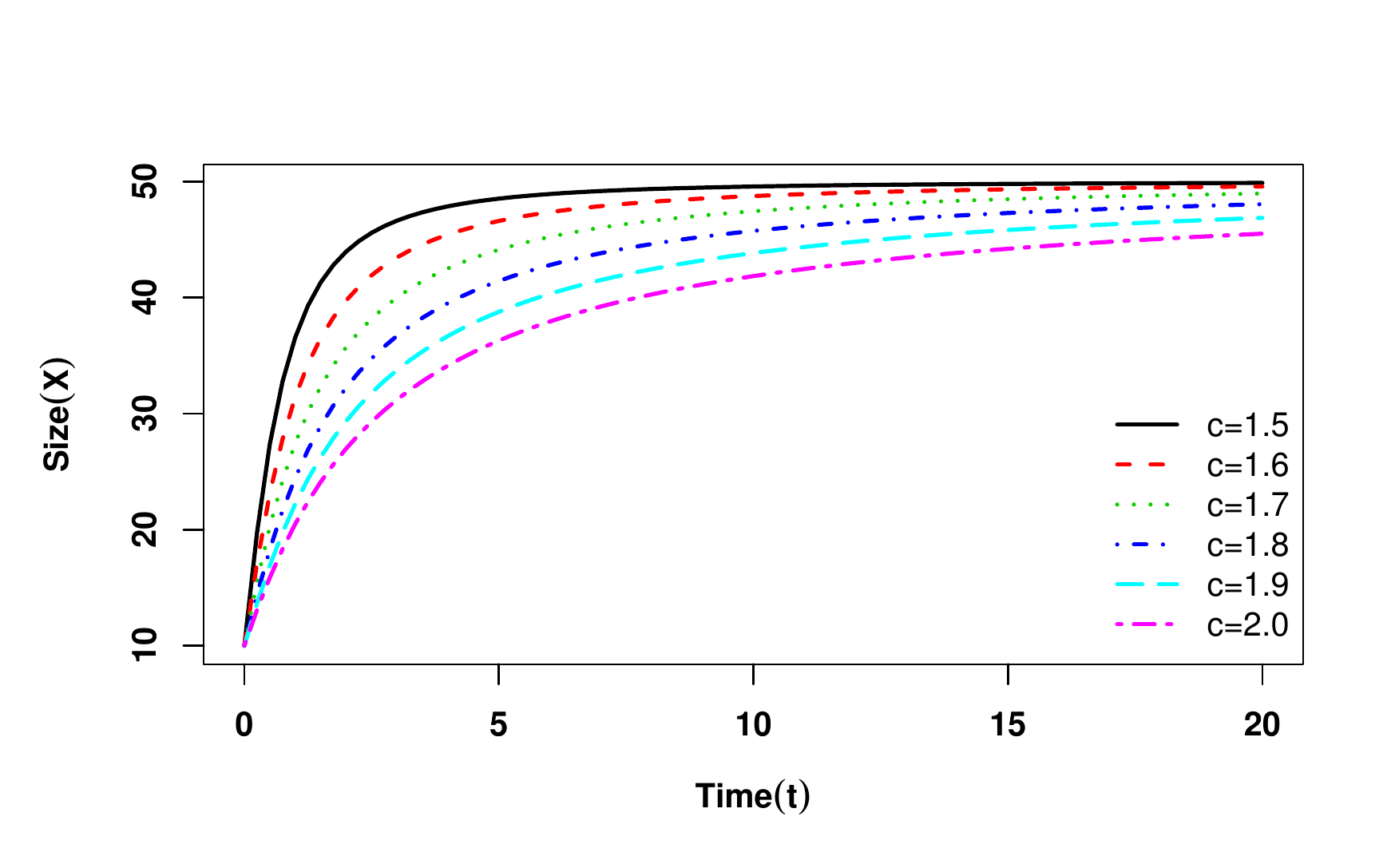}
\caption{$r=\frac{r_0}{\theta}\left(\ln{\frac{K}{X}}\right)^{c-1}$, ($r_0=0.5$, $\theta \to 0$)}
\label{subim23}
\end{subfigure}\hfill
\begin{subfigure}[t]{0.33\textwidth}
    \includegraphics[width=\linewidth]{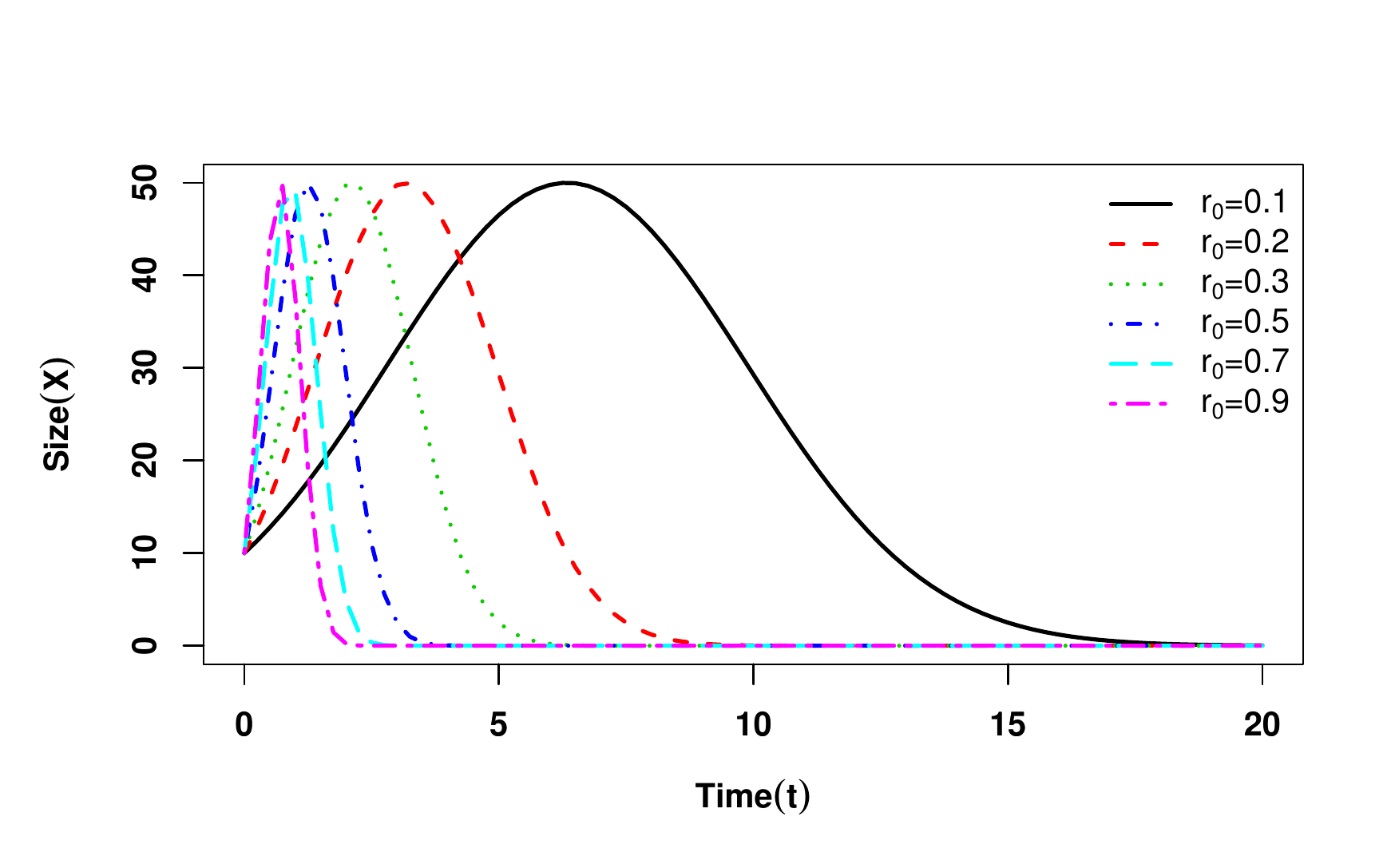}
\caption{$r=\frac{r_0}{\theta}\left(\ln{\frac{K}{X}}\right)^{-\frac{1}{2}}$, ($c=0.5$, $\theta \to 0$)}
\label{subim24}
\end{subfigure}

\caption{Size profile of the theta-logistic model for different choices of $r$ and $K$. In the upper panel, in Fig. (a), we consider $r=\frac{r_0}{\theta}$ and $\theta \to 0$, in Fig. (b), we consider $\theta \geq -1$ and $r=1$ and in Fig. (c), we consider $r$ as a polynomial function of time. In the middle panel, in Fig. (d) and (e), we consider $r$ as a linearly increasing and decreasing function of time, respectively and in Fig. (f), we consider $r$ as a polynomial function of time in the presence of $\theta$. In the lower panel (Fig. (g), (h) and (i)), we consider $r$ as a density dependent function to get Generalized Von-Bertalanffy Model, Crescenzo-Spina Model and Second-order Exponential Polynomial from theta-logistic Model, respectively. In all cases, $r_0$ be the initial values of the parameter $r$. $X_0=10$ and $K=50$ are kept fixed.}
\label{Theta_Logistic_plot}
\end{figure}

\item $r=r_0\left(\frac{K}{X}\right)^{\theta}$, ($0<\theta <1$) :\\
For this type of density dependent variation in $r$ and limitation in $\theta$ eqn. (\ref{eqn:theta_logistic_model}) transforms into Generalized Von-Bertalanffy Model \citep{von1960} for which $\displaystyle \lim_{t\to \infty} X(t)= K$  and the rate of convergence depends on $\theta$. For large value of $\theta$, $X(t)$ goes at faster rate towards $K$ (Fig.~\ref{subim22}) (Table~\ref{table4}; srl.$8$). If we take $\theta=\frac{1}{3}$, then eqn. (\ref{eqn:theta_logistic_model}) turns out into Von-Bertalanffy Model \citep{Von1949} (Table~\ref{table4}; srl.$9$).

\item $r=\frac{r_0}{\theta}\left(\ln{\frac{K}{X}}\right)^{c-1}$ ($c>0$), $\theta \to 0$ :\\
For this type of density dependent variation in $r$ and limitation in $\theta$, eqn. (\ref{eqn:theta_logistic_model}) turns out into generalized gompertz model \citep{CHAKRABORTY2017} (Table~\ref{table4}; srl.$10$). If we take $c = 1$, then generalized gompertz model turns into gompertz model \citep{Gompertz1825}. If we take $c = 0.5$, then this model turns out into second-order exponential polynomial \citep{CHAKRABORTY2017} for which the asymptotic size is zero (Fig.~\ref{subim23}) (Table~\ref{table4}; srl.$11$). If we take $c-1> 0$, then generalized gompertz model turns out into Cresenzo-Spina model \citep{CRESCENZO2016} for which $K$ is the asymptotic size (Fig.~\ref{subim24}) (Table~\ref{table4}; srl.$12$). Basically for $0 < c < 1$, $\displaystyle \lim_{t\to \infty} X(t) = 0$ always. Thus in generalized gompertz model, for $0 < c < 1$, asymptotic size is zero and for $c \geq 1$, asymptotic size is $K$.
\end{enumerate}

\subsubsection{Variation in confined exponential growth model}
In this section, we discuss various changes in the shapes of confined exponential model (eqn.~\ref{confined_exponential_model}) (also known as Monomolecular growth law) by varying the parameters in the the model. \citet{Banks1994} had already discussed about some variation of $r$ in the confined exponential model set up and found relationship with extreme minimal value distribution. Here, we explore other possible variations. In some cases, we have also obtained new growth equations as well. In the following we categorically discuss different cases. 

\begin{enumerate}
\item $r = r_0t^{c-1},$ ($c > 0$) :\\
If we take $r$ as a polynomial function of time in eqn. (\ref{confined_exponential_model}), then it turns out into Weibull distribution \citep{weibull1951}. For this distribution $K$ remains the asymptotic size (Fig.~\ref{subim25}) (Table~\ref{table5}; srl.$2$), but $X(t)$ approaches towards $K$ at faster rate for large values of $c$ or $r_0$.

\item $r=r_0\frac{K}{X}$ :\\
If we take this density dependent variation in $r$, then eqn. (\ref{confined_exponential_model}) transforms into logistic model with asymptotic size  $K$ \citep{Verhulst1838} (Table~\ref{table5}; srl.$3$).

\begin{figure}[H]
\begin{subfigure}[t]{0.33\textwidth}
    \includegraphics[width=\linewidth]{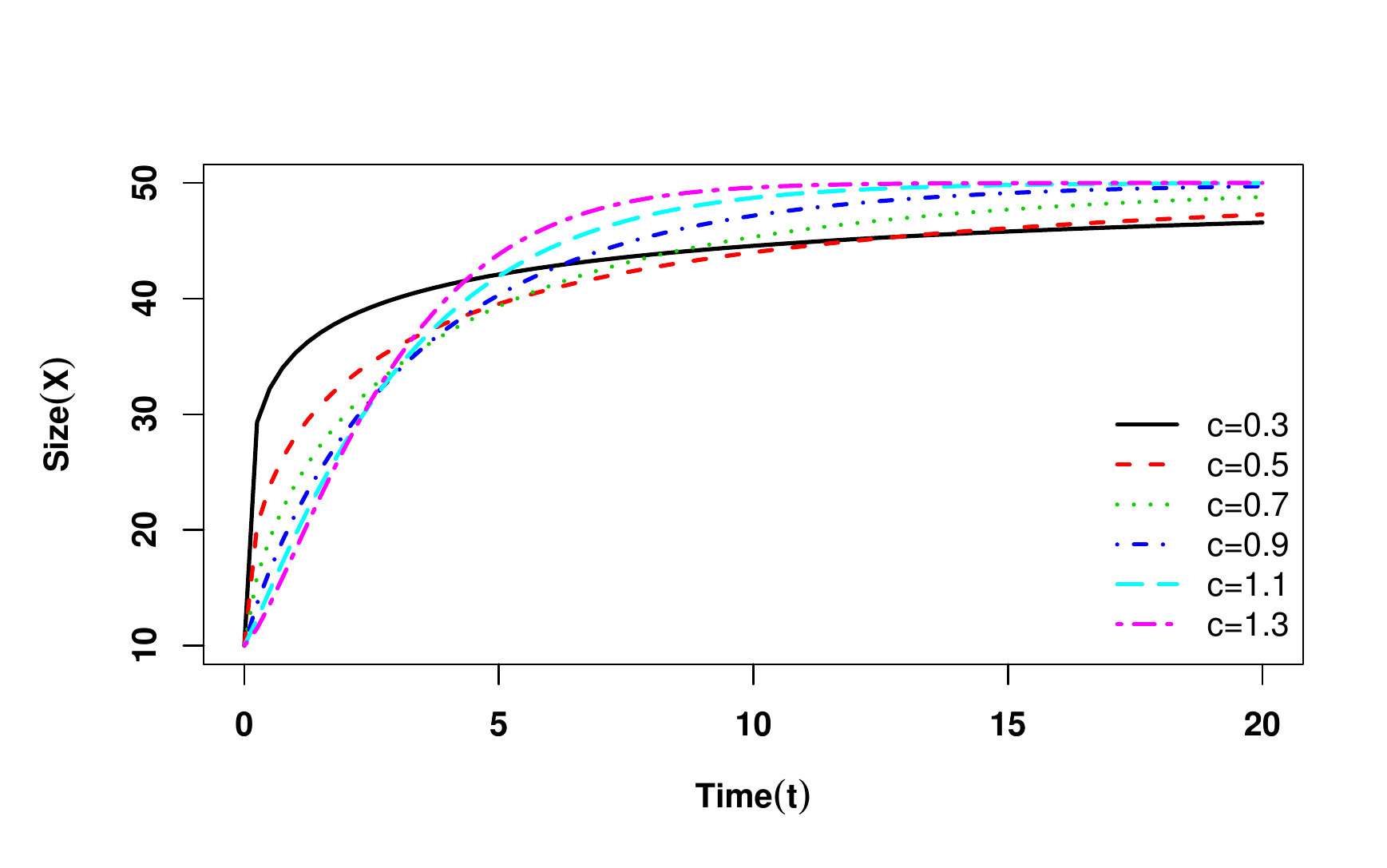}
\caption{$r=r_0t^{c-1}$, ($r_0=0.3$) }
\label{subim25}
\end{subfigure}\hfill
\begin{subfigure}[t]{0.33\textwidth}
  \includegraphics[width=\linewidth]{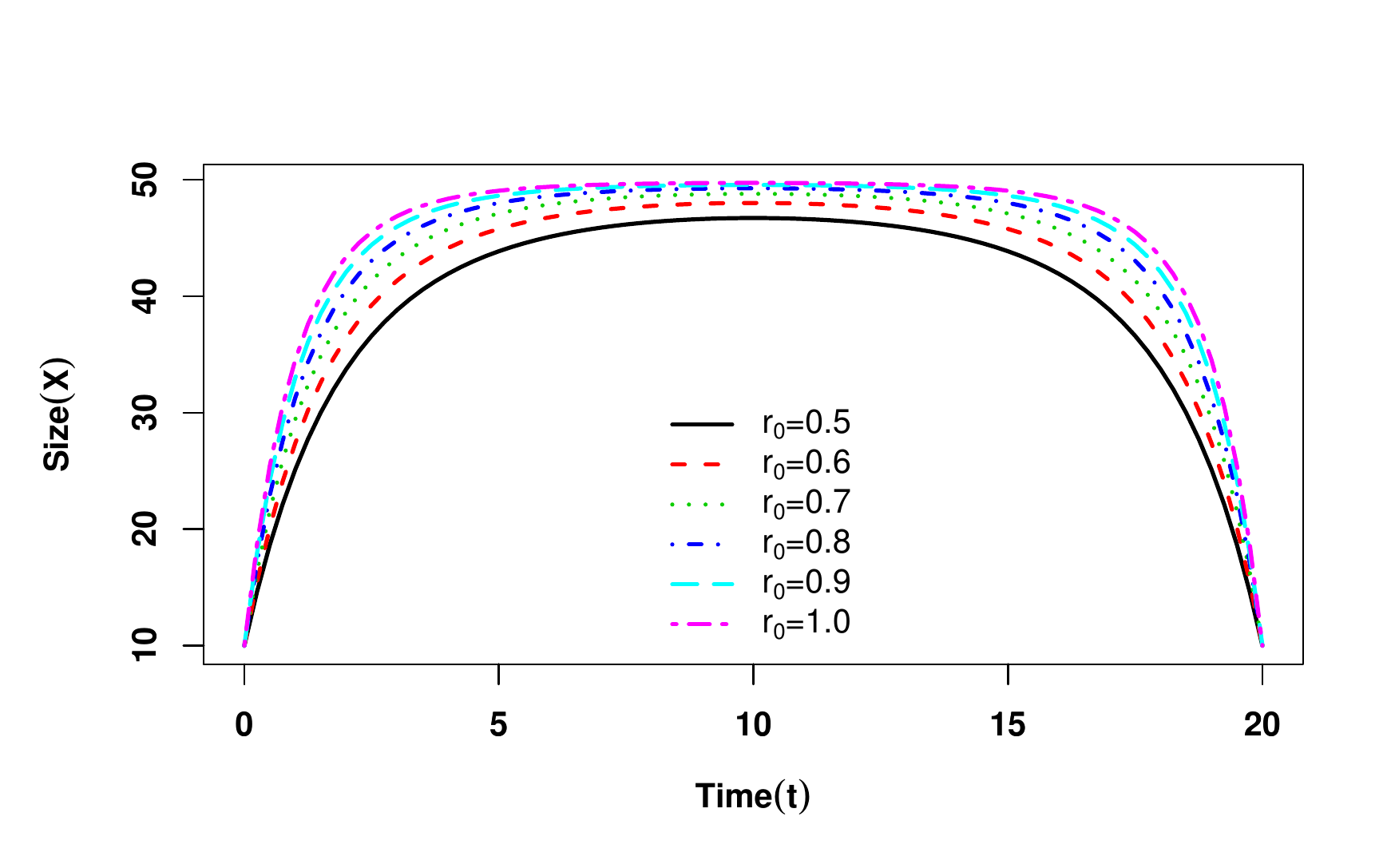}
\caption{$r=r_0(1-ct)$, ($c=0.1$) }
\label{subim26}
\end{subfigure}\hfill
\begin{subfigure}[t]{0.33\textwidth}
    \includegraphics[width=\linewidth]{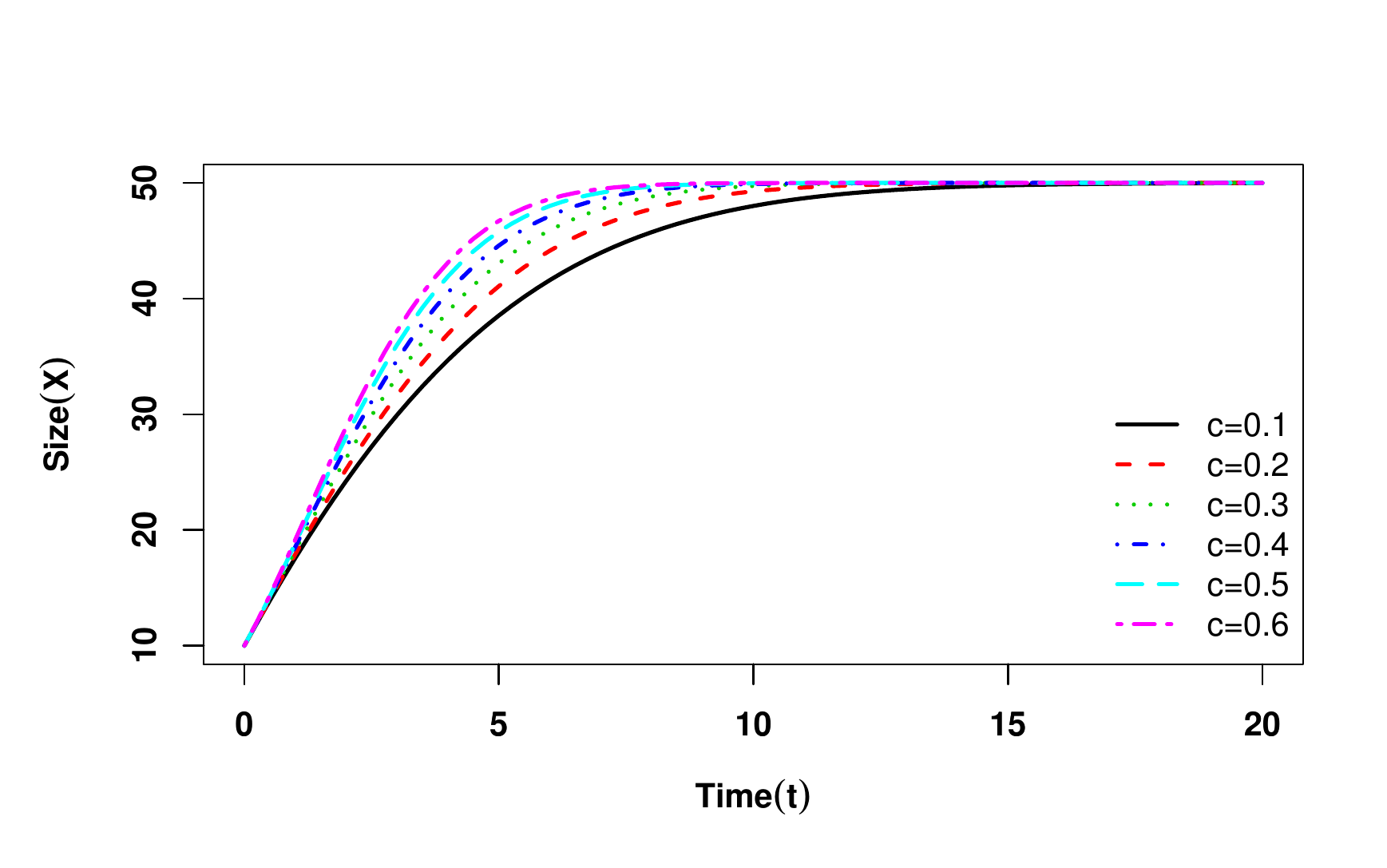}
\caption{$r=r_0(1+ct)$, ($r_0=0.2$) }
\label{subim27}
\end{subfigure}

\begin{subfigure}[t]{0.33\textwidth}
    \includegraphics[width=\linewidth]{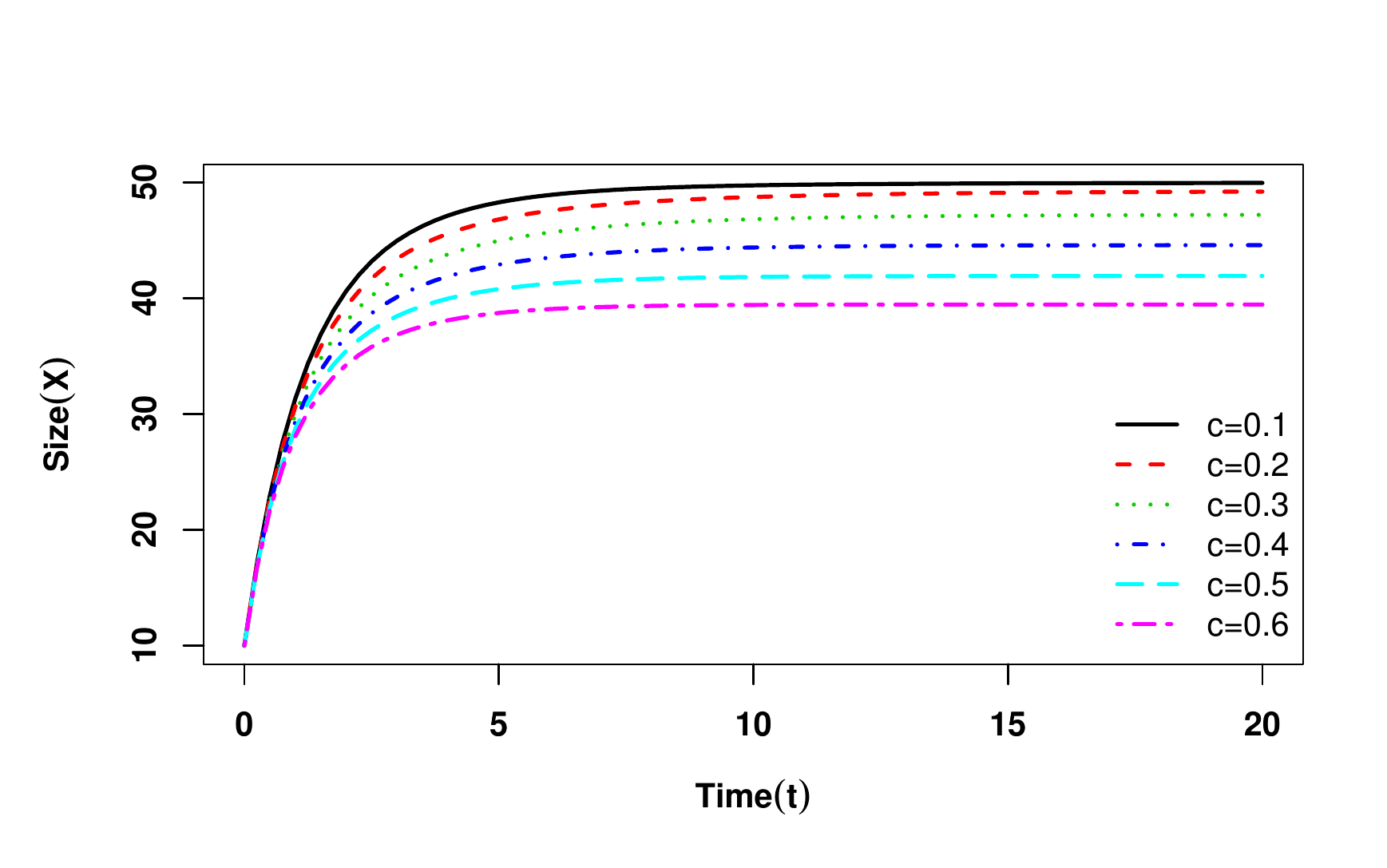}
\caption{$r=r_0e^{-ct}$, ($r_0=0.8$)}
\label{subim28}
\end{subfigure}\hfill
\begin{subfigure}[t]{0.33\textwidth}
    \includegraphics[width=\linewidth]{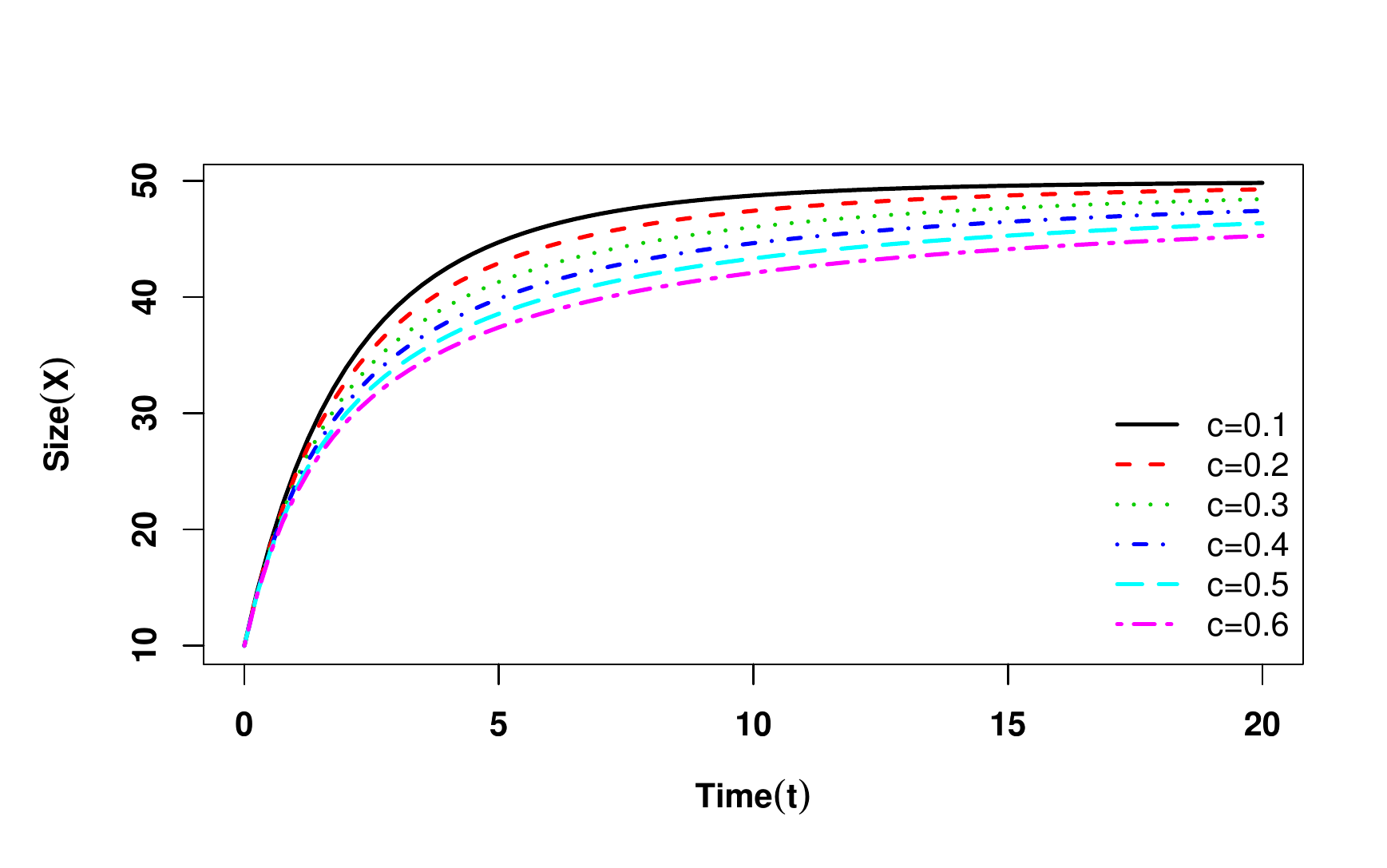}
\caption{$r=\frac{r_0}{1+ct}$, ($r_0=0.5$)}
\label{subim29}
\end{subfigure}\hfill
\begin{subfigure}[t]{0.33\textwidth}
    \includegraphics[width=\textwidth]{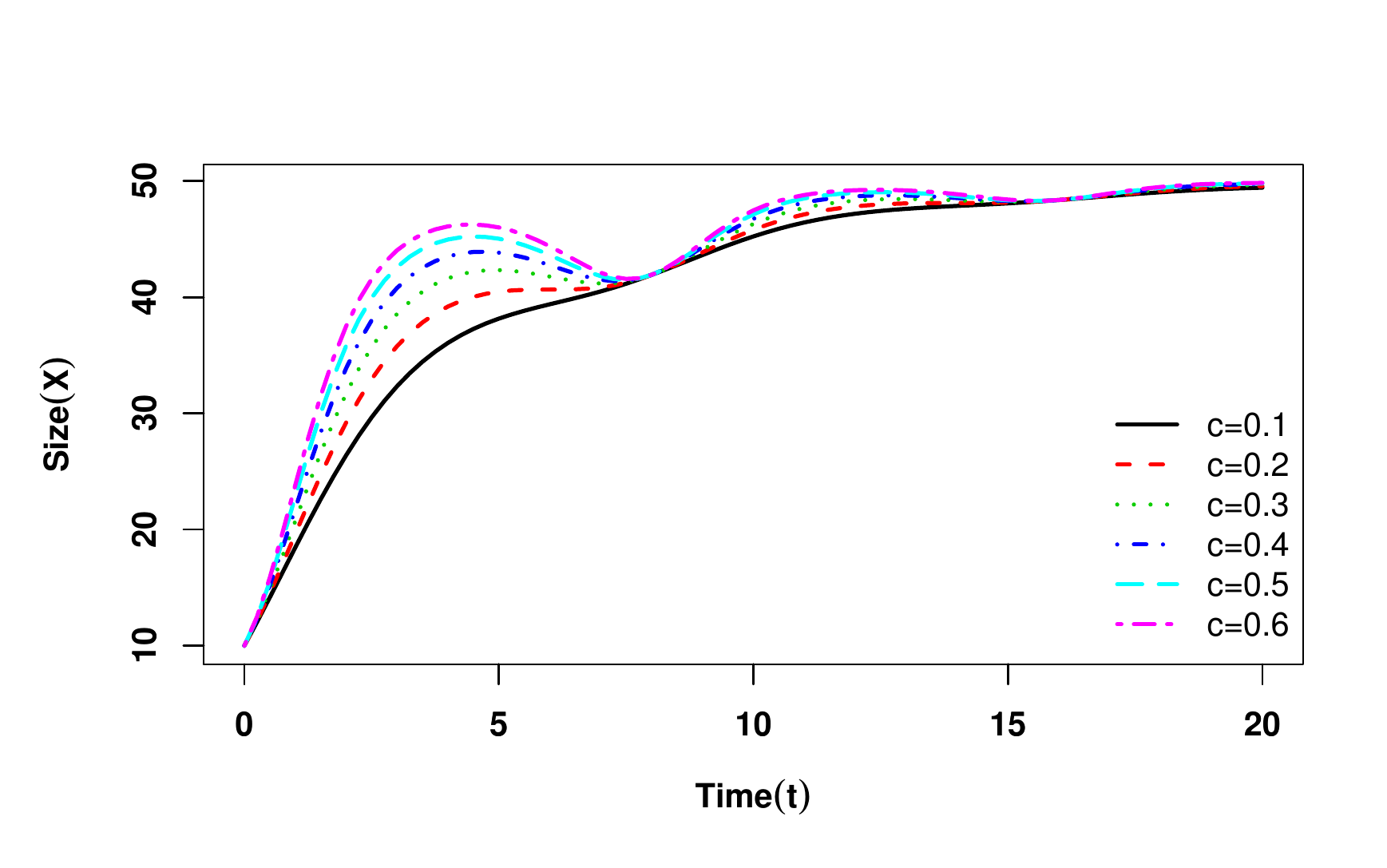}
\caption{$r=r_0+c\sin({\omega t})$, ($r_0=0.2, \omega=\frac{\pi}{4}$)}
\label{subim30}
\end{subfigure}

\begin{subfigure}[t]{0.33\textwidth}
    \includegraphics[width=\textwidth]{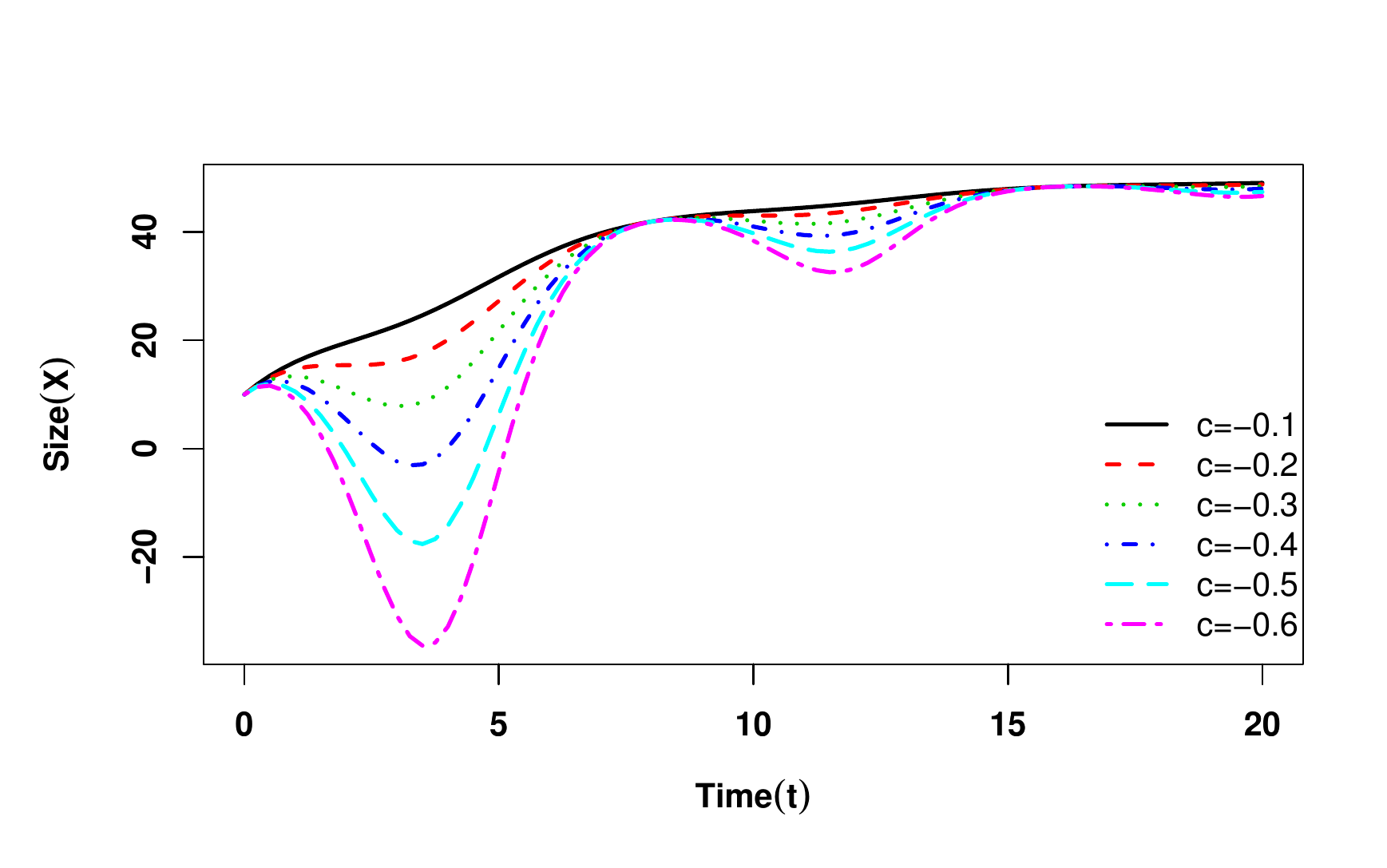}
\caption{$r=r_0+c\sin({\omega t})$, ($r_0=0.2, \omega=\frac{\pi}{4}$)}
\label{subim31}
\end{subfigure}\hfill
\begin{subfigure}[t]{0.33\textwidth}
    \includegraphics[width=\textwidth]{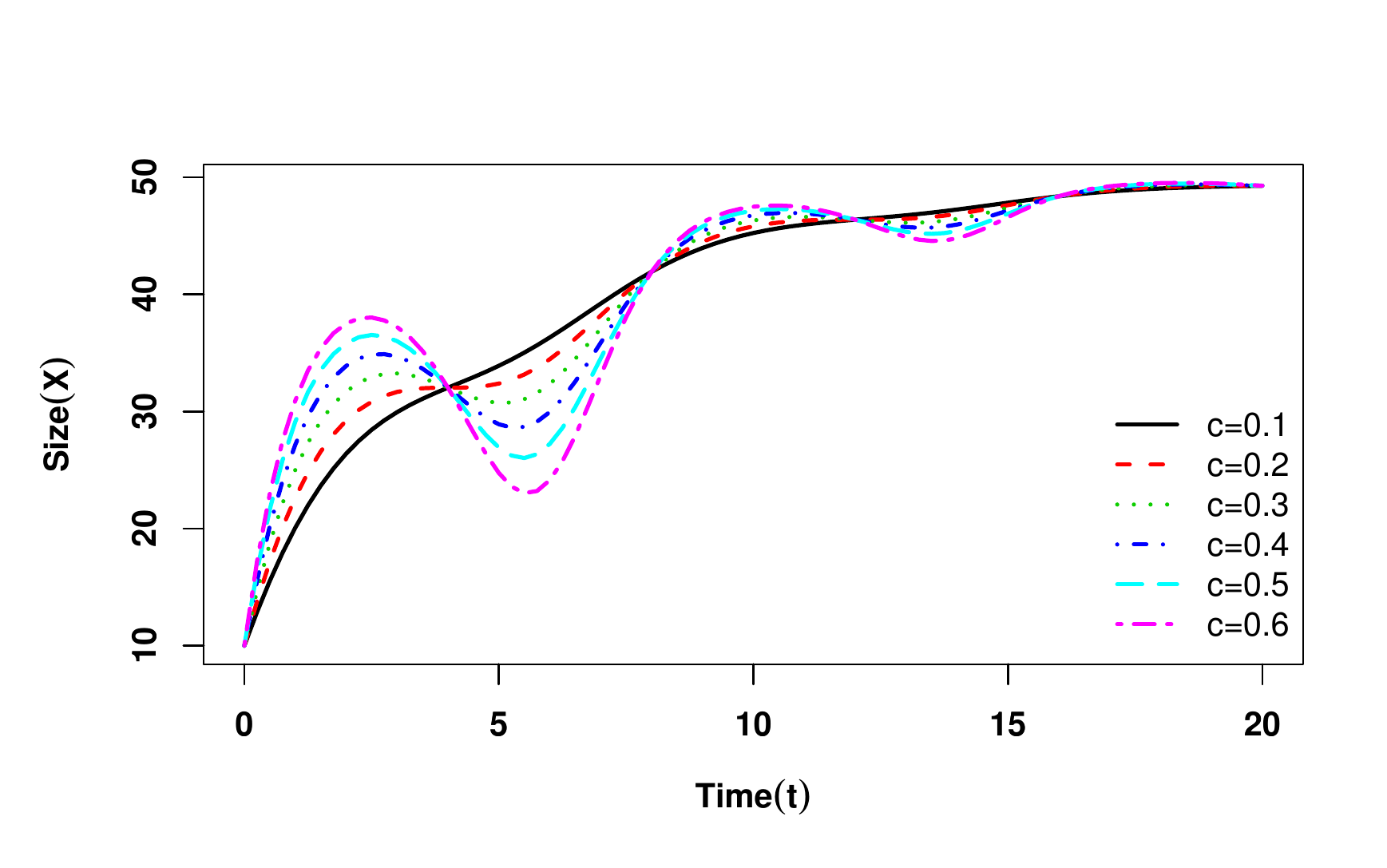}
\caption{$r=r_0+c\cos({\omega t})$, ($r_0=0.2, \omega=\frac{\pi}{4}$)}
\label{subim32}
\end{subfigure}\hfill
\begin{subfigure}[t]{0.33\textwidth}
    \includegraphics[width=\textwidth]{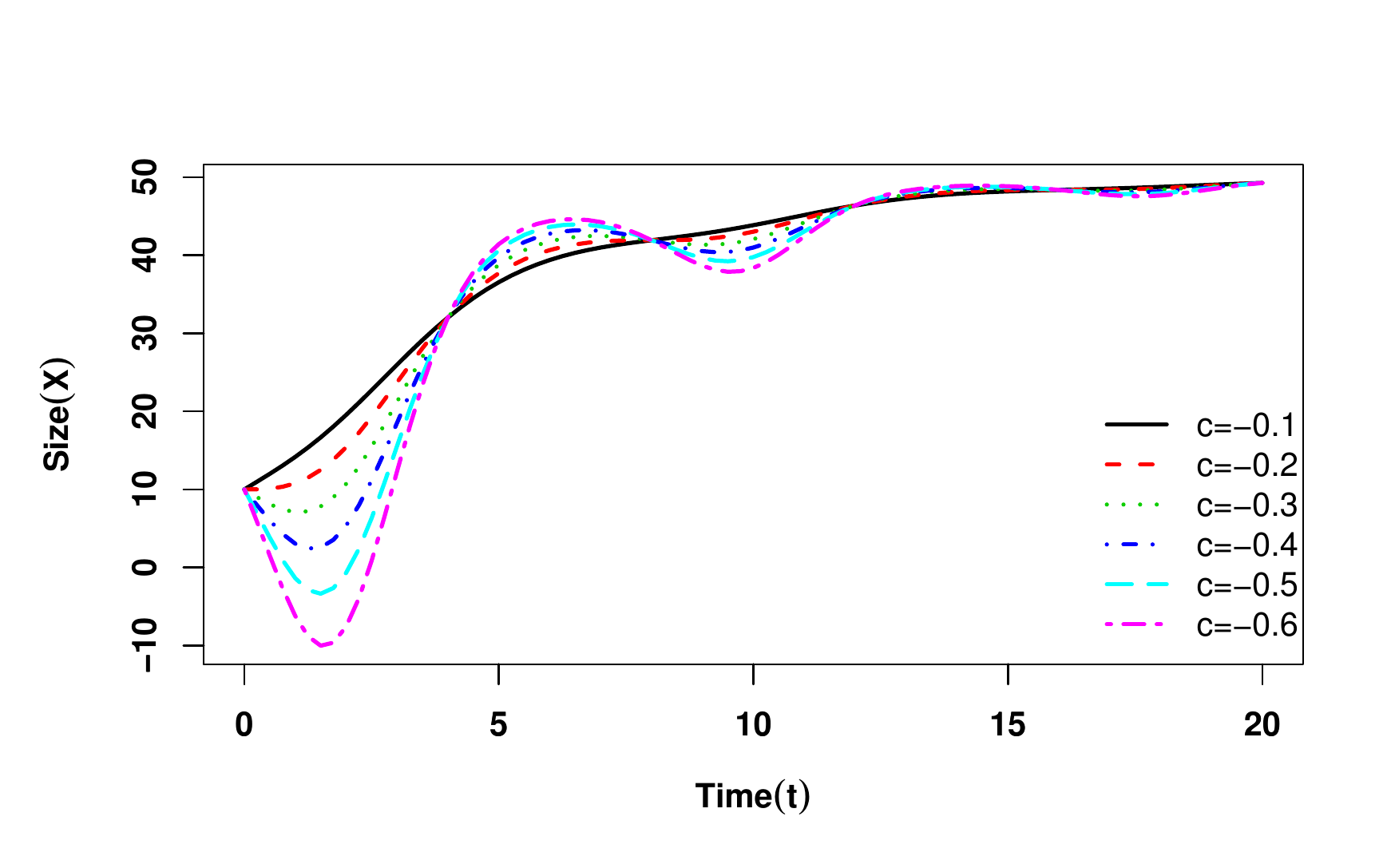}
\caption{$r=r_0+c\cos({\omega t})$, ($r_0=0.2, \omega=\frac{\pi}{4}$)}
\label{subim33}
\end{subfigure}

\begin{subfigure}[t]{0.33\textwidth}
    \includegraphics[width=\linewidth]{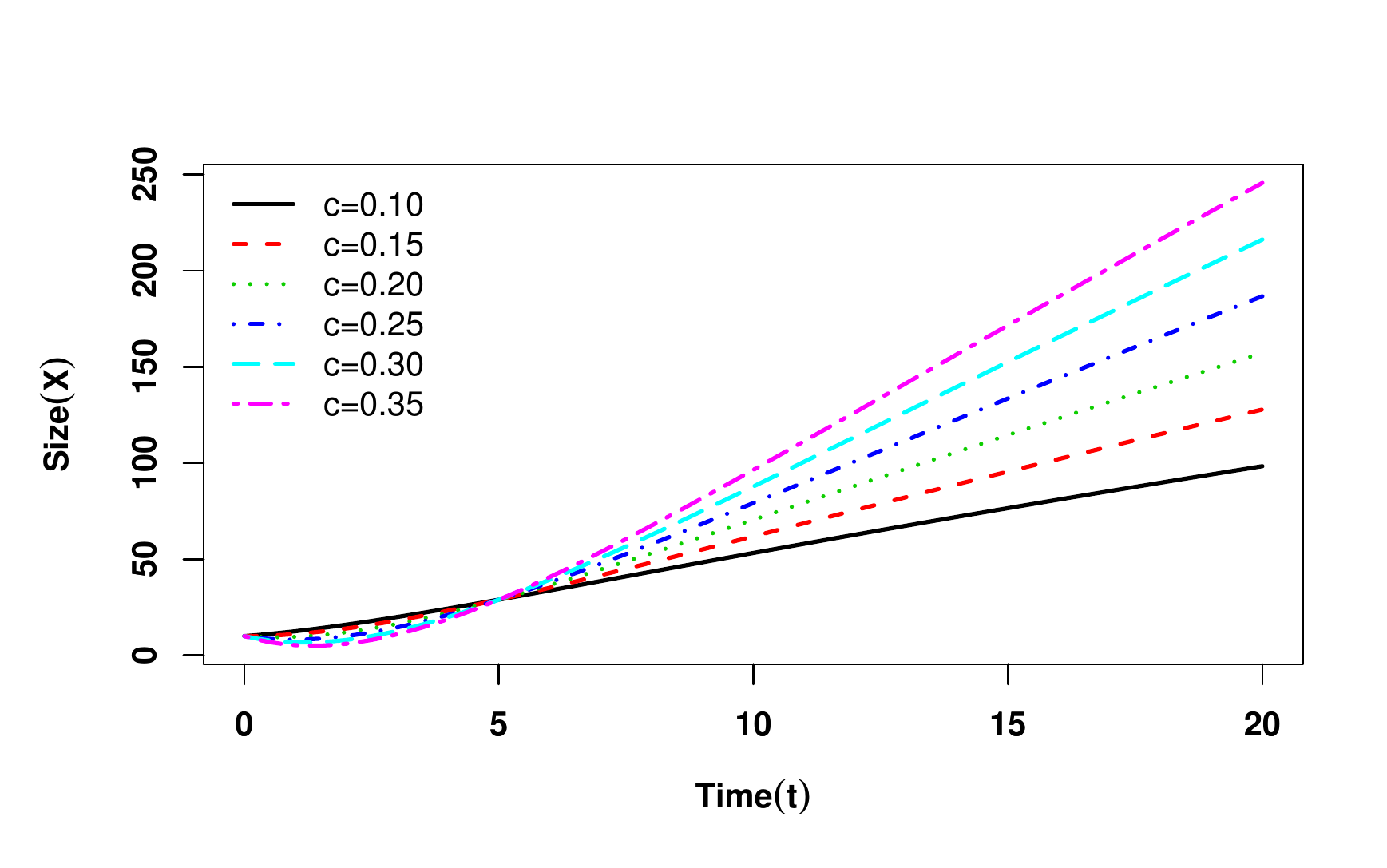}
\caption{$K=K_0(1+ct)$, ($r=0.2,K_0=40$)}
\label{subim34}
\end{subfigure}\hfill
\begin{subfigure}[t]{0.33\textwidth}
    \includegraphics[width=\linewidth]{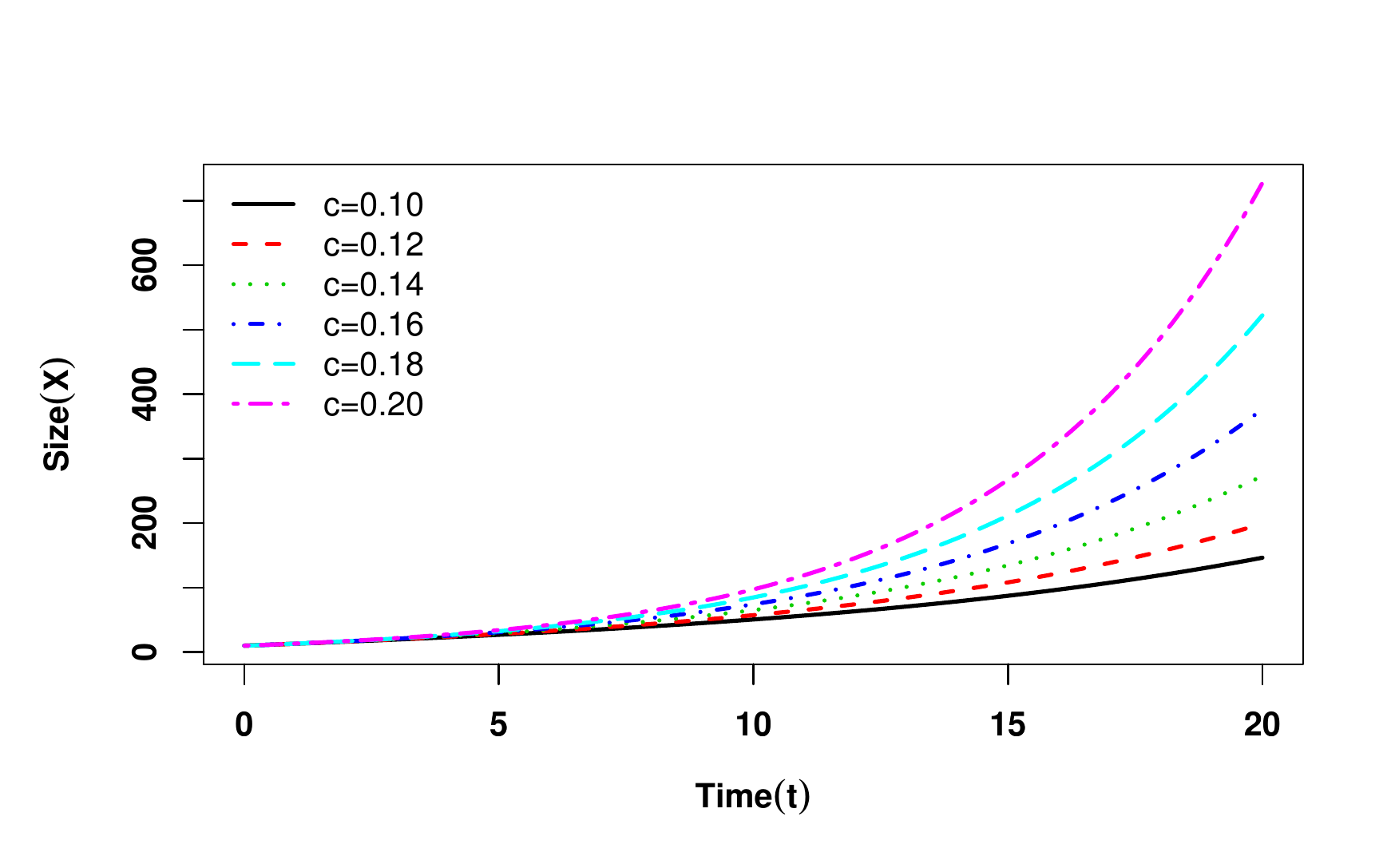}
\caption{$K=K_0e^{ct}$, ($r=0.1,K_0=40$)}
\label{subim35}
\end{subfigure}\hfill
\begin{subfigure}[t]{0.33\textwidth}
    \includegraphics[width=\linewidth]{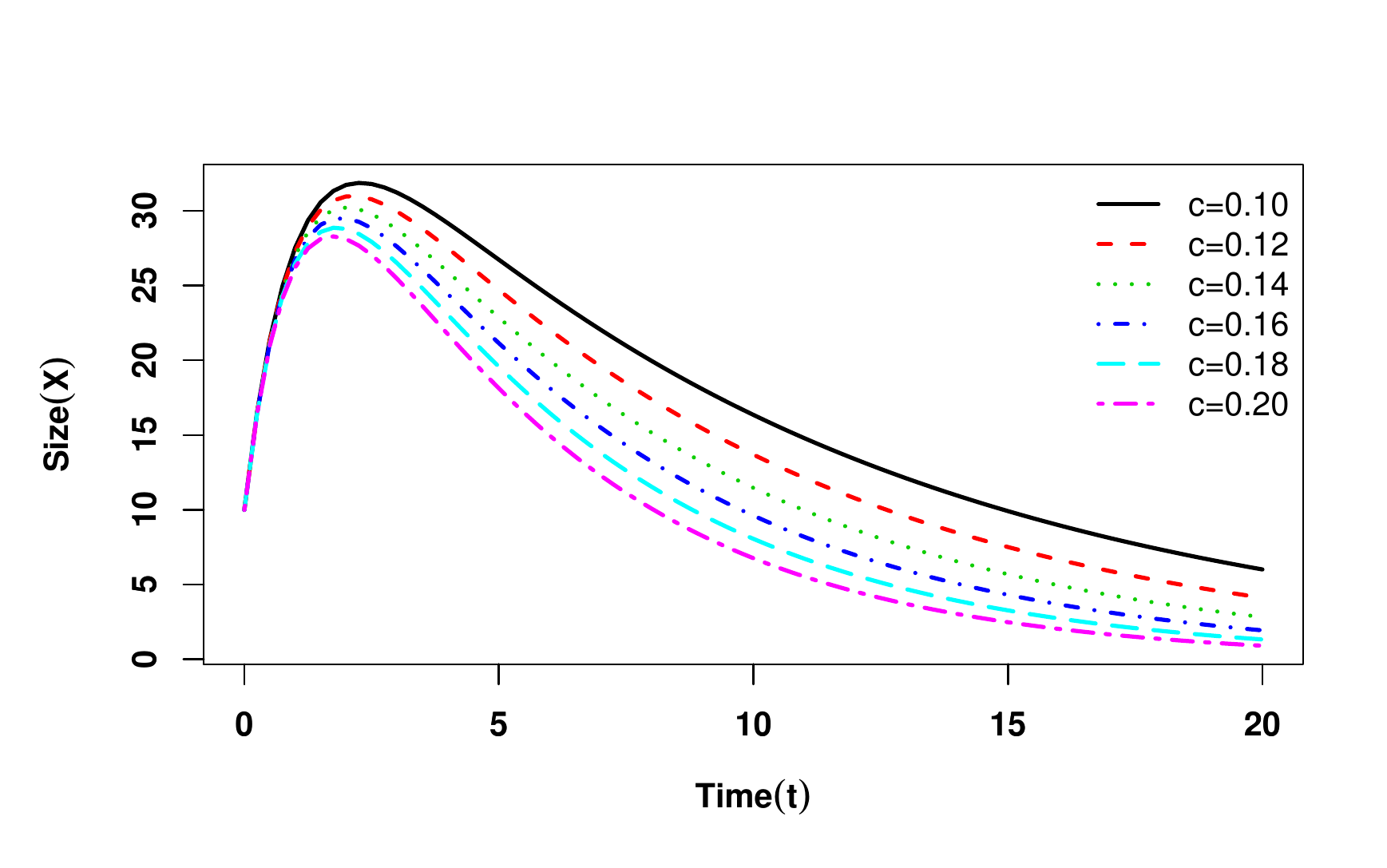}
\caption{$K=K_0e^{-ct}$, ($r=1,K_0=40$)}
\label{subim36}
\end{subfigure}

\caption{Size profile of the confined exponential model for different choices of $r$ and $K$. In the first panel, in Fig. (a) (b) and (c), we consider $r$ as a polynomial function, linearly decreasing and increasing function of time, respectively. In the second panel, in Fig. (d) we consider $r$ as an exponentially decreasing function of time, in Fig. (e) and (f), $r$ varies inversely and periodically with time, respectively. In the third panel,in Fig. (g), (h) and (i), $r$ varies periodically with time. In the forth panel (Fig. (j), (k) and (l)), we consider $K$ as a linearly increasing, exponentially increasing and decreasing function of time, respectively. In all cases, $r_0$ and $K_0$ are initial values of the parameters $r$ and $K$, respectively. $X_0$ is kept fixed at 10.} 
\label{Analytical_not_exists_plot}
\end{figure}

\item $r=r_0(1-ct)$, ($c>0$) :\\
If we take linearly decreasing time dependent function of $r$ in eqn. (\ref{confined_exponential_model}), then it turns out into a new model (given in Table~\ref{table5}; srl.$4$)). This model is very sensitive with respect to the parameter choice of $c$. The size profile first increases and then decreases (Fig.~\ref{subim26}). 

\item $r=r_0(1+ct)$, ($c>0$) :\\ 
If we take linearly increasing time dependent function of $r$ in eqn. (\ref{confined_exponential_model}), then the resulting model has asymptotic size equal to $K$ as of the original model (Fig.~\ref{subim27}). However, the convergence to the asymptotic size of $X(t)$ is much faster for large values of $c$ (Table~\ref{table5}; srl.$5$).

\item $r=r_0e^{-ct}$, ($c>0$) :\\
If $r$ decays exponentially in eqn. (\ref{confined_exponential_model}), then a new growth equation is obtained (Table~\ref{table5}; srl.$6$) whose asymptotic size is less than $K$. But $\displaystyle \lim_{t\to \infty} X(t)$ depends on the value of $c$ (Fig.~\ref{subim28}). As $c$ increases $\displaystyle \lim_{t\to \infty} X(t)$ decreases.

\item $r=\frac{r_0}{1+ct}$, ($c>0$) :\\
If we take $r$ as a inverse function of time in eqn. (\ref{confined_exponential_model}), then it turns into a new model (Table~\ref{table5}; srl.$7$). For $c>0$, $K$ is the asymptotic size and for large $c$, convergence of $X(t)$ is slow (Fig.~\ref{subim29}).

\item $r=r_0+c\sin({\omega t})$, ($\omega>0$) :\\
For this periodic variation in $r$, eqn. (\ref{confined_exponential_model}) turns into a new model in which the asymptotic size goes towards $K$ periodically with reduced amplitude (Fig.~\ref{subim30}, \ref{subim31}) (Table~\ref{table5}; srl.$8$). If we take cosine function instead of sine function, then a new model (Table~\ref{table5}; srl.$9$) is obtained having similar behaviour (Fig.~\ref{subim32}, \ref{subim33}).

\item $K=K_0(1+ct)$, ($c>0$) :\\
If we take linearly increasing time dependent function of $K$ in model (\ref{confined_exponential_model}), then it turns out into a new model (Table~\ref{table5}; srl.$10$), for which asymptotic size is $\infty$ (Fig.~\ref{subim34}) and it goes towards it at a much faster rate for bigger values of $c$.

\item $K=K_0e^{ct}$, ($c>0$) :\\
For this type of variation in $K$ (Table~\ref{table5}; srl.$11$), it is obvious that $X(t)$ diverges to $\infty$ as $t \to \infty$ (Fig.~\ref{subim35}) and at faster rate than the linearly varying $K$.\\

\item $K=K_0e^{-ct}$, ($c>0$) :\\
If we take exponentially decreasing time dependent function of $K$ in model (\ref{confined_exponential_model}), then it turns out into a new model (Table~\ref{table5}; srl.$12$), for which asymptotic size changes into zero (Fig.~\ref{subim36}) and it goes towards $X(t)$ at faster rate for large values of $c$.
\end{enumerate}

\subsection{Nonavailability of explicit expression of ISRP}
In the previous section, we have discussed the growth models in which a parameter varies continuously with time following some specific functional form. It is to be noted that for each of the case, the final differential equation (after replacing $r$ by $r(t)$) can be solved analytically or solutions are available using some special functions. In general this may not be the case and the differential equations must be solved numerically to obtain the size profile. In this section, we deal with few cases in which the analytical expression for the size variable $X(t)$ is not available. 

\begin{figure}[H]
  \begin{subfigure}{9cm}
    \includegraphics[width=9cm]{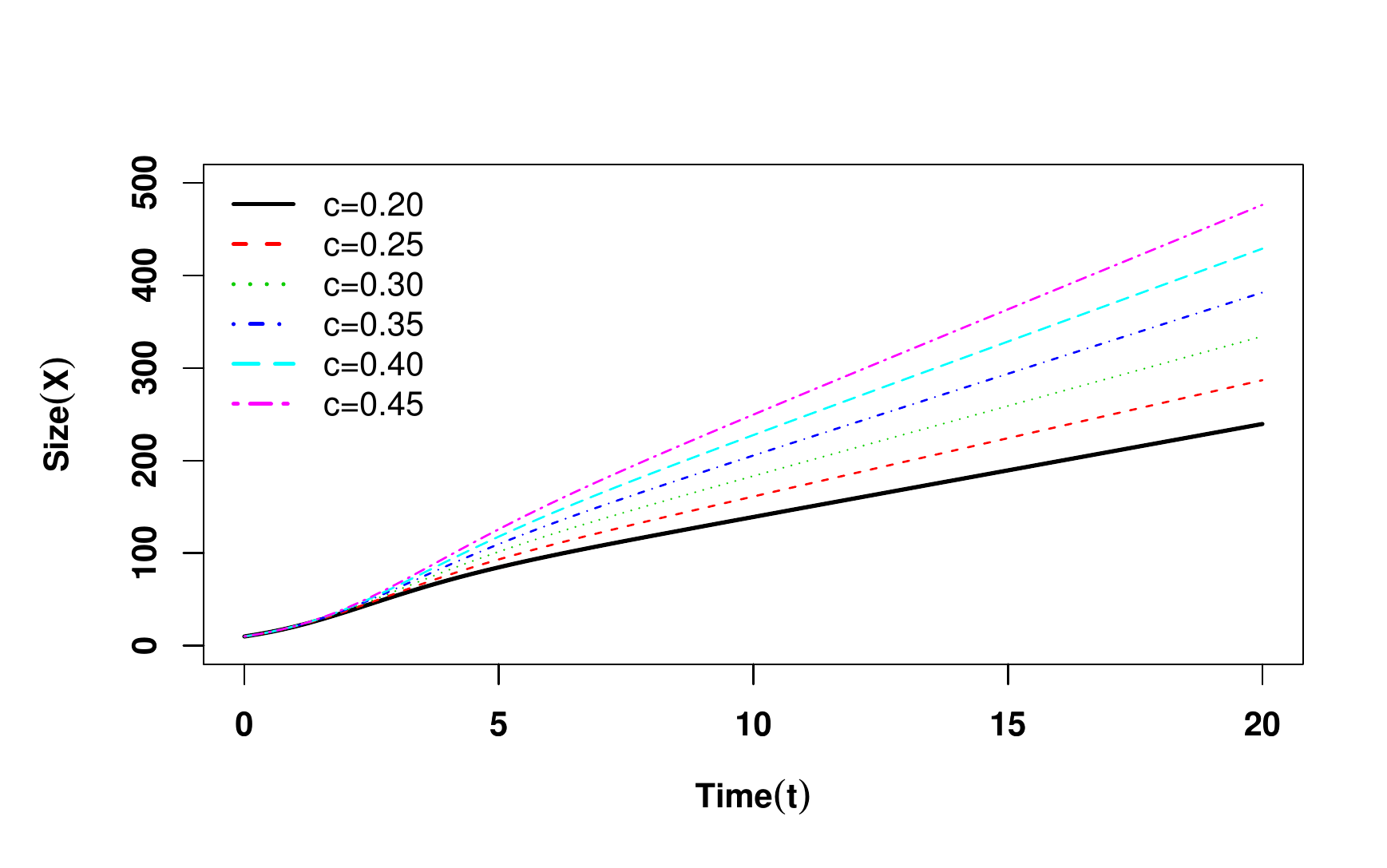}
    \caption{$K=K_0(1+ct)$, ($K_0=50$)}
    \label{subim37}
  \end{subfigure}
  \begin{subfigure}{9cm}
\includegraphics[width=9cm]{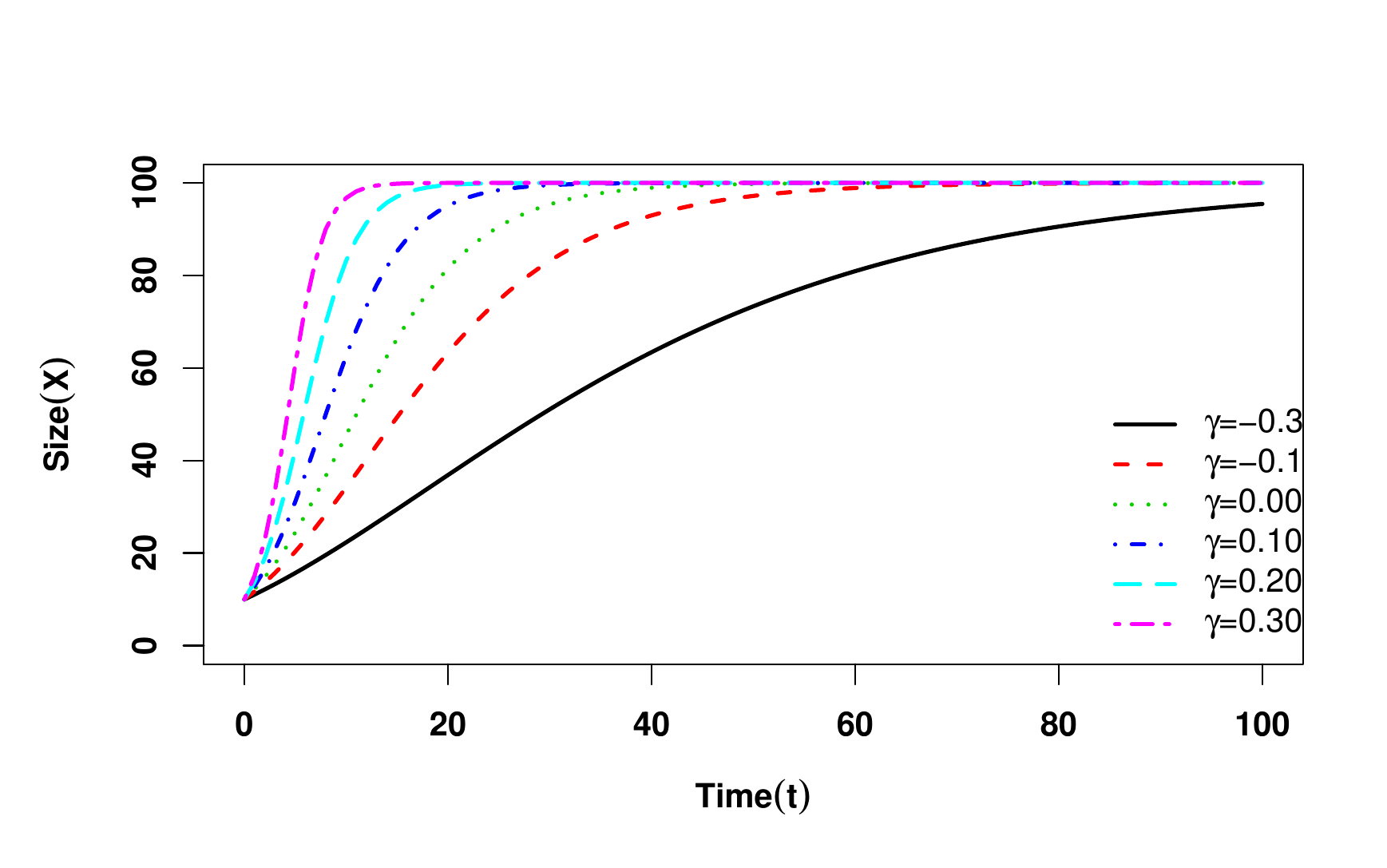}
    \caption{$r=r_0X^{\gamma}$, ($r_0=0.3$)}
    \label{subim38}
  \end{subfigure}
  \begin{subfigure}{9cm}
    \includegraphics[width=9cm]{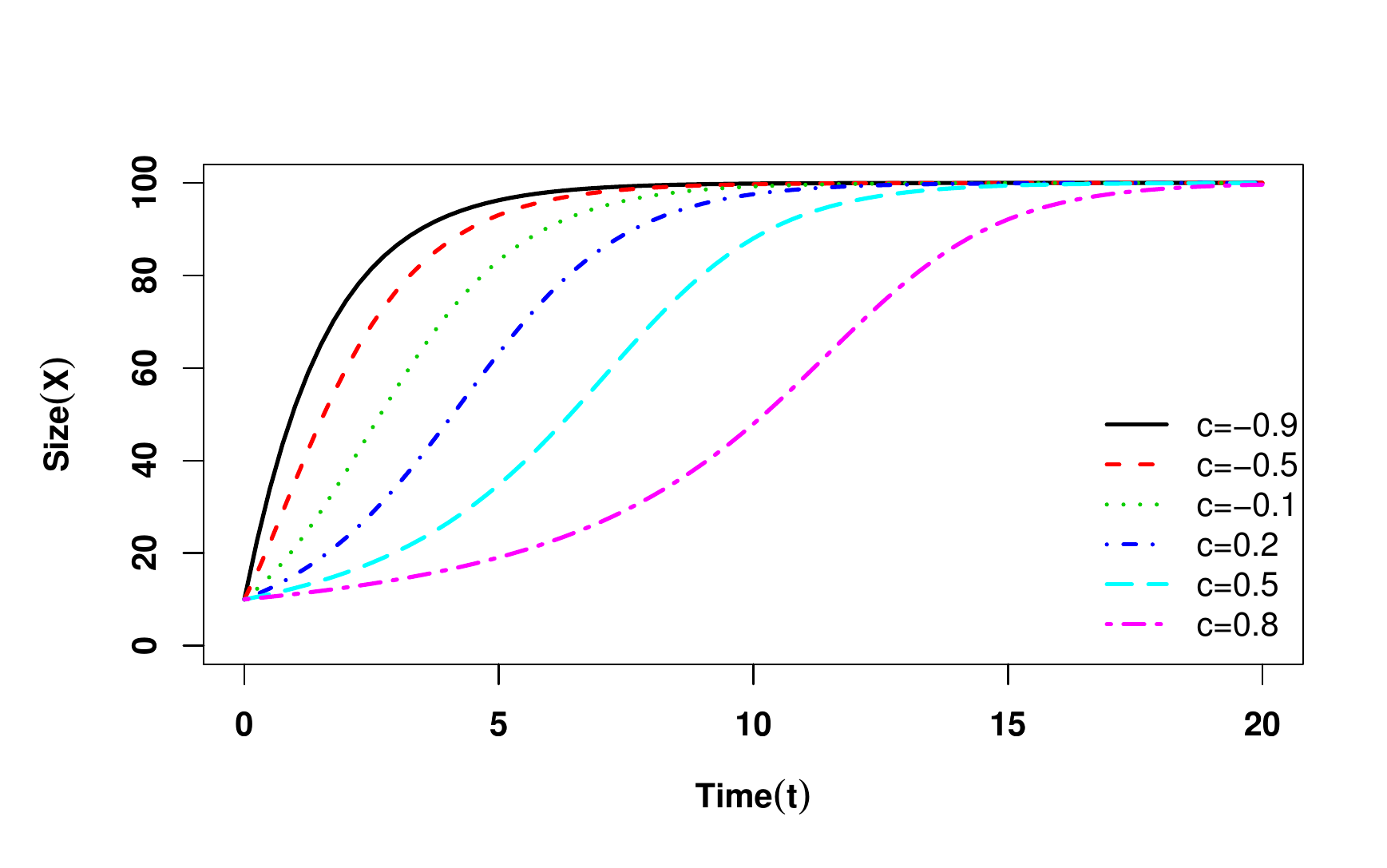}
    \caption{$r=r_0\left(\frac{X}{K}\right)^{c}$, ($r_0=0.8$), $\theta>0$}
    \label{subim39}
  \end{subfigure}
  \begin{subfigure}{9cm}
    \includegraphics[width=9cm]{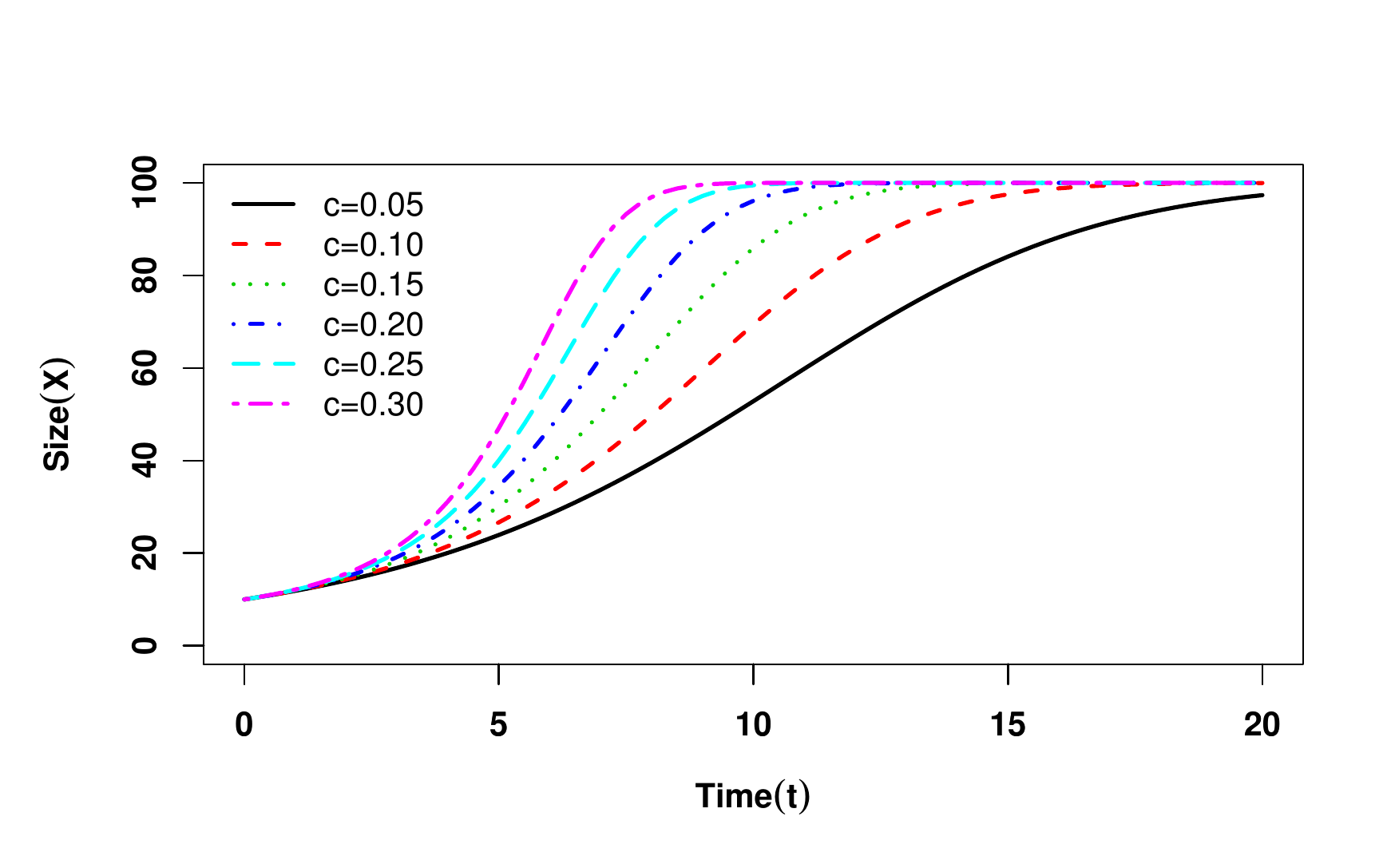}
    \caption{$r=r_0e^{ct}$, ($r_0=0.2$)}
    \label{subim40}
  \end{subfigure}
\caption{Size profile plot for continuously varying parameter in logistic and theta-logistic model. In the upper panel, in Fig. (a), we consider linearly increasing function of $K$ in Logistic model and in Fig. (b), we consider $r$ as a polynomial function of size in theta-logistic model. In lower panel, in Fig. (c), we consider $r$ as a polynomial function of $\left(\frac{X}{K}\right)$ and in Fig. (d), we consider $r$ as exponentially increasing function of time in theta-logistic model. For Fig. (a), we consider $r=1$ and for Fig. (b) to (d), we consider $K=100$. For all cases, $X_0 = 10$ is kept fixed.}
\label{Figure6}
\end{figure}

\begin{enumerate}
\item $K=K_0(1+ct)$, ($c>0$) :\\
If the carrying capacity increases linearly with time in eqn. (\ref{eqn:logistic_model}), then the analytical solution for $X(t)$ does not exists (Table~\ref{table6}; srl.$1$). From Fig.(\ref{subim37}), we conclude that $X(t) \to \infty$ as $t \to \infty$ for all values of $c>0$.

\item $r=r_0X^{\gamma}$ :\\
If we take $r$ as a polynomial function of density ($X$) in eqn. (\ref{eqn:theta_logistic_model}), then it turns into Co-operation model \citep{BHOWMICK2015} (Table~\ref{table6}; srl.$2$) with asymptotic size $K$. For large values of $\gamma$, $X(t)$ converges to $K$ at faster rate (Fig.(\ref{subim38}))  .

\item $r=r_0\left(\frac{X}{K}\right)^{c}$, $\theta>0$  :\\
If we take $r$ as a polynomial function of $\frac{X}{K}$ in eqn. (\ref{eqn:theta_logistic_model}), then the equation converts into Marusic-Bajzer model \citep{MARUSIC1993} (Table~\ref{table6}; srl.$3$). The asymptotic size in the revised model remains $K$ but for bigger values of $c$, $X(t)$ goes towards $K$ at slower rate (Fig.(\ref{subim39}))  .

\item $r=r_0e^{ct}$, $c>0$ :\\
If we take $r$ as a exponential growing function of time in eqn. (\ref{eqn:theta_logistic_model}), then this model turns into a new model  (Table~\ref{table6}; srl.$4$) whose asymptotic size is $K$ and for bigger values of $c$ $X(t)$ goes towards $K$ at faster rate (Fig.(\ref{subim40})).
\end{enumerate}

\end{document}